\documentclass[aps,pra,letterpaper,twocolumn,floats,superscriptaddress,tighten,eqsecnum,floatfix]{revtex4}
\usepackage{amssymb}
\usepackage{amsbsy}
\usepackage{amsmath}
\usepackage{graphicx}
\usepackage{graphics}
\usepackage{setspace}
\usepackage{array}
\usepackage{color}
\usepackage{fontenc}
\usepackage{textcomp}
\usepackage{rotating}
\usepackage{bm}

\DeclareMathOperator{\im}{Im}

\DeclareMathOperator{\sgn}{sgn}

\DeclareMathOperator{\Ai}{Ai}

\newcommand{\e}{\varepsilon}

\newcommand{\vex}[1]{\bm{\mathrm{#1}}}

\newcommand{\pup}[1]{{\scriptscriptstyle{({#1})}}}

\newcommand{\ket}[1]{| {#1} \rangle}
\newcommand{\bra}[1]{\langle {#1} |}

\newcommand{\bsub}{\begin{subequations}}
\newcommand{\esub}{\end{subequations}}

\newcommand{\cc}{\mathsf{c}}

\begin{document}
\title{Non-Markovian dephasing of disordered, quasi-one-dimensional fermion systems}
\author{Seth M.\ Davis}
\affiliation{Department of Physics and Astronomy, Rice University, Houston, Texas 77005, USA}
\author{Matthew S.\ Foster}
\affiliation{Department of Physics and Astronomy, Rice University, Houston, Texas 77005, USA}
\affiliation{Rice Center for Quantum Materials, Rice University, Houston, Texas 77005, USA}
\date{\today\\}

\newcommand{\be}{\begin{equation}}
\newcommand{\ee}{\end{equation}}
\newcommand{\bea}{\begin{eqnarray}}
\newcommand{\eea}{\end{eqnarray}}
\newcommand{\h}{\hspace{0.30 cm}}
\newcommand{\vs}{\vspace{0.30 cm}}
\newcommand{\n}{\nonumber}
\begin{abstract}	
As a potential window 
on transitions out of the ergodic, many-body-delocalized phase,
we study the dephasing of weakly disordered, quasi-one-dimensional fermion systems 
due to a diffusive, non-Markovian noise bath. Such a bath is self-generated by the fermions, 
via inelastic scattering mediated by short-ranged interactions. 
The ergodic 
phase can be defined by the nonzero dephasing rate, which makes transport incoherent and classical 
on long length scales.  
We calculate the dephasing of weak localization perturbatively through second order in the bath coupling,
obtaining a short-time expansion. However, no well-defined dephasing rate can be identified, 
and the expansion breaks down at long times. 
This perturbative expansion is \emph{not} stabilized by including a mean-field Cooperon ``mass'' (decay rate),
signaling a failure of the self-consistent Born approximation. 
We also consider a many-channel quantum wire where short-ranged, spin-exchange interactions coexist with 
screened Coulomb interactions. 
We calculate the dephasing rate, treating the short-ranged interactions 
perturbatively and the Coulomb interaction exactly. The latter provides a physical infrared 
regularization that stabilizes perturbation theory at long times, 
giving the first controlled calculation of quasi-1D dephasing due to diffusive noise. 
At first order in the diffusive bath coupling, we find an enhancement of the dephasing rate,
but at second order we find a \textit{rephasing} contribution.
Our results differ qualitatively from those obtained via self-consistent calculations 
commonly employed in higher dimensions.
Our results are relevant in two different contexts. First, 
in the search for precursors to many-body localization in the ergodic phase of an isolated many-fermion system. 
Second, our results provide a mechanism for the enhancement of dephasing at low temperatures in spin 
SU(2)-symmetric quantum wires, beyond the Altshuler-Aronov-Khmelnitsky result. The enhancement is possible 
due to the amplification of the triplet-channel interaction strength, and 
provides an additional physical mechanism that could contribute to 
the experimentally
observed low-temperature saturation of the dephasing time.
\end{abstract}
\maketitle

\tableofcontents

\section{Introduction}

Inelastic collisions between electrons tend to destroy quantum phase coherence in a phenomenon called \textit{dephasing}. 
Dephasing is a key physical process underlying the transition between the quantum and classical transport regimes in many-body 
fermion systems and thus is central to modern efforts in condensed matter and quantum information to understand and exploit
macroscopic quantum phenomena. 

It was understood in the 1980s that quantum interference effects in electronic systems induced by weak quenched disorder 
are governed by the dephasing timescale $\tau_{\phi}$ \cite{AAK,LeeRama,AA85}.
This timescale determines the infrared cutoff for the weak (anti) localization correction to transport,
which diverges in one or two spatial dimensions in the absence of dephasing.
The dephasing rate can be measured through the temperature dependence of the conductance \cite{AA85,AAG}.  

The dephasing rate for inelastic electron scattering mediated by dynamically screened Coulomb interactions was calculated 
exactly by Altshuler, Aronov, and Khmelnitsky (AAK) \cite{AAK}, who obtained a $\tau_\phi \sim T^{-2/3}$ power law in temperature ($T$) 
for quasi-one-dimensional (quasi-1D) many-channel wires. 
Although this result has been well-confirmed experimentally \cite{AAG}, measurements observing an anomalous low-temperature 
saturation of $\tau_\phi$ sparked a decade of controversy
\cite{Webb1,Webb2,Webb3,GZ,Natelson,AAG,VDelftPartI,VDelftPartII,ICS,KondoTheory1,KondoTheory2,KondoMeasure1,KondoMeasure2}.
Plausible explanations for the saturation include additional phase-breaking due to Kondo impurities \cite{KondoTheory1,KondoTheory2}.
The role of \emph{itinerant} electron spin-exchange scattering and its effect on dephasing \cite{NZA} in these quasi-1D wires 
was not extensively investigated at the time.

The theoretical challenges imposed by many body localization (MBL) \cite{BAA06,Gornyi05,NH,GopalParam} invite us to revisit some of these questions. 
The MBL hypothesis proposes that an interacting, disordered quantum system can undergo a nonzero-temperature transition from
the semiclassical ergodic (metallic) phase into an insulating state that fails to self-thermalize. 
In the MBL phase, local operators are long-lived and quantum coherence is not destroyed by 
dephasing \cite{BAA06,NH,GopalParam}. 
Understanding the nuances of dephasing in the ergodic phase could uncover precursors to MBL, or
even yield an analytical tool for studying the ergodic-to-MBL transition \cite{Liao18,Han18}.
Recent work has raised concerns about the feasibility of accessing this transition numerically \cite{Prosen19,MBLDefense,Panda19,PiotrSierant},
which places a renewed urgency upon identifying analytical approaches that are not limited to small system sizes. 

In this work, we revisit quantum coherence in quasi-1D fermion transport, and focus specifically on dephasing due
to short-ranged inelastic scattering. This is relevant for neutral ultracold atomic fermion systems that could be platforms for 
MBL realization \cite{Doggen20}. 
Furthermore, 
in spin SU(2)-symmetric quantum wires, inelastic electron-electron scattering is mediated by 
the combination of both
short-ranged spin-triplet exchange- 
and
Coulomb-interactions \cite{NZA}. 
MBL has been primarily investigated for fermion systems with short-ranged interactions \cite{BAA06,Gornyi05,Laumann14}. 
Inelastic scattering due to short-ranged interactions gives rise to a \emph{strongly non-Markovian}, diffusive noise kernel in the 
ergodic phase \cite{Liao17,Liao18}. By contrast, the exact solution for $\tau_\phi$ obtained by AAK relies crucially 
on the Markovian nature of the noise bath that arises from Coulomb interactions \cite{AAK}. 
As we show in this work, the Markovian case is drastically simpler than the generic case. 
An additional reason to revisit dephasing due to spin exchange interactions is the well-known but poorly understood enhancement
of the triplet channel interaction in the theory of the zero-temperature Anderson-Mott metal-insulator transition (MIT) 
\cite{BK,PunnooseFinkelstein,50YearsFinkel,QPTKravchenko}. 
An enhancement of spin exchange interactions could lead to an important role of non-Markovianity near a MIT.

The problem of dephasing weak localization by a diffusive bath is equivalent to solving a strongly-coupled, auxiliary 
quantum field theory. The upper critical spatial dimension of this theory is $d = 4$, and previous work 
using a $d = 4 - \epsilon$ expansion identified a nontrivial critical point that could signal a failure
of dephasing \cite{Liao18}. The critical point obtains from vertex corrections that are 
not captured by the standard self-consistent Born approximation (SCBA) \cite{CS,AAG,AleinerBlanter,Liao18}.
A key goal of this work is to test the veracity of the SCBA and other self-consistent approximations.

Here we present two different calculations for the dephasing of weak localization 
in a quasi-1D wire due to a diffusive bath. In both cases, we consider ``order one'' strength interactions 
and weak disorder \cite{AAK,LeeRama,AA85,AAG}. This is in contrast to the 
strongly disordered, 
weakly interacting
limit considered by Basko, Aleiner, and Altshuler (BAA) \cite{BAA06}. 
The weak-disorder limit minimizes the ``bad metal'' regime of the ergodic phase discussed by BAA, 
contracting the \emph{low-temperature} window in which the putative ergodic-to-MBL transition could occur \cite{Liao17}.
Our assumptions of weak disorder and order-one strength interactions are also in contrast 
with those of Ref.~\cite{Gornyi05}, which predicted an intermediate ``power-law hopping'' (PLH) regime between the MBL insulator and
the weakly localized metal in quasi-1D systems. As noted in Ref.~\cite{Gornyi05}, the width of the PLH regime collapses when
the short-ranged interaction strength becomes of order one. This opens up the possibility of a direct transition between
the weakly localized metal and the MBL phase. The full description of this putative transition goes well beyond
what we consider in this paper, as it would entail the calculation of both higher-order quantum conductance
corrections \emph{and their dephasing} due to the non-Markovian bath. 
Given the strongly coupled nature of the dephasing problem due to the diffusive bath \cite{Liao18} 
in the weak localization regime studied here, the crossover to, or even the existence of, the PLH regime 
for weaker interactions are not questions we are yet prepared to tackle.

First, for an isolated fermion system with short-ranged interactions, we calculate the dephasing through 
second order in the bath coupling, expanding about the un-dephased Cooperon. 
We obtain a 
short-virtual-time
expansion, unplagued by divergences (a feature unique to 1D). 
However, no well-defined dephasing rate can be identified, and the expansion breaks down at long times. 
This perturbative expansion is \emph{not} stabilized by including a mean-field Cooperon ``mass'' (decay rate),
signaling a failure of the SCBA. 
Although the expansion breaks down at long times, it contains interesting features; 
we find that the second-order term in the expansion has a positive sign and actually works \textit{against} dephasing. 
We call such a term \textit{rephasing}. 
This calculation demonstrates that the long-time behavior due to purely diffusive dephasing 
cannot be accessed perturbatively.

Second, we consider a many-channel quantum wire where short-ranged, spin-exchange interactions coexist with 
screened Coulomb interactions. 
We calculate the dephasing rate, treating the non-Markovian diffusive bath perturbatively and the 
Markovian Coulomb bath exactly via an extension of the AAK technique \cite{AAK}.
The latter provides a physical infrared regularization that, unlike the SCBA, stabilizes perturbation theory at long times. 
The expansion parameter is the dimensionless ratio of the two bath coupling strengths. At first order, the diffusive bath enhances the Markovian AAK dephasing rate. 
At second order, however, we again find a \textit{rephasing} contribution. 
Taken together with the short-time expansion result, this suggests that higher-order terms 
could have important effects in the purely diffusive limit, capable of slowing or even arresting dephasing. 
This expansion provides the first controlled calculation of the dephasing effects due to a diffusive bath
for a quasi-1D system. Our results disagree qualitatively with self-consistent schemes, 
commonly employed in higher dimensional dephasing calculations,
which we show give the incorrect dependence on the bath coupling strength. 
In particular, we show that self-consistent calculations incorrectly predict a suppression of the 
effects of the diffusive bath in the strong-coupling limit. 

Finally, we also describe how the low-temperature enhancement of the spin-triplet interaction strength \cite{BK,PunnooseFinkelstein,50YearsFinkel}
can translate into an enhancement of the AAK dephasing law $\tau_\phi \sim T^{-2/3}$ \cite{AAK},
providing a new mechanism for the apparent saturation of phase-breaking in quantum wires 
\cite{Webb1,Webb2,Webb3,GZ,Natelson,AAG,VDelftPartI,VDelftPartII,ICS,KondoTheory1,KondoTheory2,KondoMeasure1,KondoMeasure2}.

Our calculation uses nonstandard techniques and applies generally to any set of coexisting Markovian and non-Markovian noise baths, 
and so we give a pedagogical presentation.

\subsection{Outline}

This paper is organized as follows. 
Sec.~\ref{BasicDephasing} introduces the basics of dephasing and reviews the AAK solution for the Markovian case.
In Sec.~\ref{DiffusiveDephasing}, we perturbatively study dephasing due to a purely diffusive bath.
We present our results for coexisting diffusive and Coulomb (Markovian) baths in 
Sec.~\ref{CoexistingDephasingResults}, and discuss their relevance to understanding the purely diffusive limit. 
Sec.~\ref{CalculationDetails} provides an overview of the dual-bath dephasing calculation. 
This calculation exploits the Airy functions used to solve the quasi-1D Markovian problem exactly \cite{AAK}.
The series expansion for the dephasing rate is expressed in terms of amplitudes that involve sums 
of integrals over products of Airy functions; these integrals are ultimately calculated numerically. 
Finally, we discuss our results in Sec.~\ref{Conclusions}, 
including their possible implications for MBL physics and for the apparent saturation of dephasing in quantum wires.

Various technical details are relegated to Appendices. 
Appendix~\ref{Correlators} collects Gaussian correlator results for vertex operators that appear throughout this work. 
Appendix~\ref{MoreDiffusionDetails} provides additional details for the pure diffusive bath calculation in Sec.~\ref{DiffusiveDephasing}.
Appendices~\ref{MoreCoexistingDetails} and \ref{DiagramFolding} present details for the dual-bath calculation summarized in 
Secs.~\ref{CoexistingDephasingResults} and \ref{CalculationDetails}.
Appendix~\ref{MarkovianPT} applies perturbative techniques to the well-understood screened Coulomb limit for the sake of comparison.
Appendix~\ref{SeriesAcceleration} explains a ``series acceleration'' technique used to efficiently sum the Airy function amplitudes
that arise in the dual-bath calculation. 
Finally, Appendix~\ref{FieldTheory} explains an alternative field theory approach to the Markovian and non-Markovian dephasing problems,
which was exploited in the $(4-\epsilon)$ expansion calculation for the diffusive bath in Ref.~\cite{Liao18}. 
Here we highlight mathematical differences between generic dephasing and the Markovian limit. 
We also show how the AAK result can be derived as an infinite-order diagrammatic resummation.

\section{Dephasing of weak localization (review)}\label{BasicDephasing}

The weak (anti)localization correction to the dc conductivity is determined by the 
Cooperon, a propagator defined via the stochastic equation of motion
\cite{AAK,CS,AleinerBlanter,Liao17}
\begin{align}\label{CooperStoch}
	&\,
	\left\{
	\partial_\eta
	-
	\frac{D}{2}
	\nabla^2
	+
	\frac{i}{2}
	\left[
	\phi_{\text{cl}}\left(t + \frac{\eta}{2},\vex{x}\right)
	-
	\phi_{\text{cl}}\left(t - \frac{\eta}{2},\vex{x}\right)
	\right]
	\right\}
\nonumber\\
	&\,
	\qquad
	\times
	\cc^t_{\eta,\eta'}(\vex{x},\vex{x'})
	=
	\frac{D}{2}
	\delta(\eta - \eta')
	\,
	\delta^{\pup{d}}(\vex{x} - \vex{x'}).
\end{align}
In Eq.~(\ref{CooperStoch}), 
$\cc^t_{\eta,\eta'}(\vex{x},\vex{x'})$ denotes the Cooperon,
$t$ and $\{\eta,\eta'\}$ respectively denote center-of- and relative-time arguments, 
$\{\vex{x},\vex{x'}\}$ are position coordinates in $d$ spatial dimensions, 
$D$ is the classical diffusion constant due to elastic impurity scattering, 
and  
$\phi_{\text{cl}}$ is the ``classical'' component of the scalar potential 
(as opposed to the quantum component, in the Keldysh formalism \cite{Liao17,AlexAlex,KamBook}). 
$\phi_{\text{cl}}$ is a Gaussian stochastic field defined by the correlator
\begin{align}\label{BathKernel}
	\langle \phi_{\text{cl}}(\omega,\vex{k}) \, \phi_{\text{cl}}(-\omega,-\vex{k}) \rangle &\equiv \Delta(\omega,\vex{k})
\nonumber\\
	&= 
	\coth\left(\frac{\omega}{2 k_B T}\right)\rho(\omega,\vex{k}),
\end{align}
where $\Delta(\omega,\vex{k})$ is the \textit{noise kernel} (Keldysh propagator);
the spectral function for the bath is $\rho(\omega,\vex{k})$. 
The noise kernel $\Delta$ encodes real inelastic fermion-fermion scattering processes that are responsible for 
dephasing and is self-generated by thermal fluctuations of the particle 
density \cite{AAK,AAG,Liao17}. 
The Cooperon arises from interference between quantum amplitudes for 
forward- and backward-in-time propagating paths in a disordered conductor \cite{LeeRama,AA85,AAG}.
The scalar potential $\phi_{\text{cl}}$ couples to both the causal ($t + \eta/2$) and anti-causal ($t - \eta/2$)
paths, where $t$ is the global time variable and $\eta$ is a virtual time argument.

The weak (anti)localization correction to the conductivity obtains from \cite{LeeRama}
\begin{align}\label{Conductivity}
	\delta\sigma = 
	\begin{cases}
		(-4e^2/\pi \hbar) \, \mathcal{P}, & \text{WL},\\
		(+2e^2/\pi \hbar) \, \mathcal{P}, & \text{WAL},
	\end{cases}
	\;\;
	\mathcal{P} = \int\limits_0^{\infty}d\eta\ \cc(\eta),
\end{align}
where 
$\cc(\eta) \equiv \langle \cc^t_{\eta,-\eta}(\vex{x},\vex{x})\rangle$, which is independent of $t,\vex{x}$ due to the bath averaging $\langle \cdots \rangle$. 
Above, $\mathcal{P}$ is the \textit{virtual return probability} and 
W(A)L corresponds to weak (anti)localization, relevant for the case without (with) spin-orbit coupling. 
Here we focus on the spin SU(2)-symmetric case, corresponding to WL. 
Importantly, the Cooperon in the un-dephased limit (i.e.\ $\phi_{\text{cl}} = 0$) is given by 
$\cc_0(\eta) = (D/2) (4 \pi D \eta)^{-d/2}$, so that in one or two dimensions the integral in 
Eq.~(\ref{Conductivity}) diverges in the infrared, signaling Anderson localization in noninteracting systems.
However, for interacting particles, the presence of $\phi_{\text{cl}}$
generates a finite decay timescale for the bath-averaged Cooperon, 
ensuring the convergence of Eq.~(\ref{Conductivity}). 
This allows for the definition of the 
\textit{dephasing time}
\begin{equation}
\label{DephasingTimeDefinition}
    \frac{1}{\tau_{\phi}} = -\lim_{\eta \rightarrow \infty}\frac{1}{\eta} \log\big[\cc(\eta)\big].
\end{equation}
The bath-averaged Cooperon can be expressed via a Feynman path integral \cite{AAK},
\begin{widetext}
\begin{align}\label{GeneralPathIntegralUnfolded}
	\langle \cc^t_{\eta,\eta'}(\vex{x},\vex{x'}) \rangle
	=&\,
	\frac{D}{2}
	\int\limits_{\vex{r}(\eta') = \vex{x'}}^{\vex{r}(\eta) = \vex{x}}
	\mathcal{D}
	\vex{r}(\tau)
	\,
	\exp
\left[
-
	\frac{1}{2 D}
	\int\limits_{\eta'}^\eta
	d \tau
	\,
	\left[\dot{\vex{r}}(\tau)\right]^2
-
	\frac{1}{4}
	\int\limits_{\eta'}^\eta
	d \tau_1
	\int\limits_{\eta'}^\eta
	d \tau_2
	\left\{
	\begin{aligned}
	&\,
\phantom{+}\,\,\,
	\tilde{\Delta}\left[\frac{\tau_1 - \tau_2}{2},\vex{r}(\tau_1) - \vex{r}(\tau_2)\right]
\\&\,
-
	\tilde{\Delta}\left[\frac{\tau_1 + \tau_2}{2},\vex{r}(\tau_1) - \vex{r}(\tau_2)\right]
	\end{aligned}
	\right\}
\right].
\end{align}
Setting $\vex{x}'=\vex{x}, \eta'=-\eta$ and moving into ``relative time'' and ``center-of-time'' coordinates, defined by 
\begin{align}\label{Change}
	\vex{\rho}(\tau)
	\equiv
	\vex{r}(\tau) - \vex{r}(-\tau),\ \ \text{and}\ \
	\vex{R}(\tau) 
	\equiv
	\frac{1}{2}\left[\vex{r}(\tau) + \vex{r}(-\tau)\right],
\end{align}
we have 
\begin{align}\label{GeneralPathIntegralFolded-0}
	\cc(\eta)
	=&\,
	\frac{D}{2}
	\int d \vex{R_0}
	\int\limits_{\vex{R}(0) = \vex{R_0}}^{\vex{R}(\eta) = \vex{x}}
	\mathcal{D}
	\vex{R}(\tau)
	\int\limits_{\vex{\rho}(0) = \vex{0}}^{\vex{\rho}(\eta) = \vex{0}}
	\mathcal{D}
	\vex{\rho}(\tau)
	\,
	\exp
\left[
-
	\frac{1}{D}
	\int\limits_{0}^\eta
	d \tau
	\,
	\left[\dot{\vex{R}}(\tau)\right]^2
-
	\frac{1}{4D}
	\int\limits_{0}^\eta
	d \tau
	\,
	\left[\dot{\vex{\rho}}(\tau)\right]^2	
-
	S_I\left[\vex{R}(\tau),\vex{\rho}(\tau)\right]\right],
\end{align}
where the contribution to the action of the noise kernel is given by
\begin{multline}\label{NoiseBathAction}
	S_I[\vex{R}(\tau),\vex{\rho}(\tau)]
	=
	\int
	\frac{d \omega}{2\pi}
	\int
	\frac{d^d \vex{k}}{(2\pi)^d}
	\,
	\Delta(\omega,\vex{k})
	\int\limits_{0}^\eta
	d \tau_a
	\int\limits_{0}^\eta
	d \tau_b
	\,
	\left[
	e^{-i\omega(\tau_a-\tau_b)/2}-e^{-i\omega(\tau_a+\tau_b)/2}
	\right]
	\\
	\times
	e^{
	i\vex{k}
	\cdot
	\left[
	\vex{R}(\tau_a) - \vex{R}(\tau_b)
	\right]}
	\,
	\sin\left[
	\frac{\vex{k}\cdot\vex{\rho}(\tau_a)}{2}
	\right]
	\sin\left[
	\frac{\vex{k}\cdot\vex{\rho}(\tau_b)}{2}
	\right].
\end{multline}
In Eq.~(\ref{GeneralPathIntegralFolded-0}), 
$\vex{R}_0$ is the free boundary condition of the center-of-mass coordinate $\vex{R}(\tau)$ at 
$\tau = 0$. By contrast, the relative coordinate $\vex{\rho}(\tau)$ satisfies Dirichlet boundary
conditions: $\vex{\rho}(\eta) = \vex{\rho}(0) = 0$ \cite{Footnote--NonChargeNeutral}.

\end{widetext}

\subsection{Review of Markovian dephasing}\label{MarkovianDephasing}

There is a massive reduction in the complexity of the problem in the Markovian 
case of a frequency-independent bath kernel \cite{AAK}. 
In this case, $\tilde{\Delta}(t,\vex{x}) = \delta(t) \, \tilde{\Delta}_M(\vex{x})$, which removes the direct time-dependence 
of the bath in Eq.~(\ref{GeneralPathIntegralUnfolded}). 
Explicitly, if $\Delta(\omega,\vex{k}) = \Delta_M(\vex{k})$, Eq.~(\ref{NoiseBathAction}) simplifies to
\begin{align}\label{AAKAction}
	S_I[\vex{R}(\tau),&\,\vex{\rho}(\tau)] 
	\rightarrow
	S_M[\vex{\rho}(\tau)] 
\nonumber\\ 
	\equiv&\, 
	2\int\frac{d^d\vex{k}}{(2\pi)^d} \, \Delta_M(\vex{k}) \int\limits_0^\eta d\tau\ 
	\left\{\sin\left[\frac{\vex{k}\cdot\vex{\rho}(\tau)}{2}\right]\right\}^2
\nonumber\\
	=&\,
	\int\limits_{0}^\eta
	d \tau
	\left\{
	\tilde{\Delta}_M[0]
	-
	\tilde{\Delta}_M\left[\vex{\rho}(\tau)\right]
	\right\}.	
\end{align}
In the Markovian limit, the action $S_M$ has no dependence on the field $\vex{R}(\tau)$, and the $\vex{R}$-path integration 
is equal to one. 
The path integral in Eq.~(\ref{GeneralPathIntegralFolded-0}) reduces to the 
propagator for a single-particle quantum mechanics problem,	
\begin{align}
	\label{AAKReductionII}
	\cc_M(\eta)
	&=
	\frac{D}{2}
	\bra{\vex{\rho} = 0}
	e^{- \hat{h} \eta}
	\ket{\vex{\rho} = 0},
\end{align}
where we have defined the single-particle central-potential Hamiltonian
\begin{align}
\label{AAKReductionIII}
	\hat{h}
	\equiv
	-D \, \nabla_\rho^2 
	+
	\tilde{\Delta}_M[0]
	-
	\tilde{\Delta}_M[\vex{\rho}].
\end{align}	
This simplification can also be seen in a field theory approach \cite{Liao18}. 
In that framework, only a set of maximally crossed rainbow diagrams 
(a subset of the SCBA) contribute [Appendix \ref{MarkovianBath}].


\subsection{Dynamically screened Coulomb interactions}\label{CoulombDephasing}

Here we review dephasing due to dynamically screened Coulomb interactions \cite{AAK},
focusing on many-channel, quasi-1D wires.  
The noise kernel is
\begin{align}\label{AAKNoiseKernel}
	\Delta_M(\omega,k) 
	=&\, 
	-
	2\coth\left(\frac{\omega}{2k_B T}\right)
	\im\left[\frac{V_0(k)}{1-D_R^{(0)}(\omega,k)V_0(k)}\right]
\nonumber\\
	\simeq&\,
	\left(\frac{4 k_B T}{\kappa_0}\right)\frac{1}{D k^2},
\end{align}
where 
\begin{equation}
    D_R^{(0)}(\omega,k) = \frac{-\kappa_0 D k^2}{D k^2 - i\omega}
\end{equation}
is the semiclassical, retarded polarization function describing 
density diffusion in the disordered conductor,
$\kappa_0$ is the bare compressibility,  
and 
$V_0(k)$ is the bare three-dimensional Coulomb potential. 

The approximation in Eq.~(\ref{AAKNoiseKernel}) is twofold. 
First, we take $|D_R^{(0)}(\omega,k) \, V_0(k)|$ large compared to one,
due to the plasmonic (logarithmic) enhancement of $V_0(k)$ as $k \rightarrow 0$ \cite{GiamarchiBook}. 
Second, we also expand in low $\omega / k_B T$, 
\begin{align}
    \coth\left({\omega}/{2 k_B T}\right) \simeq {2 k_B T}/{\omega} + \mathcal{O}\left({\omega}/{k_B T}\right).
\end{align}
By cutting out high-frequency processes, we introduce a short-range ultraviolet cutoff for the bath. 
This is justified because interaction-mediated processes with $|\omega|$ larger than $k_B T$ contribute only to 
the conductivity via the virtual Altshuler-Aronov correction \cite{AAG}. 
A formal calculation retaining higher frequencies would not expand the $\coth$ and keep the full quantum form 
of the noise kernel. In order to avoid inconsistency, in the latter case it is necessary to also 
retain Pauli-blocking counterterms that we have dropped here \cite{VDelftPartI,VDelftPartII}. 
These terms played a role in the theoretical controversy concerning the observed low-temperature 
saturation of the dephasing rate in 1D systems \cite{Webb1,GZ,AAG,VDelftPartI}.
The expansion in Eq.~(\ref{AAKNoiseKernel}) replacing the quantum bath with classical
Johnson-Nyquist noise is that of AAK and is physically correct \cite{AAK,AAG,VDelftPartI}.

Eq.~(\ref{AAKNoiseKernel}) implies that screened Coulomb interactions can be well-approximated 
by a Markovian kernel, so that the bath-averaged Cooperon obtains from Eqs.~(\ref{AAKReductionII}) and (\ref{AAKReductionIII}). 
The effective central potential is
\begin{align}\label{EffCoulKer}
	\tilde{\Delta}_M(0) - \tilde{\Delta}_M(\rho)
	=&
	\frac{\Gamma_M}{D}
	|\rho|,
\end{align}
where we define the coupling constant
\begin{align}\label{GammaM}
	\Gamma_M \equiv \frac{2k_BT}{\kappa_0},
\end{align}
which sets an intrinsic length scale
\begin{align}\label{LengthScale}
	a \equiv \left(\frac{\Gamma_M}{D^2}\right)^{-1/3} = \left(\frac{\kappa_0 D^2}{2 k_B T}\right)^{1/3}. 
\end{align}
It follows from Eqs.~(\ref{AAKReductionIII}) and (\ref{EffCoulKer}) 
that the one-dimensional Cooperon is the imaginary-time propagator for the single-particle Hamiltonian
\begin{align}
	\hat{h} = D\left(-\frac{d^2}{d \rho^2} + \frac{|\rho|}{a^{3}}\right).
\end{align}
Diagonalizing this Hamiltonian gives the bound state energies
\begin{align}\label{Energies}
\begin{aligned}
	\e_{2n} 	&= 	- \alpha'_{n} 	\left(D/a^2\right),\\
	\e_{2n+1} 	&= 	- \alpha_{n} 	\left(D/a^2\right),
\end{aligned}
\end{align}
and the orthonormal eigenfunctions \cite{AiryBook}
\bsub\label{Eigenfunctions}
\begin{align}\label{EigenfunctionsEven}
\!\!\!
	\psi^0_{2n}(\rho;a) 
	=&\, 
	\frac{1}{\sqrt{2a}}
	\frac{|\alpha'_{n}|^{-1/2}}{\Ai(\alpha'_{n})} 
	\Ai\left(\frac{|\rho|}{a} + \alpha'_{n}\right)\!,\!\!
\\
\label{EigenfunctionsOdd}
\!\!\!
	\psi^0_{2n+1}(\rho;a) 
	=&\, 
	\frac{1}{\sqrt{2a}}
	\frac{\sgn(\rho)}{\Ai'(\alpha_{n})} 
	\Ai\left(\frac{|\rho|}{a} + \alpha_{n}\right)\!,\!\!
\end{align}
\esub
where $n \in \{0,1,2,...\}$, and $\alpha_n$ and $\alpha'_n$ respectively denote the 
$(n + 1)^{\mathrm{th}}$
(strictly negative) zero of the Airy function $\Ai(z)$ or its derivative $\Ai'(z) \equiv (d/dz)\Ai(z)$. 

The eigenfunctions in Eqs.~(\ref{EigenfunctionsEven}) and (\ref{EigenfunctionsOdd}) 
allow the explicit computation of the expectation value in Eq.~(\ref{AAKReductionII}) for $\cc_M(\eta)$ in one dimension:
\begin{align}\label{MarkovianCooperon}
	\cc_M(\eta) 
	=&\,  
	\frac{D}{4a}
	\sum_{j = 0}^\infty
		\frac{1}{|\alpha'_j|}
		\exp\left(- \frac{\eta D}{a^2}|\alpha'_j|\right)
\nonumber\\ 
	\equiv&\, 
	\frac{D}{4a}
	\,
	f_0\left(\frac{\eta D}{a^2}\right).
\end{align}
We note that $(\eta D/a^2)$ is a dimensionless time variable and that the sum in Eq.~(\ref{MarkovianCooperon}) 
is only over the even-parity energies, since the odd-parity wavefunctions vanish at the origin. 
This yields the AAK dephasing timescale
\begin{equation}\label{AAKDephasingTime}
	\tau_{\phi} 
	\simeq 
	\left[\frac{1}{D}\left(\frac{2k_B T}{ \kappa_0}\right)^2\right]^{-1/3}.
\end{equation}
Finally, the return probability is
\begin{align}\label{AAKConductivity}
	\mathcal{P} 
	=&\, 
	\int\limits_0^{\infty}
	d\eta\ \cc_M(\eta) 
	= 
	\frac{a}{4}
	\sum_{n = 0}^\infty 
	\frac{1}{(\alpha'_n)^2}
\nonumber\\
	=&\, 
	\frac{1}{4}
	\left(\frac{\kappa_0 D^2}{2 k_B T}\right)^{1/3}
	\frac{2\pi}{3^{5/6}\Gamma^2(2/3)}.
\end{align}
Via Eq.~(\ref{Conductivity}), this gives the famous AAK result for the weak localization
correction to the conductivity of a quasi-1D wire, $\delta \sigma_{\text{WL}} \propto - (k_B T)^{-1/3}$ \cite{AAK}.

The summation over Airy derivative function zeros in Eq.~(\ref{AAKConductivity}) is a known identity
\cite{AiryBook}. 
In the sequel, we will need to be able to numerically evaluate similar sums, which are slowly convergent.
To do so efficiently, we will introduce a ``series acceleration'' technique [Appendix~\ref{SeriesAcceleration}].

\section{Dephasing by a diffusive bath: Perturbation theory}\label{DiffusiveDephasing}

In this section, we attempt to understand dephasing in a setting intimately related to many-body localization
(MBL) \cite{BAA06,Gornyi05,Laumann14}. 
We study the dephasing of the ergodic phase of an MBL candidate system by perturbatively
evaluating Eq.~(\ref{GeneralPathIntegralUnfolded}) in
the presence of a diffusive bath.

Our calculation models a 1D ultracold Fermi gas with short-ranged interactions.
We focus on the many-channel version with weak disorder, so that weak localization theory applies at intermediate temperatures \cite{AA85,LeeRama,AAK,AAG}. 
Hydrodynamic modes in the ergodic phase of a dirty fermion system are generally diffusive \cite{BK}. As a result, in the case of 
short-range interactions, the noise bath governing the thermalization of the system is also diffusive \cite{NZA,Liao17}. 

The diffusive noise kernel is
\begin{align}\label{DiffusiveNoise}
	\Delta_t(\omega,k) 
	= 
	\Gamma_t
	\left(
	\frac{2 D_t k^2}{D_t^2 k^4 + \omega^2}
	\right).
\end{align}
For $|\omega| \lesssim k_B T$, this is the approximate semiclassical, diffusive 
Keldysh propagator for particle density fluctuations in the fermion gas with quenched disorder,
as arises due to short-ranged inelastic particle-particle collisions \cite{Liao17}. 
The coupling constant is
\begin{align}\label{Gammat}
	\Gamma_t 
	= 
	3\frac{\gamma_t^2 k_B T}{(1 - \gamma_t)\chi_0},
\end{align}
where $\gamma_t$ is the dimensionless interaction strength (Finkel'stein coupling parameter \cite{BK,Liao17}).

In the sequel, we will consider coexisting diffusive and Markovian baths, with the former
(latter) mediated by short-ranged spin exchange (dynamically screened Coulomb) interactions. 
In that context, $\gamma_t$ denotes the spin triplet channel interaction strength
(hence the ``$t$'' subscript), while $\chi_0$ is the bare spin susceptibility. 
For contact interactions in an ultracold Fermi gas, $\chi_0$ is the compressibility. 
The diffusion constant $D_t$ in Eq.~(\ref{DiffusiveNoise}) differs from the bare 
one entering the Cooperon [Eq.~(\ref{CooperStoch})], due to an interaction renormalization \cite{BK,Liao17},
\begin{align}\label{BetaDef}
	\beta \equiv \frac{D_t}{D} = \frac{1}{1-\gamma_t}.
\end{align}
Here and throughout this paper, we will use the symbol $\beta$ to refer to this dimensionless ratio
(and \emph{not} the inverse temperature). 
In the context of itinerant spin exchange interactions in a quantum wire, one typically has an attractive 
spin-triplet channel coupling strength $\gamma_t < 0$. 
The Stoner instability towards ferromagnetism corresponds to the limit $\gamma_t \rightarrow -\infty$ \cite{BK}.
Repulsive interactions instead give $\gamma_t > 0$; 
$\gamma_t \rightarrow 1$ corresponds to the incompressible limit \cite{BK}.

\begin{figure}[t!]
        \includegraphics[width=0.35\textwidth]{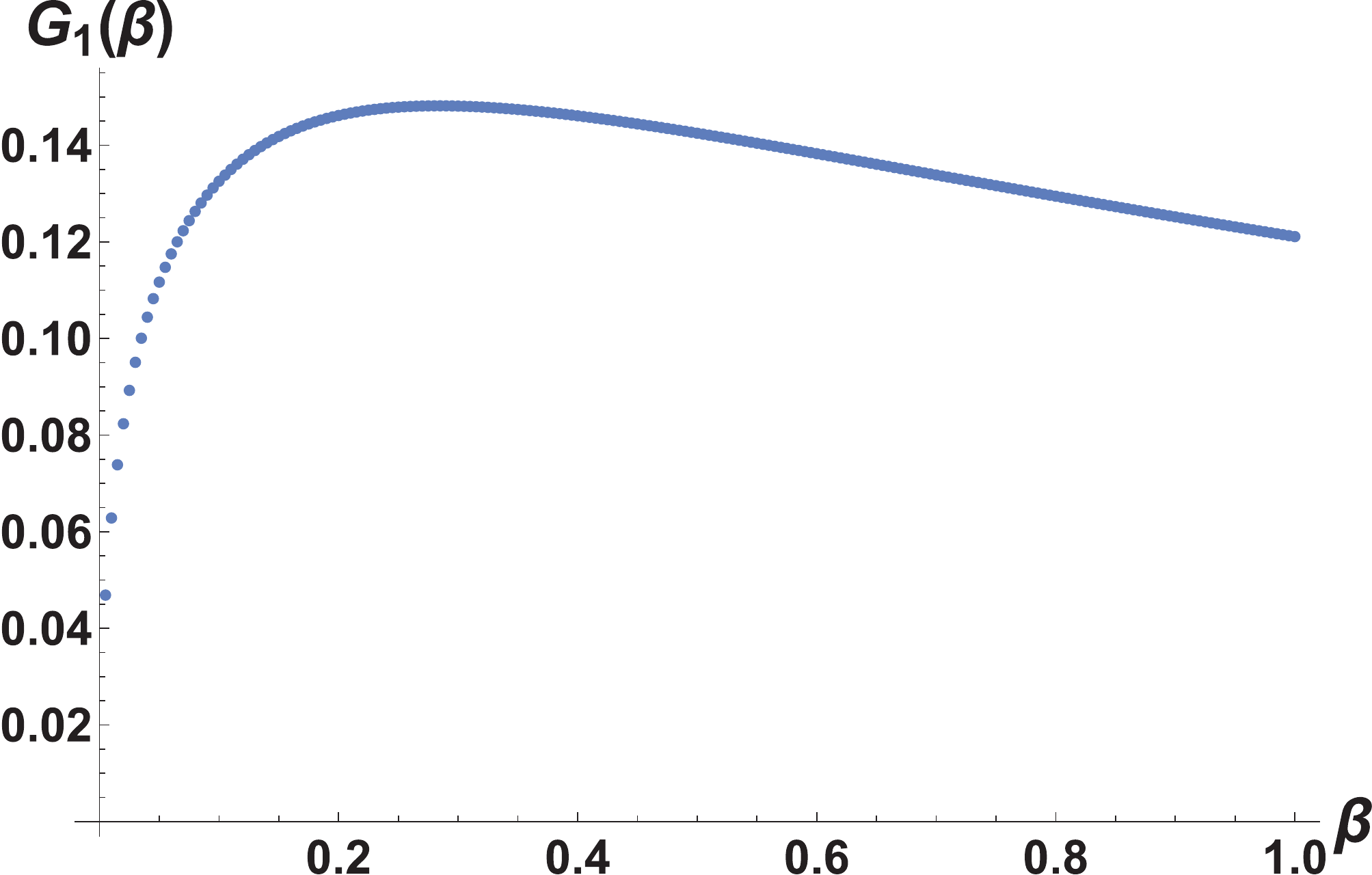}
        \caption{The coefficient function $G_1(\beta)$ for the first-order cumulant expansion
	result in Eqs.~(\ref{FirstOrderExpectationResult}) and (\ref{FirstOrderGFunc}), 
	which determines the lowest-order superexponential dephasing 
	of the Cooperon due to the diffusive bath [Eq.~(\ref{CumulantDefinition})]. 
	Here $\beta$ is the ratio of the interacting and bare diffusion constants defined by Eq.~(\ref{BetaDef}).}
        \label{FirstOrderDiffusivePlot}
\end{figure}

The AAK mapping to a single-particle problem depends crucially on the Markovian nature of the noise kernel and cannot be applied here;
the noise action in Eq.~(\ref{NoiseBathAction}) remains a nonlocal function of both the center-of-mass 
$R(\tau)$ and relative $\rho(\tau)$ coordinates.  
In this section, we present a purely perturbative calculation for dephasing due to the diffusive bath in Eq.~(\ref{DiffusiveNoise}). 
Without moving to the center-of-time and relative coordinates, 
we evaluate the Cooperon in Eq.~(\ref{GeneralPathIntegralUnfolded}) via the cumulant expansion, 
\begin{align}
\label{CumulantDefinition}
	\cc_t(\eta)
	=&\,
	\frac{D}{2}
	\int\limits_{r(-\eta) = 0}^{r(\eta) = 0}
	\mathcal{D}
	r(\tau)
	\,
	e^{
	-
	\frac{1}{2D}
	\int\limits_{-\eta}^\eta
	d \tau
	\,
	\dot{r}^2(\tau)
	-
	S_I\left[r(\tau)\right]
	}
\\
\nonumber
	=&\, 
	\cc_0(\eta)
	\,
	\exp\left[
		-
		\langle S_I \rangle_0 
		+ 
		\frac{1}{2}
		\Big(
			\left\langle S^2_I \right\rangle_0
			-
			\left\langle S_I \right\rangle_0^2
		\Big)
		+ 
		\ldots
	\right],
\end{align}
where the bare Cooperon is 
\begin{align}
	\cc_0(\eta)
	=
	(D/2)(4 \pi D \eta)^{-1/2},
\end{align}
$\langle \cdots \rangle_0$ denotes a functional average with respect to the noiseless action,
and 
the bath-induced 
interaction is
\begin{align}
\label{S_I}
	S_I[r(\tau)] =&\,
	\frac{\Gamma_t}{4}
	\int
	\frac{d k}{2\pi}
	\int\limits_{-\eta}^\eta
	d \tau_a
	\int\limits_{-\eta}^\eta
	d \tau_b
	\,\ 
	e^{
	i k
	\left[
	r(\tau_a) - r(\tau_b)
	\right]}
	\nonumber\\
	&\,
	\times
	\left[
		e^{- D_t k^2 \left|\tau_a - \tau_b\right|/2}- e^{- D_t k^2 \left|\tau_a + \tau_b\right|/2}
	\right].
\end{align}
The cumulant expansion in the bath coupling $\Gamma_t$ boils down to the computation of the 
expectation values $\langle S^n_I \rangle_0$. At first order, we only need to compute 
$\langle S_I \rangle_0$. We note that the functional average over $r(\tau)$ affects only 
the exponential factor in the top line of Eq.~(\ref{S_I}). 
Performing this average to obtain the vertex operator correlator [Appendix \ref{Correlators}]  
as well as the Gaussian integral over $k$ gives 
\begin{align}
	\label{FirstOrderExpectationResult}
	\langle S_I \rangle_0 
	=&\, 
	\left(
	\Gamma_t 
	\frac{\eta^{3/2}}{\sqrt{D}}
	\right)
	G_1(\beta),
\end{align}
where $G_1(\beta)$ is a dimensionless function of the diffusion constant ratio $\beta$ [Eq.~(\ref{BetaDef})], 
\begin{align}
	\label{FirstOrderGFunc}
	\!\!\!
	G_1(\beta) 
	= 
	\sqrt{\frac{2}{\pi}}\int\limits_{-1}^1 d\tau_a \int\limits_{\tau_a}^{1} d\tau_b 
	\left[
	\begin{aligned}
		&\, 
		g_1(\beta,\tau_a,\tau_b)^{-1/2}
	\\
		-&\,
		g_2(\beta,\tau_a,\tau_b)^{-1/2}
	\end{aligned}
    \right]\!,\!\!
\end{align}
and where the $g_{1,2}$ functions are defined by\\
\begin{align}\label{gFunctionsDefinition}
\!\!\!
\begin{aligned}
	g_1(\beta,\tau_a,\tau_b) =&\, (\beta + 1)(\tau_b-\tau_a)-\frac{1}{2}(\tau_b-\tau_a)^2,
\\    
	g_2(\beta,\tau_a,\tau_b) =&\, \beta|\tau_b+\tau_a| + (\tau_b-\tau_a)-\frac{1}{2}(\tau_b-\tau_a)^2.
\end{aligned}
\end{align}
The amplitude $G_1(\beta)$ is plotted in Fig.~\ref{FirstOrderDiffusivePlot}. Since $G_1(\beta)$ is positive, 
the action of the bath is to suppress (dephase) the Cooperon with increasing virtual time $\eta$. 
However, Eqs.~(\ref{FirstOrderExpectationResult}) and (\ref{CumulantDefinition}) predict a 
superexponential dampening of the Cooperon, and thus do not define a finite dephasing rate. 

\begin{figure}[b!]
	\includegraphics[width=0.27\textwidth]{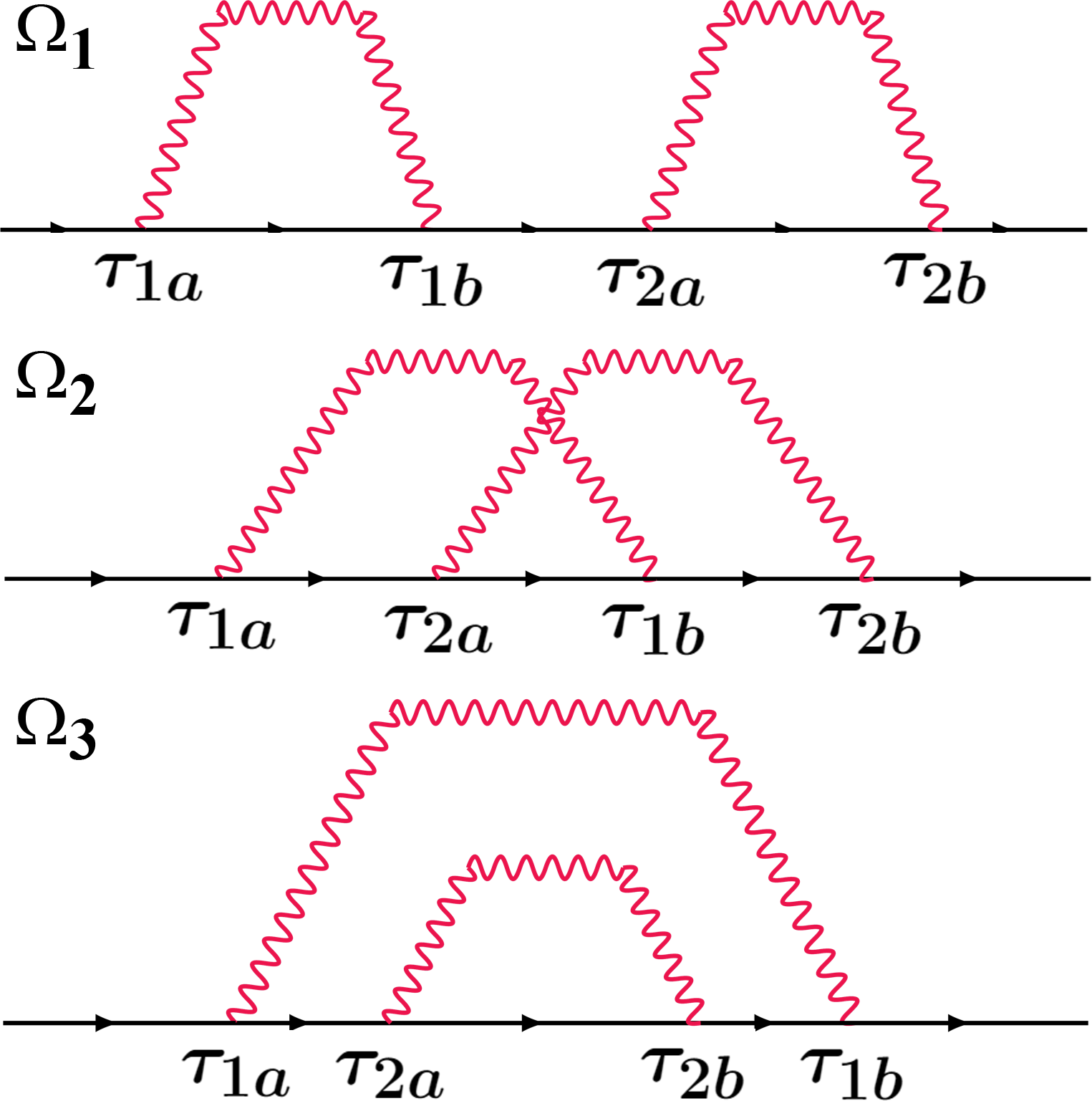}
        \caption{Diagrams giving the three topologically distinct contributions to 
	the Cooperon due to the diffusive noise bath at second order in perturbation theory. 
	These correspond to the time-ordering sectors $\{\Omega_{1,2,3}\}$ in Eq.~(\ref{SecondOrderTimeSectors}). 
	From top to bottom, we have 
	``sector 1'' $\Omega_1$, 
	``sector 2'' $\Omega_2$,
	and 
	``sector 3'' $\Omega_3$; 
	we also refer to these amplitudes as 
	``double,'' ``crossed,'' and ``nested,'' respectively. 
	We show in Appendix \ref{Correlators} how the form of the correlator depends 
	on the topology of the time-ordering.}
        \label{SecondOrderDiagrams}
\end{figure}

To better interpret the first-order result, we continue the calculation to second order, 
requiring the evaluation of the expectation value
\begin{widetext}
\begin{align}\label{S_I^2}
	\left\langle S_I^2\right\rangle_0
	=&\,
	\frac{\Gamma_t^2}{16}
	\int\frac{d k_1}{2\pi}
	\int\frac{d k_2}{2\pi}
\nonumber\\
	&\,
	\times
	\int\limits_{-\eta}^\eta
	d \tau_{1a}
	\int\limits_{-\eta}^\eta
	d \tau_{1b}
	\int\limits_{-\eta}^\eta
	d \tau_{2a}
	\int\limits_{-\eta}^\eta
	d \tau_{2b}	
	\,
	\left[
	e^{- D_t k_1^2 \left|\tau_{1a}-\tau_{1b}\right|/2} - e^{- D_t k_1^2 \left|\tau_{1a}+\tau_{1b}\right|/2}
	\right]
	\left[
	e^{- D_t k_2^2 \left|\tau_{2a}-\tau_{2b}\right|/2} - e^{- D_t k_2^2 \left|\tau_{2a}+\tau_{2b}\right|/2}
	\right]
\nonumber\\
	&\,
	\times
	\left\langle
	e^{
	ik_1
	\left[
	r(\tau_{1a}) - r(\tau_{1b})
	\right]}
	e^{
	ik_2
	\left[
	r(\tau_{2a}) - r(\tau_{2b})
	\right]}
	\right\rangle_0.
\end{align}
\end{widetext}
The 4-point vertex function correlator on the last line of this equation
is evaluated in closed form in Appendix~\ref{Correlators}. 
We note that Eq.~(\ref{S_I^2}) is invariant under the symmetries 
$\tau_{1a} \leftrightarrow \tau_{1b}$, 
$\tau_{2a} \leftrightarrow \tau_{2b}$, 
and 
$(\omega_1,k_1,\tau_{1a},\tau_{1b}) \leftrightarrow (\omega_2,k_2,\tau_{2a},\tau_{2b})$. 
This 8-fold symmetry group leaves 3 distinct topological classes of the 24 distinct time-orderings, 
and we may thus reduce to the case where 
$\tau_{1a}<\tau_{1b}$,
$\tau_{2a}<\tau_{2b}$,
and 
$\tau_{1a}<\tau_{2a}$. 
We define the three inequivalent time-sectors $\Omega_s$ to be
\begin{align}\label{SecondOrderTimeSectors}
	\Omega_s 
	\equiv
	\begin{cases}
	\{ \tau_{1a}<\tau_{1b}<\tau_{2a}<\tau_{2b}\} & s = 1,\\
	\{ \tau_{1a}<\tau_{2a}<\tau_{1b}<\tau_{2b}\} & s = 2,\\
	\{ \tau_{1a}<\tau_{2a}<\tau_{2b}<\tau_{1b}\} & s = 3.
	\end{cases}
\end{align}
These correspond to the three diagrams shown in 
Fig.~\ref{SecondOrderDiagrams}. The form of the functional average 
depends upon the topological class of the time-ordering.  

Using the vertex operator correlator from Appendix~\ref{Correlators}, the momentum integrals in Eq.~(\ref{S_I^2}) can be obtained in closed form.
The final result can be expressed as follows, 
\begin{align}\label{SecondOrderGs}
	\langle S^2_I\rangle_0 
	= 
	\Gamma_t^2\frac{\eta^3}{D}
	\left[	
		G^{(1)}_2(\beta) 
		+ 
		G^{(2)}_2(\beta) 
		+ 
		G^{(3)}_2(\beta) 
	\right],
\end{align}
where $G^{(s)}_2$ is a dimensionless function corresponding to the topological sector $s$. 
These functions are defined explicitly in Appendix~\ref{MoreDiffusionDetails}, as parametric
integrals over rescaled $\{\tau_{1a},\tau_{1b},\tau_{2a},\tau_{2b}\}$ variables [Eq.~(\ref{S_I^2})].

The final result for the diffusive bath-averaged Cooperon, computed through second order in the cumulant expansion, 
is given by 
\begin{align}\label{FinalAnswerDiffusive}
\!\!\!\!
	\frac{\cc_t(\eta)}{\cc_0(\eta)} 
	=&\, 
	\exp
	\!
	\left[
	\begin{aligned}
		&
		-
		\left(\Gamma_t\frac{\eta^{3/2}}{\sqrt{D}}\right) G_1\left(\beta\right)
	\\
		&
		+ 
		\frac{1}{2}
		\left(\Gamma_t\frac{\eta^{3/2}}{\sqrt{D}}\right)^2 G^{(T)}_2\left(\beta\right) 
		+
		\mathcal{O}(\Gamma_t^3)
	\end{aligned}
	\right]\!\!,\!\!
\end{align}
where
\begin{align}\label{GT2Def}
	G^{(T)}_2\left(\beta\right) 
	\equiv&\, 
	G_2^{(1)}\left(\beta\right)
	+
	G_2^{(2)}\left(\beta\right)
	+
	G_2^{(3)}\left(\beta\right)
	-
	[G_1\left(\beta\right)]^2.
\end{align}
The cumulant expansion in Eq.~(\ref{FinalAnswerDiffusive}) is well-defined because 
the integrals that determine the amplitude functions 
$G_1(\beta)$ and $G^{(T)}_2(\beta)$ are free of divergences in one dimension. 
(In 2D or higher, the cumulant expansion is plagued by UV divergences; these can be regularized
by self-consistency, but see below.)
Nevertheless, no finite dephasing rate as in Eq.~(\ref{DephasingTimeDefinition})
can be identified, because the expansion is a series in powers of $(\Gamma_t \sqrt{\eta^3/D})$.
Moreover, this series evidently breaks down for long virtual times $\eta \gtrsim (D / \Gamma_t^2)^{1/3}$,
signaling a failure of perturbation theory. The Cooperon is needed for arbitrarily large $\eta$, 
in order to compute the weak localization correction in Eq.~(\ref{Conductivity}). 

\begin{figure}[t!]
	\includegraphics[width=0.35\textwidth]{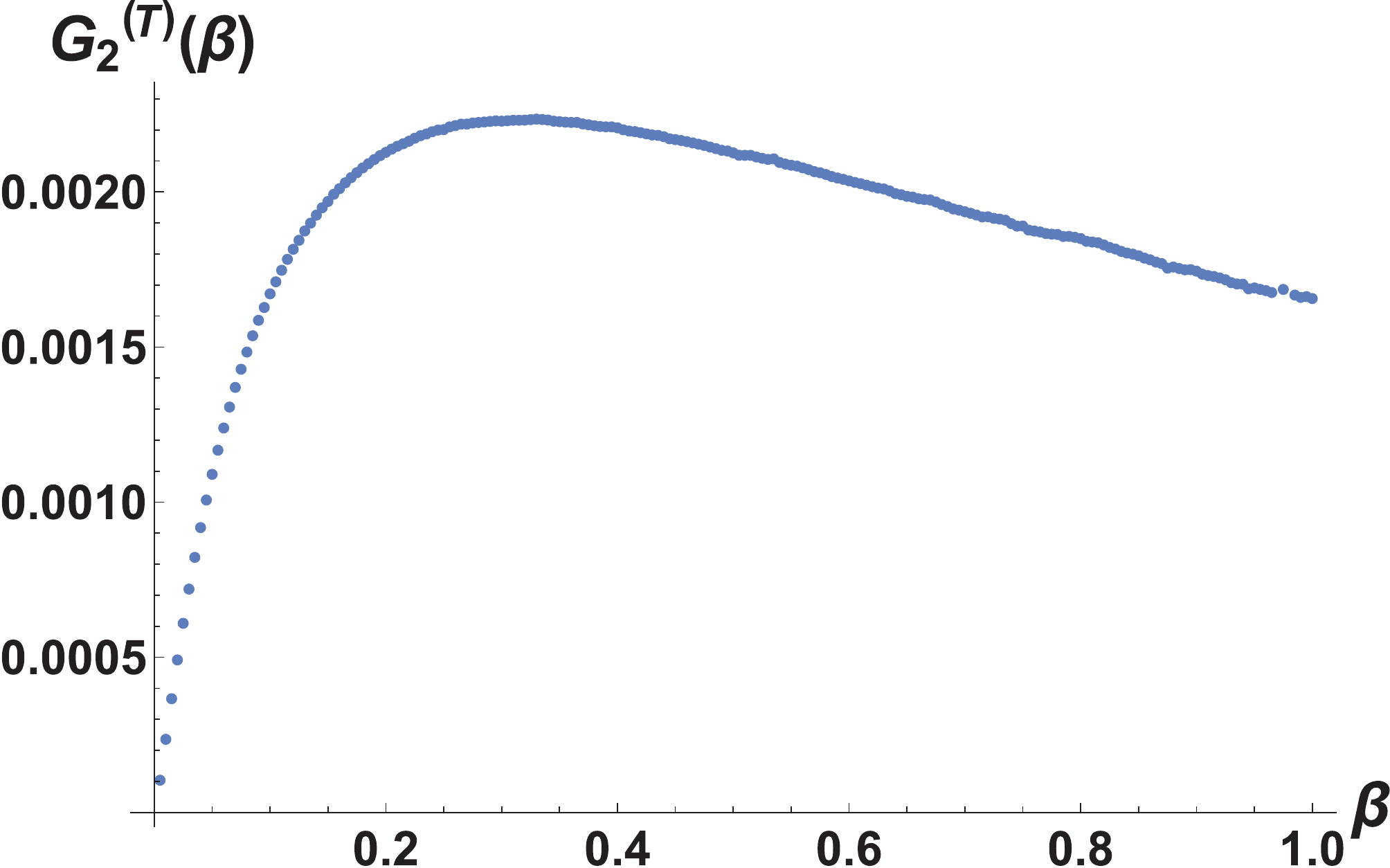}
	\caption{
	The coefficient function $G_2^{(T)}(\beta)$ for the second-order cumulant expansion
	result in Eqs.~(\ref{FinalAnswerDiffusive}) and (\ref{GT2Def}).
	Since $G_2^{(T)}(\beta) > 0$, this shows that the second-order cumulant 
	gives a superexponential \emph{rephasing} contribution to the Cooperon in Eq.~(\ref{FinalAnswerDiffusive}),	
	due to the interaction with the diffusive bath. 
	Here $\beta$ is the ratio of the interacting and bare diffusion constants defined by Eq.~(\ref{BetaDef}).
	See Fig.~\ref{SecondOrderBreakdownPlot} for the plot of the individual components of Eq.~(\ref{GT2Def}) that
	yield the total coefficient.} 
	\label{SecondOrderTotalPlot}
\end{figure}

\begin{figure}[b!]
	\includegraphics[width=0.35\textwidth]{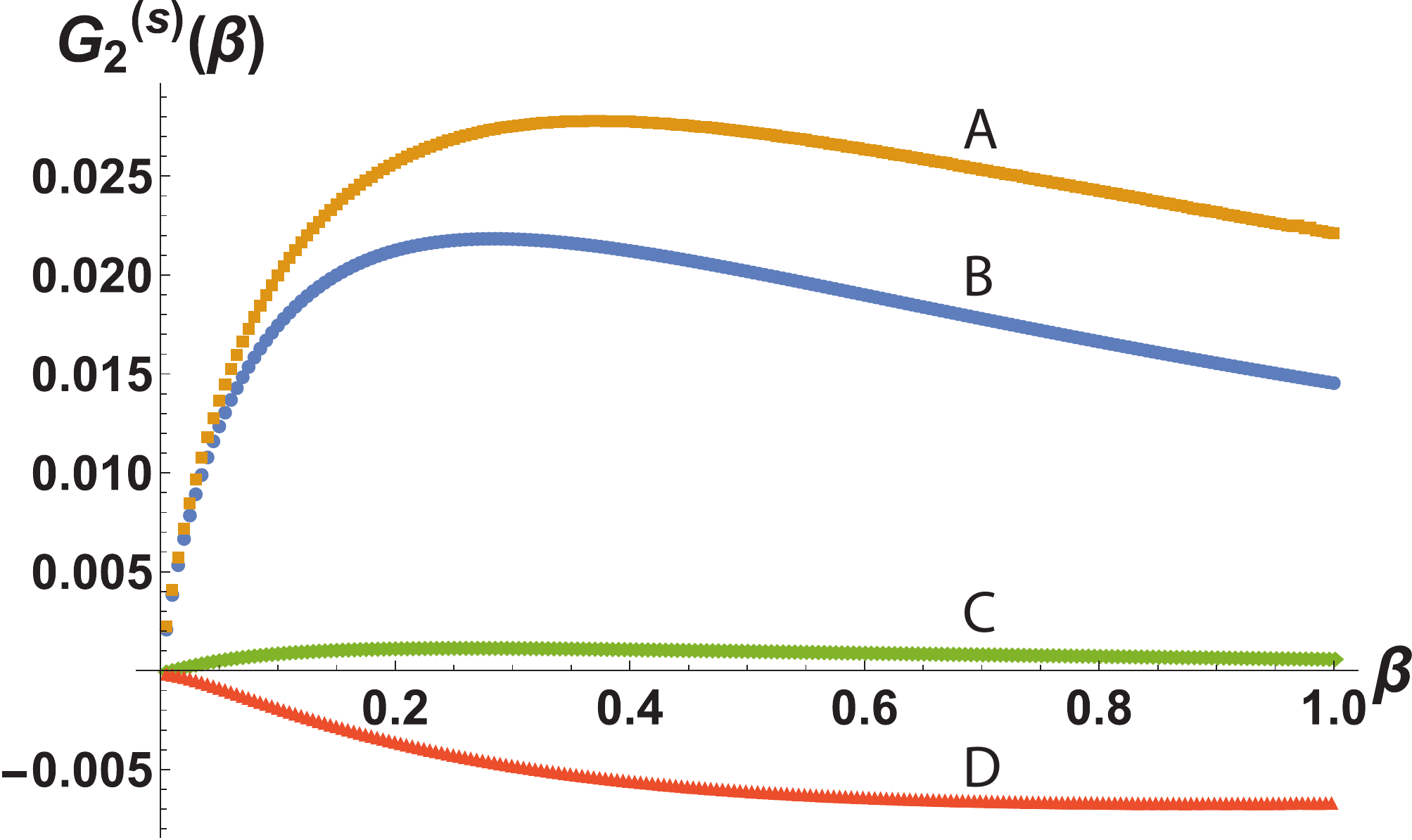}
	\caption{
	Plot of the individual contributions to the total second-order coefficient function $G_2^{(T)}(\beta)$ 
	defined by Eq.~(\ref{GT2Def}), plotted in Fig.~\ref{SecondOrderTotalPlot}. The second order contribution 
	to the cumulant expansion in Eq.~(\ref{FinalAnswerDiffusive}) gives a net \emph{rephasing}, due to the combination
	of the competing terms that do not exactly cancel out. 
	The contributions are indicated as  
	A: double diagram, $G_2^{(1)}\left(\beta\right)$, 
	B: the square of first order term, $[G_1\left(\beta\right)]^2$, 
	C: the crossed diagram, $G_2^{(2)}\left(\beta\right)$, 
	and 
	D: the nested diagram, $G_2^{(3)}\left(\beta\right)$
	[see Fig.~\ref{SecondOrderDiagrams}].}
	\label{SecondOrderBreakdownPlot}
\end{figure}

The most interesting aspect of the result in Eq.~(\ref{FinalAnswerDiffusive}) is the \emph{sign} of the
second-order correction. As shown in Fig.~\ref{SecondOrderTotalPlot}, 
the net coefficient $G^{(T)}_2(\beta)$ is positive and nonzero for $\beta > 0$. 
Fig.~\ref{SecondOrderBreakdownPlot} shows that this is due to a 
competition between terms coming from the different topological sectors, which do not cancel
the square of the first-order coefficient.
The range plotted in Figs.~\ref{SecondOrderTotalPlot} and \ref{SecondOrderBreakdownPlot} 
corresponds to interparticle scattering 
due to attractive interactions $\gamma_t < 0$ [Eq.~(\ref{BetaDef})]; $G^{(T)}_2(\beta)$ remains
positive and nonzero for $\beta > 1$, corresponding to repulsive interactions. 
We conclude that at second order, the interaction of the Cooperon with the diffusive bath
gives a net \emph{rephasing} contribution, and this quickly overwhelms the first-order
dephasing result when $\eta \gtrsim (D / \Gamma_t^2)^{1/3}$. The ultimate fate
of the Cooperon at long times requires a nonperturbative treatment of the diffusive bath.

We note that the self-consistent Born approximation (SCBA) is \textit{not} sufficient to stabilize these results. 
In the SCBA, one 
sums the set of all non-crossing diagrams 
(as defined using an alternative field theory language, see Appendix~\ref{FieldTheory}).
This yields the self-consistent equation \cite{Liao18}
\begin{align}
\label{SCBA1}
	\tau_{SCBA}^{-1}
	= 
	2
	\int
	\frac{d \omega}{2\pi}
	\int
	\frac{d^d \vex{k}}{(2\pi)^d}
	\frac{\Delta_t(\omega,k)}{D k^2 - i\omega + \tau_{SCBA}^{-1}}.
\end{align}
Evaluating this in 1D, using Eq.~(\ref{DiffusiveNoise}) gives 
\begin{align}
\label{SCBA2}
	\tau_{SCBA}^{-1}
	=
	\left(\frac{\Gamma_t^2}{D + D_t}\right)^{1/3}
	\propto
	(k_B T)^{2/3},
\end{align}
identical to the temperature dependence obtained by AAK \cite{AAK} for the Markovian screened Coulomb bath
[Eq.~(\ref{AAKDephasingTime})]. However, adding the ``mass'' $\tau_{SCBA}^{-1}$ to the bare Cooperon 
$\cc_0(\eta)$ merely appends the decaying exponential prefactor $\exp(-\eta/\tau_{SCBA})$ to the 
path integral in Eq.~(\ref{CumulantDefinition}). At large virtual times $\eta \rightarrow \infty$, 
linear dephasing is overpowered by the second-order 
terms contributing to Eq.~(\ref{FinalAnswerDiffusive}) that are neglected in the SCBA, 
and still give a nonzero contribution proportional to $\eta^3$.

These results can be compared with those of AAK for the screened Coulomb Markovian bath, reviewed in 
Sec.~\ref{CoulombDephasing}. In that case there is an exact solution [Eq.~(\ref{MarkovianCooperon})]. 
However, one could instead employ a perturbative calculation similar to the one presented above. 
Performing the cumulant expansion for the Markovian bath [Appendix \ref{MarkovianPT}], 
one finds the same power-law behavior in $\eta$ seen above in Eq.~(\ref{FinalAnswerDiffusive}) \cite{VDelftPartI}. 
The $\eta^{3/2}$-dependence is generic to perturbing around the bare Cooperon in 1D, 
and is not tied to the diffusive character of the bath.

For the diffusive bath, we find that every order in the cumulant expansion is 
governed by a competition between many dephasing and rephasing terms. Our second-order result in 
Eq.~(\ref{FinalAnswerDiffusive}) demonstrates that rephasing diagrams may dominate at any given order. 
By contrast, the cumulant expansion for the Markovian bath [Appendix \ref{MarkovianPT}] yields a single term 
at each order. This difference in complexity can also be seen in the field theory description [Appendix \ref{FieldTheory}].

The path integral Eq.~(\ref{GeneralPathIntegralUnfolded}) gives a strongly coupled field theory \cite{Liao18} governing 
the dephasing of a system with a diffusive noise bath.
The bare cumulant expansion breaks down after short virtual times, so that a nonperturbative technique is required 
to characterize the dephasing of the system.
However, the SCBA is not sufficient to stabilize the theory against additional perturbative corrections. 
In the next two sections, we employ an additional Markovian noise bath (which we treat exactly) as an infrared regularization; 
this stabilizes the perturbation theory for the diffusive, non-Markovian bath at long virtual times 
[see Eq.~(\ref{FinalCoexistCooperon})]. 

Finally, we note that the calculation presented in this section can also be carried out in a field theory formalism, 
and that there is a well-defined mapping between the Feynman diagrams there and the different contributions seen here in the 
cumulant expansion. This connection is described in Appendix~\ref{FieldTheory}.

\section{Dephasing due to combined diffusive and Markovian baths: Results}\label{CoexistingDephasingResults}

We demonstrated in Sec.~\ref{DiffusiveDephasing} that a naive perturbative treatment of the diffusive noise bath
modulating the Cooperon in Eq.~(\ref{GeneralPathIntegralUnfolded}) is insufficient to determine the
dephasing time [Eq.~(\ref{DephasingTimeDefinition})]. We argued that the standard partial summation of 
perturbation theory [the self-consistent Born approximation (SCBA)] does not stabilize the calculation against
neglected perturbative corrections.  
More work is required to understand dephasing in such a weakly disordered, quasi-1D fermion system
with pure short-ranged interactions. 

To understand the effects of the diffusive noise bath, in this section we study coexisting interactions. 
In particular, we consider the diffusive kernel [Eq.~(\ref{DiffusiveNoise})] in parallel with the 
Markovian kernel in Eq.~(\ref{EffCoulKer}). 
This scenario corresponds to a quasi-1D, many-channel quantum wire with spin SU(2) symmetry,
possessing both long-ranged Coulomb and short-ranged, spin exchange interactions.
These interactions are respectively associated to the charge and spin density hydrodynamic modes,
and each gives rise to its own noise bath that interacts with the Cooperon \cite{NZA,BK,Liao17}. 

We treat the Markovian bath exactly, extending the AAK solution reviewed in Sec.~\ref{CoulombDephasing},
whilst simultaneously employing the cumulant expansion [Eq.~(\ref{CumulantDefinition})] for the diffusive bath.
We find that the Markovian bath provides a physical infrared regularization of the terms computed
in the perturbative expansion for the diffusive bath. Unlike the bare expansion presented in Sec.~\ref{DiffusiveDephasing} or the SCBA, 
this regularization stabilizes perturbation theory at long virtual times. This is due to nontrivial cancellations between higher-order terms, 
detailed in Sec.~\ref{CalculationDetails}, with no analogue in the bare expansion [Eq.~(\ref{FinalAnswerDiffusive})].

\begin{figure}[t!]
	\includegraphics[width=0.35\textwidth]{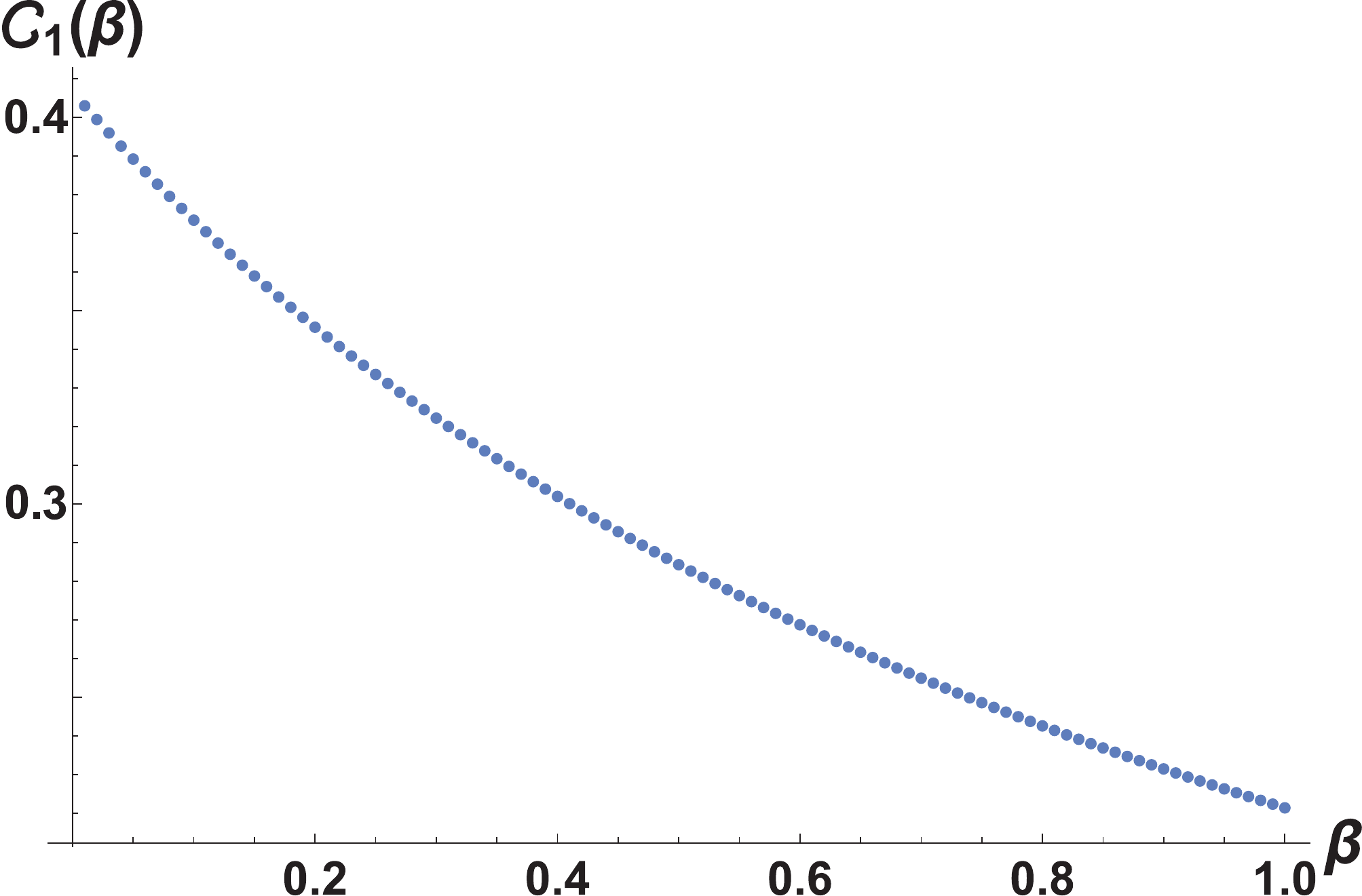}
        \caption{The coefficient function $\mathcal{C}_1(\beta)$ for the first-order cumulant expansion
	result in Eq.~(\ref{FinalCoexistCooperon}), which describes dephasing due to inelastic spin-triplet 
	exchange scattering, in the presence of an additional screened-Coulomb Markovian bath. 
	Here $\beta$ is the ratio of the interacting and bare diffusion constants defined by Eq.~(\ref{BetaDef}).
	The coefficient $\mathcal{C}_1(\beta)$ is expressed in terms of a slowly-converging infinite sum in Eq.~(\ref{FirstOrderLinear}).
	In order to reliably approximate the result, here we have used the series acceleration technique 	
	described in Appendix~\ref{SeriesAcceleration}. 
	We plot the coefficient for $0 < \beta \leq 1$, which corresponds to ferromagnetic exchange interactions
	[$\gamma_t < 0$ in Eqs.~(\ref{Gammat}) and (\ref{BetaDef})]. 
	Since $\mathcal{C}_1(\beta) > 0$, the first-order correction in Eq.~(\ref{FinalCoexistCooperon}) enhances the dephasing
	rate of the pure Markovian result in $\cc_M(\eta)$.}
	\label{FirstOrderCoexisitingPlot}
\end{figure}

\begin{figure}[b!]
	\includegraphics[width=0.35\textwidth]{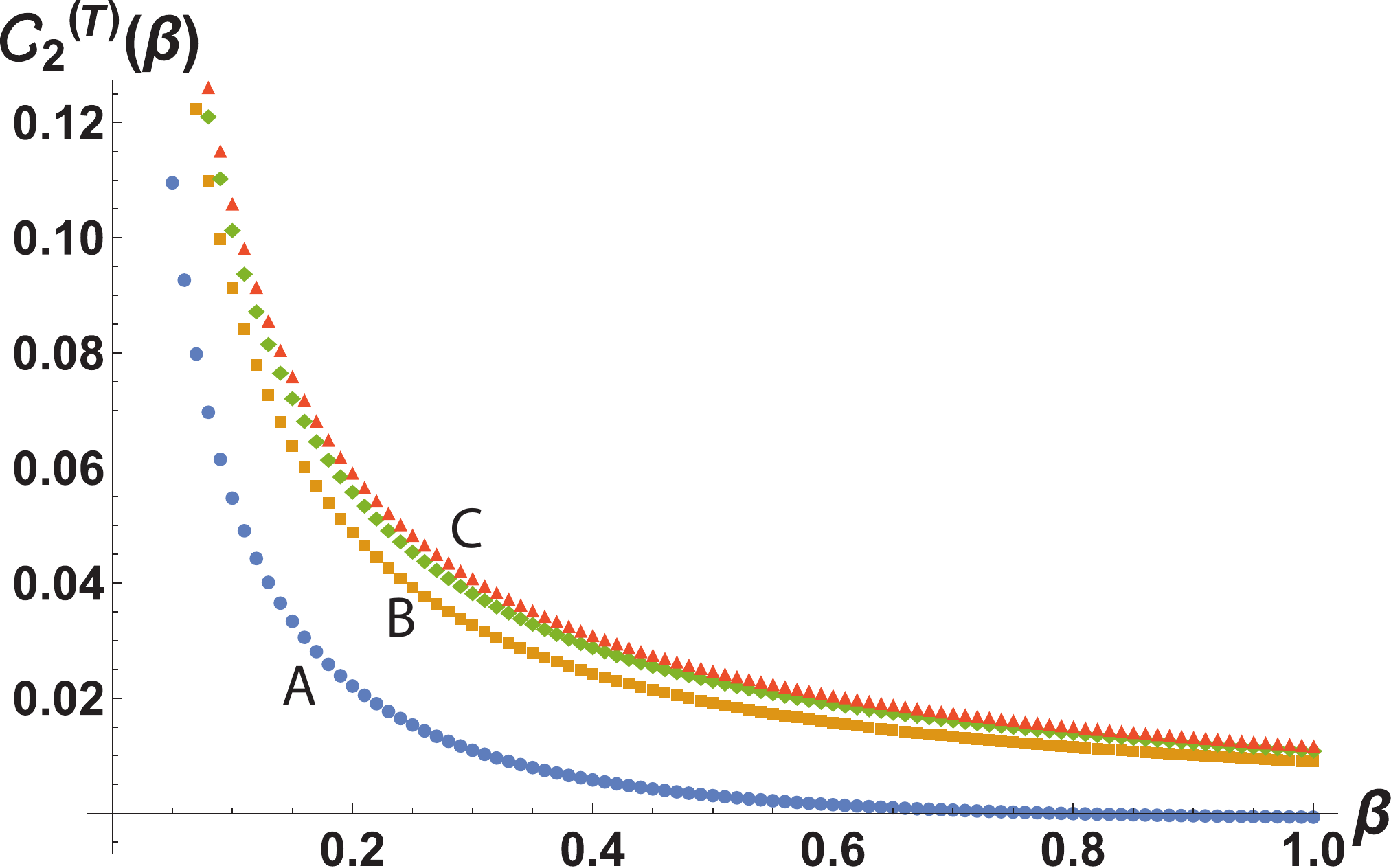}
        \caption{The coefficient function $\mathcal{C}^{(T)}_2(\beta)$ for the second-order cumulant expansion 	
	result in Eq.~(\ref{FinalCoexistCooperon}). This coefficient is expressed as the combination of terms $\{\mathcal{C}_{j}(\beta)\}$
	shown in Eq.~(\ref{CT2Def}). The individual terms in the latter equation arise from the three different 
	time-ordering sectors depicted in Fig.~\ref{SecondOrderDiagrams}, minus the square of the first-order result. 
	The terms $\mathcal{C}_1(\beta)$ and $\mathcal{C}_2(\beta)$ appear already at first order [see Fig.~\ref{FirstOrderCoexisitingPlot}],
	while the rest obtain exclusively from the second-order diagrams. 
	To evaluate the total coefficient in Eq.~(\ref{CT2Def}), the terms $\mathcal{C}_1$ and $\mathcal{C}_2$ are each 
	approximated using the series acceleration technique described in Appendix~\ref{SeriesAcceleration}. 
	The remaining terms $\{\mathcal{C}_4,\mathcal{C}_7,\mathcal{C}_8\}$ 
	are each given by a triply-infinite summation over Airy eigenfunction energy levels [Eq.~(\ref{Energies})]. 
	Here they are estimated by series truncation after summing the first $N_{\e}$ energy levels. 
	We plot $\mathcal{C}^{(T)}_2(\beta)$ for several values of $N_{\e}$ to show convergence. 
	A: $N_{\e} = 1$, 
	B: $N_{\e} = 2$, 
	($N_{\e} = 3$ unlabeled), 
	C: $N_{\e} = 4$. 
	We see that $\mathcal{C}^{(T)}_2(\beta) > 0 $, so that the second-order contribution to the dephasing rate in Eq.~(\ref{FinalCoexistCooperon})
	is \emph{negative}. In other words, the second-order contribution is \emph{rephasing}. 
	This is similar to the second-order correction to the short-time expansion for the pure diffusive bath 
	calculation [Eq.~(\ref{FinalAnswerDiffusive}) and Fig.~\ref{SecondOrderTotalPlot}].}
	\label{SecondOrderCoexisitingPlot}
\end{figure}

At first order, we find that the diffusive bath enhances the dephasing rate, but at second order we again find a 
positive \textit{rephasing} contribution. Interestingly, this exactly parallels the short-time expansion 
for the purely diffusive bath in Eq.~(\ref{FinalAnswerDiffusive}). By contrast to that calculation,
the results here obtain in the limit of large virtual times $\eta \rightarrow \infty$. 
In the latter limit, the dual bath-averaged Cooperon can be cast in the form
\begin{align}\label{FinalCoexistCooperon}
	\!\!\!
	\frac{\cc(\eta)}{\cc_M(\eta)} 
	=&\,
	A(\beta)
	\,
	\exp\left\{
		-
		\frac{\eta D}{a^2}
		\left[
		\begin{aligned}
			&
			\left(\frac{\Gamma_t}{\Gamma_M}\right)
			\mathcal{C}_1\left(\beta\right) 
			\\&
			- 
			\left(\frac{\Gamma_t}{\Gamma_M}\right)^2
			\mathcal{C}^{(T)}_2\left(\beta\right)
			\\&
			+
			\mathcal{O}\left(\frac{\Gamma_t}{\Gamma_M}\right)^3
		\end{aligned}
		\right]
	\right\},
\end{align}
where the prefactor $A(\beta)$ and rate coefficients 
$\mathcal{C}_1\left(\beta\right)$ and $\mathcal{C}^{(T)}_2\left(\beta\right)$ are 
dimensionless functions of the diffusion constant ratio $\beta$, defined by Eq.~(\ref{BetaDef}). 
In Eq.~(\ref{FinalCoexistCooperon}), $\cc_M(\eta)$ is the exact result for the 
Markovian-dephased Cooperon in Eq.~(\ref{MarkovianCooperon}), while the coupling strengths 
$\Gamma_M$ and $\Gamma_t$ for the Coulomb Markovian and spin-triplet diffusive baths were defined by 
Eqs.~(\ref{GammaM}) and (\ref{Gammat}), respectively. 

The rate coefficient functions $\mathcal{C}_1(\beta)$ and $\mathcal{C}^{(T)}_2\left(\beta\right)$ are both 
positive, so that the former (latter) enhances (suppresses) the dephasing relative to the  
Markovian result $\cc_M(\eta)$. 
Similar to the second-order coefficient in the bare expansion for the diffusive bath [Eqs.~(\ref{FinalAnswerDiffusive}) and (\ref{GT2Def})],
the second-order rate coefficient obtains from a combination of several terms, 
corresponding to contributions from the three different time-ordering topologies explicated
in Eqs.~(\ref{SecondOrderTimeSectors}) and Fig.~\ref{SecondOrderDiagrams}, minus the square of the first order result.
The net result can be expressed by the combination 
\begin{align}\label{CT2Def}
	\mathcal{C}^{(T)}_2\left(\beta\right) 
	\equiv 
	\mathcal{C}_4\left(\beta\right)
	-
	\mathcal{C}_1\left(\beta\right) \, \mathcal{C}_2\left(\beta\right)
	+
	\mathcal{C}_7\left(\beta\right)
	+
	\mathcal{C}_8\left(\beta\right),
\end{align}
where the components $\{\mathcal{C}_j(\beta)\}$ are precisely defined in Appendix~\ref{MoreFirstAndSecondOrderDetails}.
Each of the functions $\{\mathcal{C}_j(\beta)\}$ can be expressed through one or more infinite summations over
the Airy bound states that solve the Markovian problem [Eqs.~(\ref{Energies}) and (\ref{Eigenfunctions})]; summands 
are given by integrals over matrix elements involving these eigenfunctions.  
Figs. \ref{FirstOrderCoexisitingPlot} and \ref{SecondOrderCoexisitingPlot} show that the sign of these 
functions give net dephasing and rephasing contributions at first and second order, respectively. 
We note that the dimensionless perturbative parameter in the dual-bath cumulant expansion Eq.~(\ref{FinalCoexistCooperon}) 
is the ratio of the bath coupling constants $(\Gamma_t/\Gamma_M)$, which is independent of temperature
(unless further renormalization of the triplet coupling strength is taken into account---see Sec.~\ref{Sec:TripletEnh}), 
in contrast to the pure diffusive bath expansion in Eq.~(\ref{FinalAnswerDiffusive}).

The fact that the second-order correction in Eq.~(\ref{FinalCoexistCooperon}) is rephasing is a main result of this section. 
This is similar to the pure diffusive bath result in Eq.~(\ref{FinalAnswerDiffusive}), except that Eq.~(\ref{FinalCoexistCooperon}) 
is well-defined in the long-virtual-time limit $\eta \rightarrow \infty$, and gives a valid dephasing time via Eq.~(\ref{DephasingTimeDefinition}). 
The advent of rephasing corrections due to the diffusive bath beyond first order was anticipated by the RG study in Ref.~\cite{Liao18},
which located a nontrivial fixed point in a $d = 4 - \epsilon$ expansion. The fixed point arises due to vertex corrections
neglected in the SCBA [Eq.~(\ref{SCBA1})] that suppress the Cooperon-bath coupling strength. Such corrections 
are absent in the Markovian case [Appendix~\ref{MarkovianBath}].


\subsection{Comparison with self-consistent calculations}

In Sec.~\ref{DiffusiveDephasing}, we argued that the perturbative calculation of dephasing due to the purely diffusive bath is \textit{not} 
stabilized by the SCBA [Eqs.~(\ref{SCBA1}) and (\ref{SCBA2})]. 
Here we discuss how the long-virtual-time result for the combined baths obtained in Eq.~(\ref{FinalCoexistCooperon}) compares to self-consistent calculations. 

On physical grounds (but see below), low frequencies $|\omega| < \tau^{-1}_{\phi}$ are not expected to contribute to dephasing \cite{CS,AleinerBlanter,NZA}. 
This motivates the self-consistent truncation of first-order perturbation theory, 
\begin{align}
\label{FrequencyCutting}
	\frac{1}{\tau_{\phi}} 
	= 
	2
	\int
		\frac{dk}{2\pi}
	\int\limits^\infty_{\tau^{-1}_{\phi}}
		\frac{d\omega}{2\pi}
	\frac{2 D k^2}{\left[(Dk^2)^2+\omega^2\right]}
	\,
	\Delta(\omega,k),
\end{align}
where $\Delta(\omega,k)$ is the noise kernel [\emph{cf}.\ the SCBA in Eq.~(\ref{SCBA1})]. 

We emphasize that Eq.~(\ref{FrequencyCutting}) is an ad-hoc prescription; 
for the case of the Markovian bath, it \emph{artificially introduces non-Markovianity} into the effective 
bath kernel by excising the IR. 
By contrast, the dephasing problem itself, stated in Eqs.~(\ref{CooperStoch}) and (\ref{BathKernel}), obtains from the fully controlled
many-body perturbation theory \cite{AAG} or Finkel'stein nonlinear sigma model \cite{Liao17}. 
For the quasi-1D case, dephasing of the Cooperon in Eq.~(\ref{GeneralPathIntegralUnfolded}) 
due to either the Coulomb [Eq.~(\ref{AAKNoiseKernel})] or diffusive [Eq.~(\ref{DiffusiveNoise})] baths is equivalent to a UV-convergent,
but strongly coupled quantum field theory \cite{Liao18}. 
The exact solution for the Markovian case exploits the AAK transfer matrix procedure reviewed in 
Sec.~\ref{MarkovianDephasing}. For the Coulomb interaction this exact result gives the same temperature dependence 
as the self-consistent Eq.~(\ref{FrequencyCutting}). As we illustrate below, however, the self-consistent 
calculation fails for the dual bath calculation, due to the sensitive dependence on the cutting procedure. 
Fortunately, our perturbative expansion using the AAK Markovian bath to regularize the diffusive one gives the fully controlled, 
long-time result (through second order) in Eq.~(\ref{FinalCoexistCooperon}).
  
Applying the self-consistent Eq.~(\ref{FrequencyCutting})
to the 1D dual-bath Cooperon studied in this (and the next) section, we find the result
\begin{align}
\label{FrequencyCuttingDualBathResult}
	\!\!
	\frac{1}{\tau_{\phi}} 
	=&\, 
	\frac{2 D}{\pi^{2/3} a^2}
	\left[
		1
		+
		\left(\frac{\Gamma_t}{\Gamma_M}\right)
		H(\beta)			
	\right]^{2/3}
	\nonumber
\\
	=&\,
	\frac{2 D}{\pi^{2/3} a^2}
\left[
\begin{aligned}
	&\,
	1 
	+ \frac{2}{3}\left(\frac{\Gamma_t}{\Gamma_M}\right) H(\beta)
	\\&\,
	-\frac{1}{9}\left(\frac{\Gamma_t}{\Gamma_M}\right)^2 \left[H(\beta)\right]^2
	+\mathcal{O}\left(\frac{\Gamma_t}{\Gamma_M}\right)^3
\end{aligned}
\right]\!\!,\!\!\!
\end{align}
where 
\begin{align}\label{H(beta)}
	H(\beta) \equiv \frac{\sqrt{\beta}}{(1+\sqrt{\beta})(1+\beta)}.
\end{align}
This result shares some features with Eq.~(\ref{FinalCoexistCooperon}), in particular the appearance of a second-order rephasing contribution. 
However, the numerical prefactor is incorrect for the pure Coulomb contribution, and more importantly the coefficient function at $n^{\mathrm{th}}$ order
$[H(\beta)]^n$ is qualitatively incorrect. The function $H(\beta)$ vanishes as $\beta^{1/2}$ in the $\beta \rightarrow 0$ limit; this corresponds to
strong interaction coupling $\gamma_t \rightarrow -\infty$ [Eq.~(\ref{BetaDef})].   
By contrast, the coefficient functions $\mathcal{C}_1(\beta)$ and $\mathcal{C}^{(T)}_2\left(\beta\right)$ 
shown in Figs.~\ref{FirstOrderCoexisitingPlot} and \ref{SecondOrderCoexisitingPlot} asymptote to nonzero values as $\beta \rightarrow 0$. 
This discrepancy is an order-of-limits issue; the $\eta \rightarrow \infty$ and $\beta \rightarrow 0$ limits do not commute. 
The self-consistent result in Eq.~(\ref{FrequencyCuttingDualBathResult}) obtains at first order in perturbation theory,
valid at best for short virtual times $\eta \leq 1/\tau_\phi$. 

On the other hand, the result in Eq.~(\ref{FinalCoexistCooperon}) with coefficient functions $\{C_1(\beta),\mathcal{C}^{(T)}_2\left(\beta\right)\}$ 
obtains only in the large $\eta$-limit. 
Since $\beta \propto D_t$, the effect of the diffusive bath must vanish at $\beta = 0$, see Eq.~(\ref{S_I}).
Eq.~(\ref{FinalCoexistCooperon}) is 
valid for $D \eta / a^2 \gtrsim 1/\beta$.
The controlled result in Eq.~(\ref{FinalCoexistCooperon}) is particularly relevant 
in the context of the RG enhancement of the spin exchange interaction $\gamma_t$ near a MIT 
\cite{BK,PunnooseFinkelstein,50YearsFinkel,QPTKravchenko}.

Another key point is that the second-order rephasing term in the self-consistent result obtained
in Eq.~(\ref{FrequencyCuttingDualBathResult}) depends sensitively on the cutting scheme, 
due to the strong infrared divergence in Eq.~(\ref{FrequencyCutting}). 
Instead of applying Eq.~(\ref{FrequencyCutting}) simultaneously to the Markovian and diffusive baths,
we can instead use the AAK result for the dephasing time due to the Coulomb bath, 
$(\tau^{AAK}_\phi)^{-1} = |\alpha'_0| D/a^2$ [Eq.~(\ref{MarkovianCooperon})] 
to cut the frequency integral for the correction due to the diffusive bath. This gives 
\begin{align}
	\left(\frac{1}{\tau_{\phi}}\right)_{diff} 
	=&\,
	\frac{4}{\pi}
	\left(\frac{\Gamma_t}{\sqrt{2D}}\right)
	H(\beta)
	\sqrt{\tau^{AAK}_\phi}
\nonumber
\\
	=&\, 
	\frac{2^{3/2}}{\pi |\alpha'_0|^{1/2}}
	\left(\frac{D}{a^2}\right)
	\left(\frac{\Gamma_t}{\Gamma_M}\right) H(\beta),
	\label{FrequencyCuttingScheme2}
\end{align}
with no second-order rephasing correction. 

We conclude that the sensitive dependence on the infrared makes self-consistency even qualitatively
incorrect for the dephasing of quasi-1D systems due to a diffusive bath, with or without an additional regularizing Markovian bath.

\section{Dephasing due to combined diffusive and Markovian baths: Calculation}\label{CalculationDetails}

\subsection{General method}\label{GeneralDetails}

In this section we provide an overview of the calculation leading to the dual-bath result in Eq.~(\ref{FinalCoexistCooperon}). 
This arises due to inelastic electron-electron scattering mediated by both screened Coulomb and spin-triplet exchange interactions, 
encoded respectively in the Markovian AAK bath [Eq.~(\ref{AAKNoiseKernel})] and the diffusive (non-Markovian) bath [Eq.~(\ref{DiffusiveNoise})].  
We employ the same cumulant expansion as in Sec.~\ref{DiffusiveDephasing}, expanding perturbatively in the diffusive bath whilst 
treating the Markovian bath exactly. The latter requires that we work in terms of the relative- $\rho(\tau)$ and center-of-time $R(\tau)$ 
coordinates [see Eq.~(\ref{NoiseBathAction})].  
We define $\cc_M(\eta)$ to be the exact bath-averaged Cooperon in the pure Markovian limit, given by 
Eqs.~(\ref{AAKReductionII}) and (\ref{MarkovianCooperon}).

The dual-bath-averaged Cooperon is expressed as the path integral
\begin{widetext}
\begin{align}\label{GeneralPathIntegralFolded}
	\cc(\eta)
	=&\,
	\frac{D}{2}
	\int d R_0
	\int\limits_{R(0) = R_0}^{R(\eta) = x}
	\mathcal{D}
	R(\tau)
	\int\limits_{\rho(0) = 0}^{\rho(\eta) = 0}
	\mathcal{D}
	\rho(\tau)
	\,
	\exp
\left[
\begin{aligned}
&-
	\int\limits_{0}^\eta
	d \tau
	\,
	\left\{
	\frac{1}{D}
	\left[\dot{R}(\tau)\right]^2
+
	\frac{1}{4D}
	\left[\dot{\rho}(\tau)\right]^2	
+
	\frac{\Gamma_M}{D} 
	\big|\rho(\tau)\big|
	\right\}
\\	
&-    S_I\left[R(\tau),\rho(\tau)\right]
\end{aligned}	
\right],
\end{align}
where $S_I$ is as in Eq.~(\ref{NoiseBathAction}), with $\Delta(\omega,k) \rightarrow \Delta_t(\omega,k)$ given by 
Eq.~(\ref{DiffusiveNoise}). 

We let $\langle\cdots\rangle_0^R$ and $\langle\cdots\rangle_0^{\rho}$ denote the averages with respect to the 
noiseless $R(\tau)$ 
and 
Markovian-bath-averaged $\rho(\tau)$ actions, respectively. 
As in Sec.~\ref{DiffusiveDephasing}, the cumulant expansion boils down to the calculation of expectation values of powers 
of the perturbing action. In general, to do an $n^{th}$ order calculation, we must evaluate
\begin{align}
    \nonumber
    \langle
    S_I^n
    \rangle_0
    =&
	\Gamma_t^n
	\int\frac{dk_1}{2\pi}
	\cdots
	\int\frac{dk_n}{2\pi}
	\,
	\int\limits_{0}^\eta
	d \tau_{1a}
	\int\limits_{0}^\eta
	d \tau_{1b}
	\cdots
	\int\limits_{0}^\eta
	d \tau_{na}
	\int\limits_{0}^\eta
	d \tau_{nb}
\\
\nonumber
&\,
\times
	\left[
	e^{-D_t k^2|\tau_{1a}-\tau_{1b}|/2}-e^{-D_t k^2(\tau_{1a}+\tau_{1b})/2}
	\right]
	\times
	\cdots
	\times
	\left[
	e^{-D_t k^2|\tau_{na}-\tau_{nb}|/2}-e^{-D_t k^2(\tau_{na}+\tau_{nb})/2}
	\right]
\\
\nonumber
&\,
\times
	\left\langle
	\exp\left[
	ik_1
	\bigg(
	R(\tau_{1a}) - R(\tau_{1b})
	\bigg)
	\right]
	\times
	\cdots
	\times
	\exp\left[
	ik_n
	\bigg(
	R(\tau_{na}) - R(\tau_{nb})
	\bigg)
	\right]
	\right\rangle_0^R
\\
\label{SIn}
&\,
\times
	\left\langle
	\sin\left[
	\frac{k_1\rho(\tau_{1a})}{2}
	\right]
	\sin\left[
	\frac{k_1\rho(\tau_{1b})}{2}
	\right]
	\times
	\cdots
	\times
	\sin\left[
	\frac{k_n\rho(\tau_{na})}{2}
	\right]
	\sin\left[
	\frac{k_n\rho(\tau_{nb})}{2}
	\right]
	\right\rangle_0^\rho,
\end{align}
where the elementary frequency integrations have already been carried out via 
\begin{align}\label{FrequencyIntegration}
	\int\frac{d \omega}{2\pi}
	\left(\frac{2 D_t k^2}{D_t^2 k^4 + \omega^2}\right)
	\left[ 
		e^{-i\omega(\tau_a-\tau_b)/2}
		-
		e^{-i\omega(\tau_a+\tau_b)/2}
	\right] 
	=&\, 
	\exp\left[-D_t k^2\left(\frac{|\tau_b-\tau_a|}{2}\right)\right] - \exp\left[-D_t k^2 \left(\frac{\tau_b+\tau_a}{2}\right)\right]
\nonumber\\
	\equiv&\, 
	\tilde{T}_1(D_t,k,\tau_{a},\tau_{b}) - \tilde{T}_2(D_t,k,\tau_{a},\tau_{b}).
\end{align}
We have two functional averages to perform:
\bsub\label{DualBathVertexCorrs}
\begin{align}
\label{RVertexCorr}
	F^n_{R}(k's,\tau's) 
	\equiv&\, 	
	\left\langle
	\exp\left[
	i
	k_1
	\,
	\bigg(
	R(\tau_{1a}) - R(\tau_{1b})
	\bigg)
	\right]
	\times
	\cdots
	\times
	\exp\left[
	i
	k_n
	\,
	\bigg(
	R(\tau_{na}) - R(\tau_{nb})
	\bigg)
	\right]
	\right\rangle_0^R
\\
\label{rhoVertexCorr}
	F^n_{\rho}(k's,\tau's) 
	\equiv&\,
	\left\langle
	\sin\left[
	\frac{k_1\rho(\tau_{1a})}{2}
	\right]
	\sin\left[
	\frac{k_1\rho(\tau_{1b})}{2}
	\right]
	\times
	\cdots
	\times
	\sin\left[
	\frac{k_n\rho(\tau_{na})}{2}
	\right]
	\sin\left[
	\frac{k_n\rho(\tau_{nb})}{2}
	\right]
	\right\rangle_0^\rho.
\end{align}
\esub
\end{widetext}
As in Sec.~\ref{DiffusiveDephasing}, 
the path integral expectation values will have nontrivial dependencies on the ordering of the time variables. 
In general, there are $(2n)!$ such orderings, corresponding to the permutation group $S_{2n}$ acting on the time variables. 
However, the integral is preserved under a subgroup of order $(n! \cdot 2^n)$, generated by the $n$ exchange operations 
$\tau_{ja} \leftrightarrow \tau_{jb}$ and the $n!$ operations that permute 
$(k_j,\tau_{ja},\tau_{jb}) \rightarrow (k_{\sigma(j)},\tau_{\sigma(j)a},\tau_{\sigma(j)b}),$ for $\sigma \in S_n$. 
With these symmetries in mind, we can restrict the $\tau$-integration region so that 
$
	\tau_{1a} < \tau_{2a} < \cdots <\tau_{na},
$ 
and $\tau_{j a} < \tau_{j b}$ for each $j\in\{1,...,n\}.$ We thus only need to consider $(2n)!/(n! \cdot 2^n)$ 
\textit{topologically distinct} time ordering sectors, which are in direct, one-to-one correspondence with the 
topologically distinct diagrams contributing to the Cooperon at $n^{th}$ order in the field theory description [see Appendix \ref{FieldTheory}]. 
We define these regions in $\tau$-space as 
$\{\Omega^n_s\}$ [with $1 \leq s \leq (2n)!/(n!\cdot2^n)$], 
generalizing the second order decomposition in Eq.~(\ref{SecondOrderTimeSectors}) [see also Fig.~\ref{SecondOrderDiagrams}]. 
This folding of the integration region produces a leading factor of $(n! \cdot 2^n)$.

The Gaussian functional average over $R(\tau)$ in Eq.~(\ref{RVertexCorr})
follows from Wick's theorem, since the vertex operator products 
appearing in it are charge-neutral [see Appendix~\ref{Correlators}]. 
While the $\rho(\tau)$ expectation value in Eq.~(\ref{rhoVertexCorr})
cannot be evaluated similarly due to the confining potential from the Markovian bath in Eq.~(\ref{GeneralPathIntegralFolded}), 
we can instead express it as an expansion in terms of the eigenfunctions $\{\psi^0_j(\rho;a)$\} in 
Eqs.~(\ref{EigenfunctionsEven}) and (\ref{EigenfunctionsOdd}). 
This gives $2n+1$ distinct summations over the eigenenergies [Eq.~(\ref{Energies})], 
and introduces the matrix elements 
\begin{align}\label{MatrixElement1}
	S_{ij}(k;a) 
	\equiv&\, 
	\langle \e_i | \sin\left[\frac{k\hat{\rho}}{2}\right] | \e_j \rangle
\nonumber\\
	=&\, 
	\int\limits_{-\infty}^{\infty} 
	d\rho\ 
	\,
	\psi^0_i(\rho;a) 
	\,
	\sin\left(\frac{k\rho}{2}\right)
	\,
	\psi^0_j(\rho;a).
\end{align}
Eq.~(\ref{MatrixElement1}) vanishes unless one of the eigenfunctions is even and the other is odd,
a parity selection rule. 
When evaluating Eq.~(\ref{rhoVertexCorr}), we need to keep in mind the Dirichlet boundary conditions 
$\rho(0) = \rho(\eta) = 0$. Since the odd-parity eigenfunctions in Eq.~(\ref{EigenfunctionsOdd}) vanish
at the origin, when Eq.~(\ref{rhoVertexCorr}) is evaluated by inserting $2n+1$ resolutions of the identity,
the first and last energies in the Trotterization must have even parity.  
Moreover, the parity of $\e_j$ must correspond to the parity of $j$, for all $j \in \{0,1,2,\ldots,2n\}$.  
We give the explicit general form of the $\rho$-correlator in Eq.~(\ref{rhoVertexCorr})
in Appendix~\ref{MoreGeneralDetails}.

We can simplify our calculation by scaling our integration variables to make them dimensionless,
\begin{align}
\label{VariableScaling}
	(\tau,\rho,k) 
	\rightarrow&\, 
	\left(
		\eta \, \tau,
		a \, \rho,
		\frac{k}{a}
	\right),
\end{align}
where $\eta$ is the external virtual time argument of the Cooperon in Eq.~(\ref{GeneralPathIntegralFolded}), 
and 
$a$ denotes the characteristic dephasing length scale for the Markovian screened-Coulomb problem
[Eq.~(\ref{LengthScale})]. 
This leaves
\begin{align}
	\langle S_1^n \rangle_0 
	=&\, 
	\Gamma_t^n 
	\left(\frac{\Gamma_M}{D^2}\right)^{n/3}
	\eta^{2n}
	\times
	\big[\text{dimensionless integrals}\big]
\nonumber\\
	\label{Scalingfs}
	\equiv&\, 
	\left(\frac{\Gamma_t}{\Gamma_M}\right)^n
	\,
	f_n(z,\beta),
\end{align}
where $f_n$ is a function only of the 
dimensionless external virtual time variable 
\begin{align}\label{zDef}
	z \equiv (\eta D / a^2),
\end{align}
and of the diffusion constant ratio $\beta$ [Eq.~(\ref{BetaDef})]. 
The scaling procedure shows that the control parameter in the cumulant expansion 
is the ratio of the bath  coupling strengths
\begin{align}
    \frac{\Gamma_t}{\Gamma_M} = \frac{3\kappa_0\gamma_t^2}{2 \chi_0(1 - \gamma_t)}.
\end{align}
\begin{widetext}
The scaling also sends the matrix elements to dimensionless functions
\begin{align}\label{MatrixElement2}
	S_{ij}(k;a) 
	\rightarrow 
	S_{ij}(k/a;a) 
	\equiv 
	\tilde{S}_{ij}(k) 
	=  
	\frac{1}{\sqrt{|\alpha'_i|}}\frac{1}{\Ai(\alpha'_i)\Ai'(\alpha_j)}
	\int\limits_0^{\infty}d\rho\ \, \Ai\big(\rho+\alpha'_i\big) \, \Ai\big(\rho+\alpha_j\big) \, \sin\left(\frac{k\rho}{2}\right),
\end{align}
where $i,j$ correspond to even and odd-parity energies, respectively. For a given pair of levels, 
$\tilde{S}_{ij}(k)$ can be efficiently computed numerically as a function of the dimensionless momentum parameter $k$.
This is facilitated by the superexponential fall off of the Airy functions with positive argument, 
$\Ai(x \gg 1) \sim x^{-1/4} \, \exp\left[-(2/3) \, x^{3/2}\right]$.

Our strategy is then as follows.
After computing the correlators in Eq.~(\ref{DualBathVertexCorrs}),
Eq.~(\ref{SIn}) requires the evaluation of 
$n$ momentum integrations 
and 
$2n$ integrations over the $\tau$ variables, the latter of which are partitioned into $(2n)!/(n!\cdot2^n)$ 
topological sectors $\{\Omega^n_s\}$. Note that each frequency integration produces 2 terms [dubbed ``$\tilde{T}_1$'' and ``$\tilde{T}_2$'' in Eq.~(\ref{FrequencyIntegration})], 
so that the full expression has $2^n$ terms. The $\tau$ integrations are elementary and we carry them out in closed form, 
defining the ``$T$-kernels'' to be 
\begin{align}\label{GeneralTKernel}
	T^s_{p_1 p_2 \cdots p_n}(z,\beta,\{k_i\},\{\sigma_i\}) 
	\equiv
	\frac{z^{2n}}{f_0(z)}
	\int\limits_{\Omega^n_s}d^{2n}\vex{\tau}
	&\,
	\tilde{T}_{p_1}(z\beta,k_1,\tau_{1a},\tau_{1b})
	\,
	\times
	\cdots
	\times
	\,
	\tilde{T}_{p_n}(z\beta,k_n,\tau_{na},\tau_{nb})
\nonumber\\
	&\, 
	\times
	F^n_{R}\big(\{k_i\},\{\tilde{\tau}_i\}\big) 
	\exp\left\{-z\left[\sigma_{2n} + \sum_{i=1}^{2n}\tilde{\tau}_i(\sigma_{i-1}-\sigma_{i})\right]\right\},
\end{align}
\end{widetext}
where $p_m \in \{1,2\}$ for $1 \leq m \leq n$, 
$f_0(z)$ is the pure Markovian-dephased Cooperon amplitude defined via Eq.~(\ref{MarkovianCooperon}), 
and $\tilde{\tau}_j$ gives the time-ordering of the $2n$ $\{\tau_{m \alpha}\}$ variables ($\alpha \in \{a,b\}$). 
The parameters $\{\sigma_i\}$ are dimensionless eigenenergies for the pure Markovian problem [Eq.~(\ref{Energies})];
these are 
Airy prime zeroes $\sigma_i \in \{\alpha'_j\}$ (for $i \in \{0,2,\ldots,2n\}$) 
or 
Airy zeroes $\sigma_i \in \{\alpha_j\}$ (for $i \in \{1,3,\ldots,2n-1\}$). 

At this point there are $n$ momentum integrations left over the $T$-kernels and the matrix elements in 
Eq.~(\ref{MatrixElement2}). We tabulate the matrix elements as functions of $k$ ahead of time and 
compute the final momentum integrations numerically. This gives a numerical function of $z$ [Eq.~(\ref{zDef})]
for each sector and choice of energies $\{\sigma_i\}$. 
The final result from which the $\eta$-dependence of the expectation value $\left\langle S_I^n \right\rangle_0$
can be extracted requires summing over $(2 n + 1)$ energy arguments over the appropriate Airy or Airy prime zeroes.


\subsection{First order}\label{FirstOrderDetails}

We now specialize to the first order calculation $\left\langle S_I \right\rangle_0$ and carry out the time integrations to get 
\begin{widetext}
\begin{align}\label{FirstOrderInitialExpression}
	f_1(z,\beta) =
	\sum_{i_0,i_1,i_2 = 0}^\infty	
	\frac{2}{\sqrt{|\alpha'_{i_0}\alpha'_{i_2}|}}
	\int
	\frac{d k}{2\pi}\,
	\tilde{S}_{i_1i_2}(k)
	\,
	\tilde{S}_{i_0i_1}(k)
	\left[
		T_1(z,\beta,k^2,\alpha'_{i_2},\alpha_{i_1},\alpha'_{i_0})
		-
		T_2(z,\beta,k^2,\alpha'_{i_2},\alpha_{i_1},\alpha'_{i_0})
	\right],
\end{align}
where $f_1$ is defined via Eq.~(\ref{Scalingfs}), 
and 
where the ``$T$-kernels'' introduced in Eq.~(\ref{GeneralTKernel}) take the first-order forms
\bsub
\label{TKernelsFirstOrder}
\begin{align}
	\label{T1Def}
	T_1(z,\beta,k^2,\sigma_2,\sigma_1,\sigma_0) 
	=&\, 
	\frac{1}{f_0(z)}
	\left(\frac{1}{\sigma_2-\sigma_1+\theta k^2}\right)
	\left[
		\left(\frac{e^{z\sigma_1} e^{- z\theta k^2}-e^{z\sigma_0}}{\sigma_0-\sigma_1 + \theta k^2}\right)
		+
		\left(\frac{e^{z\sigma_2}-e^{z\sigma_0}}{\sigma_2-\sigma_0}\right)
	\right],
\\
	\label{T2Def}
	T_2(z,\beta,k^2,\sigma_2,\sigma_1,\sigma_0) 
	=&\, 
	\frac{1}{f_0(z)}
	\left(\frac{1}{\sigma_2-\sigma_1+\theta k^2}\right)
	\left[
		\left(\frac{e^{z\sigma_1}e^{-\theta zk^2}-e^{z\sigma_0}e^{-\beta z k^2}}{\sigma_0-\sigma_1 + (\theta-\beta) k^2}\right)
		+
		\left(\frac{e^{z\sigma_2}-e^{z\sigma_0}e^{-\beta  zk^2}}{\sigma_2-\sigma_0 + \beta k^2}\right)
	\right],
\end{align}
\esub
\end{widetext}
where $\theta \equiv 1/4 + \beta/2$ and $f_0(z)$ is given by Eq.~(\ref{MarkovianCooperon}).
We may now finish the evaluation of Eq.~(\ref{FirstOrderInitialExpression}) by numerically integrating in $k$, 
using our analytical expressions for $T_{1,2}$ and tabulated values of the matrix elements $\tilde{S}_{ij}(k)$ as functions of $k$. 
Each triplet of Airy and Airy-prime zero labels $\{i_0,i_1,i_2\}$ contributes a summand to $f_1(z,\beta)$ 
in Eq.~(\ref{FirstOrderInitialExpression}).

\begin{table}[b!]
    \centering
    \begin{tabular}{|c||c|c|c|c|}
    \hline
    \multicolumn{5}{|c|}{First order contributions} \\
    \hline
    Term & $i_0=i_2=0$ & $i_0>i_2=0$ & $i_2>i_0=0$ & Else\\
        \hline
        $T_1$ & $\begin{aligned}zC_1(\beta,i_1)\\-C_2(\beta,i_1)\end{aligned}$    & $C_3(\beta,i_0,i_1)$ &  $C_3(\beta,i_2,i_1)$ & 0\\
        \hline
        $T_2$ &  $\begin{aligned}C_5(\beta,0,i_1)\\
        -
	D_1(z;\beta,i_1)\end{aligned}$&  $C_5(\beta,i_0,i_1)$ & 0 & 0\\
        \hline
    \end{tabular}
	\caption{We tabulate the asymptotic, large $z \rightarrow \infty$ form 
	of non-vanishing contributions to the first-order dual-bath cumulant amplitude in Eq.~(\ref{FirstOrderInitialExpression}), 
	retaining all contributions decaying slower than $1/z$.
	Here $z$ is the dimensionless virtual time argument of the Cooperon [Eq.~(\ref{zDef})]. 
	$T_1$ and $T_2$ correspond to the two terms in Eq.~(\ref{FirstOrderInitialExpression}), 
	defined by Eq.~(\ref{TKernelsFirstOrder}). The indices $\{i_0,i_1,i_2\}$ refer to the three 
	distinct sums over the energy eigenvalues [Airy or Airy-prime zeroes, Eq.~(\ref{Energies})] in 
	Eq.~(\ref{FirstOrderInitialExpression}). 
	The $\{C_i,D_1\}$ functions are defined in Appendix~\ref{MoreFirstAndSecondOrderDetails}, and are independent of 
	$z$, with the exception of $D_1$, which decays asymptotically as $z^{-1/2}$ for large $z$.
	The table indicates that the first-order dephasing rate comes solely from the $i_0 = i_2=0$ energy terms in $T_1$.}
	\label{FirstOrderTable}
\end{table}

\begin{figure}[b!]
    \includegraphics[width=0.40\textwidth]{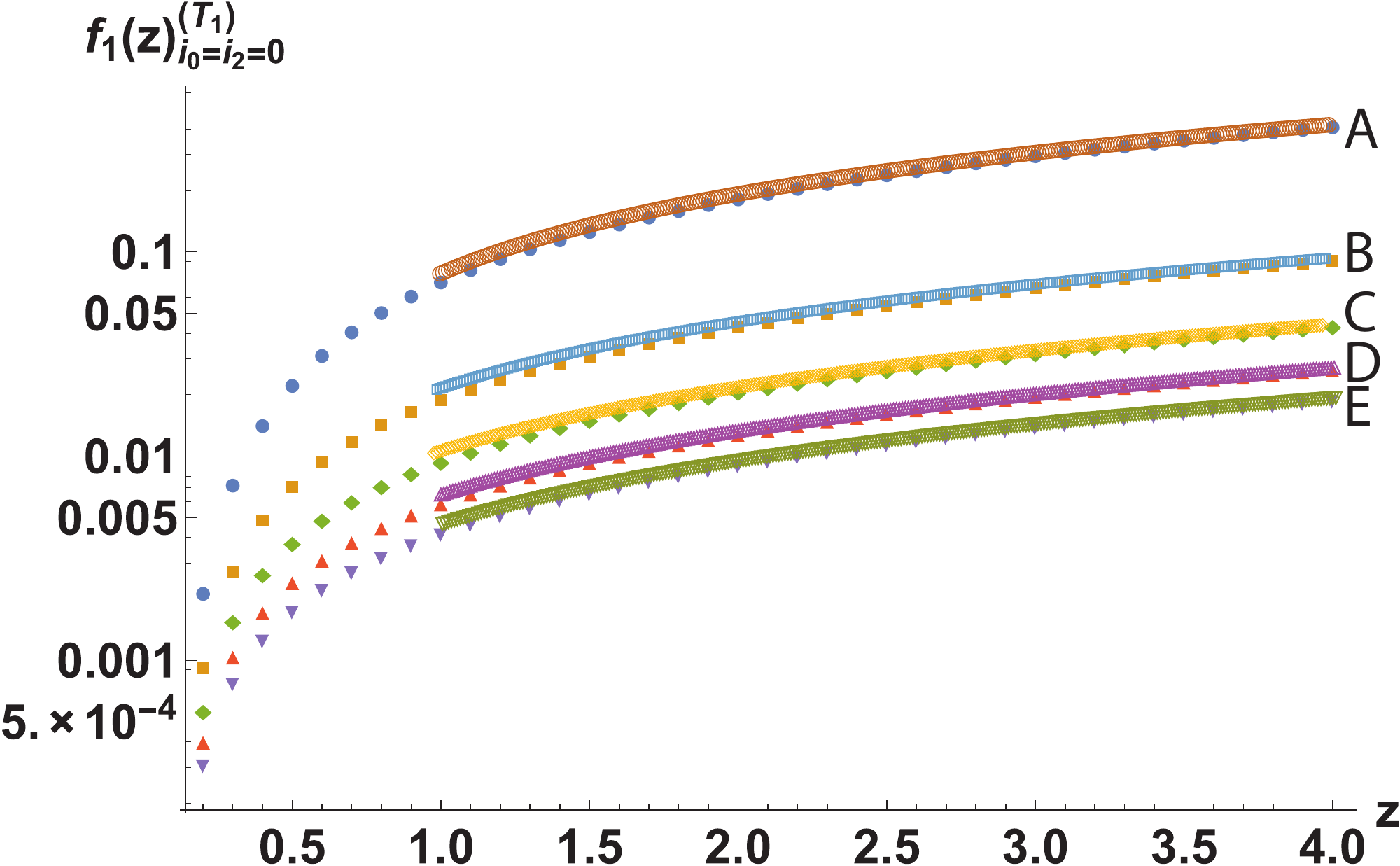}
	\caption{Comparison of full numerical calculations with the asymptotic, 
	large-virtual-time $z \rightarrow \infty$ approximation to the 
	first-order dual-bath cumulant amplitude in Eq.~(\ref{FirstOrderInitialExpression}).
	Here we define $f_1(z)_{i_0=i_2=0}^{(T_1)}$ to be the contribution to $f_1$ [Eqs.~(\ref{Scalingfs}) and (\ref{FirstOrderInitialExpression})]
	from $T_1$ [Eq.~(\ref{T1Def})], at energy $i_0=i_2=0$, with $i_1$ left unspecified.
	This plot demonstrates that the asymptotic form for this $T_1$ contribution 
	with Airy prime summand labels pinned at $i_0 = i_2 = 0$, as specified in Table~\ref{FirstOrderTable}, is correct. 
	Above, $i_j$ gives the energy level of the $j^{th}$ energy and $z = \eta D/a^2$. 
	In this plot, we vary the Airy summand label $i_1$ from $0$ to $4$, and we set $\beta = 1$ [Eq.~(\ref{BetaDef})]. 
	A: $i_1 = 0$, 
	B: $i_1 = 1$, 
	C: $i_1 = 2$, 
	D: $i_1 = 3$, 
	E: $i_1 = 4$. 
	The solid curves are the asymptotic approximation in Table~\ref{FirstOrderTable}, $zC_1(1,i_1)-C_2(1,i_1)$. 
	The symbols obtain from the full numerical integration of Eq.~(\ref{FirstOrderInitialExpression}), using the 
	exact expression for $T_1(z,1,k^2,\alpha'_0,\alpha_{i_1},\alpha'_0)$ from Eq.~(\ref{T1Def}).	
	}
    \hfill
    \label{T1GroundComparison}
\end{figure}

The described numerical procedure produces the correct $z$-dependence for any set of these three energy labels,  
but it turns out that only a small subset of possible energy combinations are non-decaying in the large-$z$ (virtual time) limit. 
In particular, the only contributions that grow with $z$ are linear terms that arise when both the initial and final energies 
are at the ground state, $i_0 = i_2 = 0$. 
The linear terms arise from the limit
\begin{align}
	(T_1)_{i_0 = i_2 = 0} 
	\rightarrow 
	\frac{z |\alpha'_0|}{\alpha'_{0} - \alpha_{i_1} + \theta k^2} + \text{(const.)}, 
\end{align}
which gives the total asymptotic contribution to the expectation value in Eq.~(\ref{FirstOrderInitialExpression}) as
\begin{align}\label{FirstOrderLinear}
	f_1(z,\beta) 
	\rightarrow&\, 
	z \, \mathcal{C}_1(\beta) 
	\equiv 
	z \, \sum_{i_1 = 0}^\infty C_1(\beta,i_1)
\nonumber\\
	\equiv&\, 
	z \, 
	\sum_{i_1 = 0}^\infty
	\int \frac{dk}{2 \pi}
	\,
	\frac{
		2 \left[\tilde{S}_{i_10}(k)\right]^2
	}{
	(\alpha'_0 - \alpha_{i_1} + \theta k^2)
	}.
\end{align}
We note that the sum in Eq.~(\ref{FirstOrderLinear}) is slowly converging but can be 
numerically estimated by series acceleration [Appendix~\ref{SeriesAcceleration}]. 

\begin{figure}[b!]
    \centering
    \includegraphics[width=0.40\textwidth]{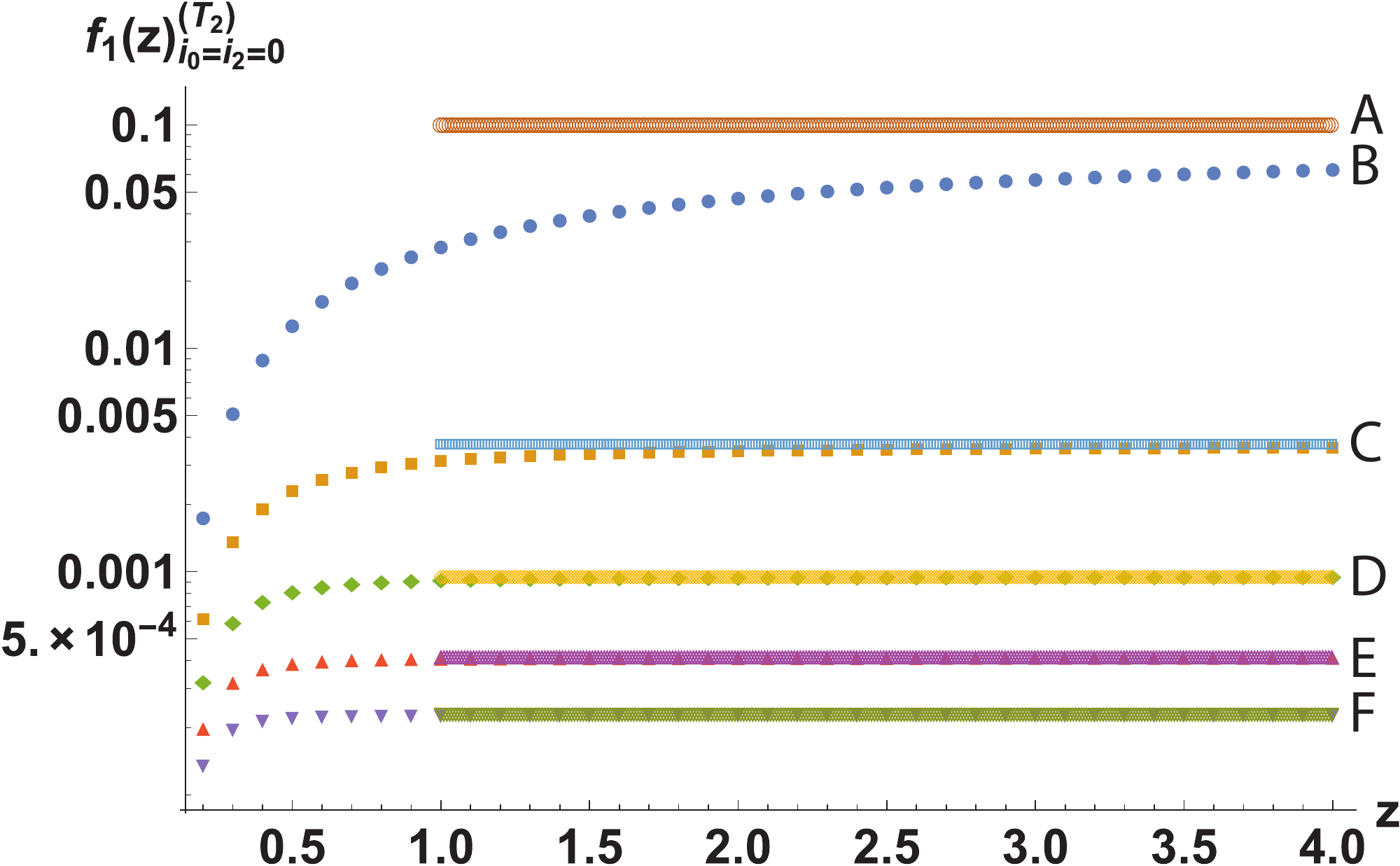}
	\caption{Comparison of full numerical calculations with the asymptotic, 
	large-virtual-time $z \rightarrow \infty$ approximation to the 
	first-order dual-bath cumulant amplitude in Eq.~(\ref{FirstOrderInitialExpression}).
	Here we define $f_1(z)_{i_0=i_2=0}^{(T_2)}$ to be the contribution to $f_1$ [Eqs.~(\ref{Scalingfs}) and (\ref{FirstOrderInitialExpression})]
	from $T_2$ [Eq.~(\ref{T2Def})], at energy $i_0=i_2=0$, with $i_1$ left unspecified.
	This plot demonstrates that the asymptotic form for this $T_2$ contribution 
	with Airy prime summand labels pinned at $i_0 = i_2 = 0$, as specified in Table~\ref{FirstOrderTable}, is correct. 
	Above, $i_j$ gives the energy level of the $j^{th}$ energy and $z = \eta D/a^2$. 
	In this plot, we vary the Airy summand label $i_1$ from $0$ to $4$, and we set $\beta = 1$ [Eq.~(\ref{BetaDef})]. 
	A,B: $i_1 = 0$, 
	C: $i_1 = 1$, 
	D: $i_1 = 2$, 
	E: $i_1 = 3$, 
	F: $i_1 = 4$. 
	The solid curves are the asymptotic approximation given by $C_5(\beta,0,i_1)$, [see Table~\ref{FirstOrderTable} and the second line of Eq.~(\ref{FullFirstOrder})]. 
	The symbols obtain from the full numerical integration of Eq.~(\ref{FirstOrderInitialExpression}), using the 
	exact expression for $T_2(z,1,k^2,\alpha'_0,\alpha_{i_1},\alpha'_0)$ from Eq.~(\ref{T2Def}).
	We note that the convergence speed is determined by the decay of $D_1(z;\beta,i_1) \sim z^{-1/2}$.
	}
    \label{T2GroundComparison}
    \hfill
\end{figure}

The full asymptotic analysis is summarized in Table~\ref{FirstOrderTable}. 
Contributions decaying slower than $1/z$ will be important for the second-order calculation, 
so we carefully retain them all in Table~\ref{FirstOrderTable}.
Explicit expressions for the coefficients listed in this table appear 
in Appendix~\ref{MoreFirstAndSecondOrderDetails}. 
Defining
\begin{align}\label{CSum}
	\mathcal{C}_j(\beta)
	&\equiv 
	\sum_{i's}C_j(\beta,i's),
\end{align}
\begin{align}\label{D1Sum}
	\mathcal{D}_1(z;\beta)
	&\equiv
	\sum_{i's}D_1(z;\beta,i's),
\end{align}
and summing over all energies, we have the final asymptotic formula for the first order expectation value
\begin{align}\label{FullFirstOrder}
	\langle S_I \rangle_0 
	\simeq 	
	\left(\frac{\Gamma_t}{\Gamma_M}\right)
	\left[
	\begin{aligned}
		z\mathcal \, \mathcal{C}_1(\beta) - \mathcal{C}_2(\beta) + 2\mathcal{C}_3(\beta)
		\\ 
		- \mathcal{C}_5(\beta)  +  \mathcal{D}_1(z;\beta)
	\end{aligned}
	\right].
\end{align}
We find that $C_1(\beta,i_1)$ and $C_2(\beta,i_1)$ are strictly positive, 
while $C_3(\beta,i_0,i_1)$ and $C_5(\beta,i_2,i_1)$ 
can alternate in sign [Appendix~\ref{MoreFirstAndSecondOrderDetails}]. 
We plot the asymptotic ``$T_1$'' and ``$T_2$'' contributions to $f_1(z,\beta)$
[the first and second lines inside the square brackets of Eq.~(\ref{FullFirstOrder}), see Table~\ref{FirstOrderTable}]
in Figs.~\ref{T1GroundComparison} and \ref{T2GroundComparison}, and compare these to the direct numerical
integration of Eq.~(\ref{FirstOrderInitialExpression}) using the full expressions for $T_{1,2}$ in 
Eq.~(\ref{TKernelsFirstOrder}).

\subsection{Second order overview}\label{SecondOrderOverview}

We evaluate the second order contribution to the cumulant expansion for the dual-bath
model along the lines explained above, although there are complications not seen at first order. 
As before, we obtain the $R$-correlator in Eq.~(\ref{RVertexCorr}) in closed form. 
Here, however, the $R$-correlator breaks into three terms for the three distinct topological sectors
in Eq.~(\ref{SecondOrderTimeSectors}), pictured in Fig.~\ref{SecondOrderDiagrams}; explicit 
expressions appear in Appendix~\ref{Correlators}.  
We again use the Airy-eigenfunction expansion to Trotterize the $\rho$-correlator Eq.~(\ref{rhoVertexCorr}), 
which here gives a 5-fold summation over the energy states. 
The $\tau$-integrations in Eq.~(\ref{SIn}) are again elementary, 
but here each topological sector $\Omega_s$ has 4 ``$T$-kernel'' [Eq.~(\ref{GeneralTKernel})] terms, 
$\{T^s_{11},T^s_{12},T^s_{21},T^s_{22}\}$. 
In terms of the $T$-kernels, the second-order amplitude function [see Eq.~(\ref{Scalingfs})] is
\begin{widetext}
\begin{align}
\label{SecondOrderInitial}
	f_2(z,\beta) =
	\sum_{i_0,\cdots,i_4 = 0}^\infty
	\frac{8}{\sqrt{|\alpha'_{i_0}\alpha'_{i_4}|}}
	\int
	\frac{d k_1}{2\pi}
	\int
	\frac{d k_2}{2\pi}
	\left\{
	\begin{aligned}
		\left[T^1_{11}-T^1_{21}-T^1_{12}+T^1_{22}\right]
		\tilde{S}_{i_4i_3}(k_2)\tilde{S}_{i_3i_2}(k_2)\tilde{S}_{i_2i_1}(k_1)\tilde{S}_{i_1i_0}(k_1)
	\\
		+
		\left[T^2_{11}-T^2_{21}-T^2_{12}+T^2_{22}\right]
		\tilde{S}_{i_4i_3}(k_2)\tilde{S}_{i_3i_2}(k_1)\tilde{S}_{i_2i_1}(k_2)\tilde{S}_{i_1i_0}(k_1)
	\\
		+
		\left[T^3_{11}-T^3_{21}-T^3_{12}+T^3_{22}\right]
		\tilde{S}_{i_4i_3}(k_1)\tilde{S}_{i_3i_2}(k_2)\tilde{S}_{i_2i_1}(k_2)\tilde{S}_{i_1i_0}(k_1)
    \end{aligned}
    \right\},
\end{align}    
\end{widetext}
where 
$	
	T^s_{jk} 
	= 
	T^s_{jk}(z,\beta;k_1,k_2;\alpha'_{i_4},\alpha_{i_3},\alpha'_{i_2},\alpha_{i_1},\alpha'_{i_0}).
$ 

Using analytical forms for the 12 second-order $T$-kernels and tabulated values for the matrix elements $\tilde{S}_{ij}(k)$
[Eq.~(\ref{MatrixElement2})], we can numerically perform the momentum integrations. This produces the functional 
dependence on the dimensionless virtual time $z$ [Eq.~(\ref{zDef})] for a given topological sector and set of energy levels
$\{\alpha'_{i_4},\alpha_{i_3},\alpha'_{i_2},\alpha_{i_1},\alpha'_{i_0}\}$. 
As in the first order case, we can extract simple expressions for the asymptotic behavior in the large $z \rightarrow \infty$ 
limit. The important contributions at second order are summarized in Table~\ref{SecondOrderTable}.

\begin{table*}[t!]
\centering
\begin{tabular}{|c||c|c|c|c|c|}
    \hline
    \multicolumn{6}{|c|}{Second order contributions} \\
    \hline
    Term & $i_0=i_2=i_4=0$ & $i_0>i_2=i_4=0$ & $i_2>i_0=i_4=0$ & $i_4>i_0=i_2=0$ & else\\
    \hline
    \hline
    $T^1_{11}$ & $\begin{aligned}
    z^2C_1(\beta,i_1)C_1(\beta,i_3)\\
    -2z\left[
    \begin{aligned}
    C_1(\beta,i_1)C_2(\beta,i_3)\\
    +C_1(\beta,i_3)C_2(\beta,i_1)
    \end{aligned}
    \right]
    \end{aligned}$
    & $2z C_1(\beta,i_3)C_3(\beta,i_0,i_1)$ & $2z C_4(\beta,i_1,i_2,i_3)$ & $2z C_1(\beta,i_1)C_3(\beta,i_4,i_3)$ & $\mathcal{O}(1)$\\
    \hline
    $T^1_{21}$ &  $\begin{aligned}
    2z C_1(\beta,i_3) C_5(\beta,0,i_1)\\
    - C_1(\beta, i_3) D_2(z;\beta,i_1)
    \end{aligned}$ & $2z C_1(\beta,i_3) C_5(\beta,i_0,i_1)$   & $\mathcal{O}(1)$ & $\mathcal{O}(1)$ & $\mathcal{O}(1)$\\
    \hline
    $T^1_{12}$ &  $\begin{aligned}-2z C_1(\beta,i_1) D_1(z;\beta,i_3)\\ + C_1(\beta,i_1) D_2(z;\beta,i_3)\end{aligned}$  & $\mathcal{O}(1)$ & $\mathcal{O}(1)$ & $\mathcal{O}(1)$ & $\mathcal{O}(1)$\\
    \hline
    $T^1_{22}$ &  $\mathcal{O}(1)$  &  $\mathcal{O}(1)$ & $\mathcal{O}(1)$ & $\mathcal{O}(1)$ & $\mathcal{O}(1)$\\
    \hline
    \hline
    $T^2_{11}$ &  $2z C_7(\beta,i_1,0,i_3)$  &  $\mathcal{O}(1)$   & $2z C_7(\beta,i_1,i_2,i_3)$ & $\mathcal{O}(1)$ & $\mathcal{O}(1)$\\
    \hline
    $T^2_{21}$ &  $\mathcal{O}(1)$  &  $\mathcal{O}(1)$   & $\mathcal{O}(1)$ & $\mathcal{O}(1)$ & $\mathcal{O}(1)$\\
    \hline
    $T^2_{12}$ &  $\mathcal{O}(1)$  &  $\mathcal{O}(1)$   & $\mathcal{O}(1)$ & $\mathcal{O}(1)$ & $\mathcal{O}(1)$\\
    \hline
    $T^2_{22}$ &  $\mathcal{O}(1)$  &  $\mathcal{O}(1)$   & $\mathcal{O}(1)$ & $\mathcal{O}(1)$ & $\mathcal{O}(1)$\\
    \hline
    \hline
    $T^3_{11}$ &  $2z C_8(\beta,i_1,0,i_3)$  &  $\mathcal{O}(1)$   & $2z C_8(\beta,i_1,i_2,i_3)$ & $\mathcal{O}(1)$ & $\mathcal{O}(1)$\\
    \hline
    $T^3_{21}$ &  $\mathcal{O}(1)$  &  $\mathcal{O}(1)$   & $\mathcal{O}(1)$ & $\mathcal{O}(1)$ & $\mathcal{O}(1)$\\
    \hline
    $T^3_{12}$ &  $\mathcal{O}(1)$  &  $\mathcal{O}(1)$   & $\mathcal{O}(1)$ & $\mathcal{O}(1)$ & $\mathcal{O}(1)$\\
    \hline
    $T^3_{22}$ &  $\mathcal{O}(1)$  &  $\mathcal{O}(1)$   & $\mathcal{O}(1)$ & $\mathcal{O}(1)$ & $\mathcal{O}(1)$\\
    \hline
\end{tabular}
\caption{
	We tabulate the asymptotic, large $z \rightarrow \infty$ form 
	of non-vanishing contributions to the second-order dual-bath cumulant amplitude in Eq.~(\ref{SecondOrderInitial}).
	Here $z$ is the dimensionless virtual time argument of the Cooperon [Eq.~(\ref{zDef})]. 
	The terms labeled $\{T^s_{j k}\}$ correspond to the amplitudes arising from the twelve terms in Eq.~(\ref{SecondOrderInitial}).
	The indices $\{i_0,i_1,i_2,i_3,i_4\}$ refer to the five distinct sums over the energy eigenvalues [Airy or Airy-prime zeroes, Eq.~(\ref{Energies})] in 
	Eq.~(\ref{SecondOrderInitial}). 
	The functions $\{C_i\}$ are defined in Appendix~\ref{MoreFirstAndSecondOrderDetails} and are independent of 
	$z$. Two additional amplitudes $D_1$ and $D_2$ depend on $z$ through asymptotic powers laws with exponents given by $-1/2$ and $1/2$, respectively;
	however, the contributions of these two amplitudes cancel out. 
	From the table, we see that the $i_0 = i_2 = i_4 = 0$ terms from $T^1_{11}$ 
	give quadratic $z^2$ contributions that exactly cancel the square of the first order terms 
	in the cumulant expansion, 
	Table~\ref{FirstOrderTable} and Eq.~(\ref{FullFirstOrder}). 
	Several other nontrivial cancellations take place between the various contributions. 
	Finally, we see that additional linear terms arise from the $T^s_{11}$ terms in the $s = \{2,3\}$ topological sectors
	[see Fig.~\ref{SecondOrderDiagrams} for the diagrammatic definition of the latter].
}
\label{SecondOrderTable}
\end{table*}

\begin{figure}[b!]    
    \includegraphics[width=0.4\textwidth]{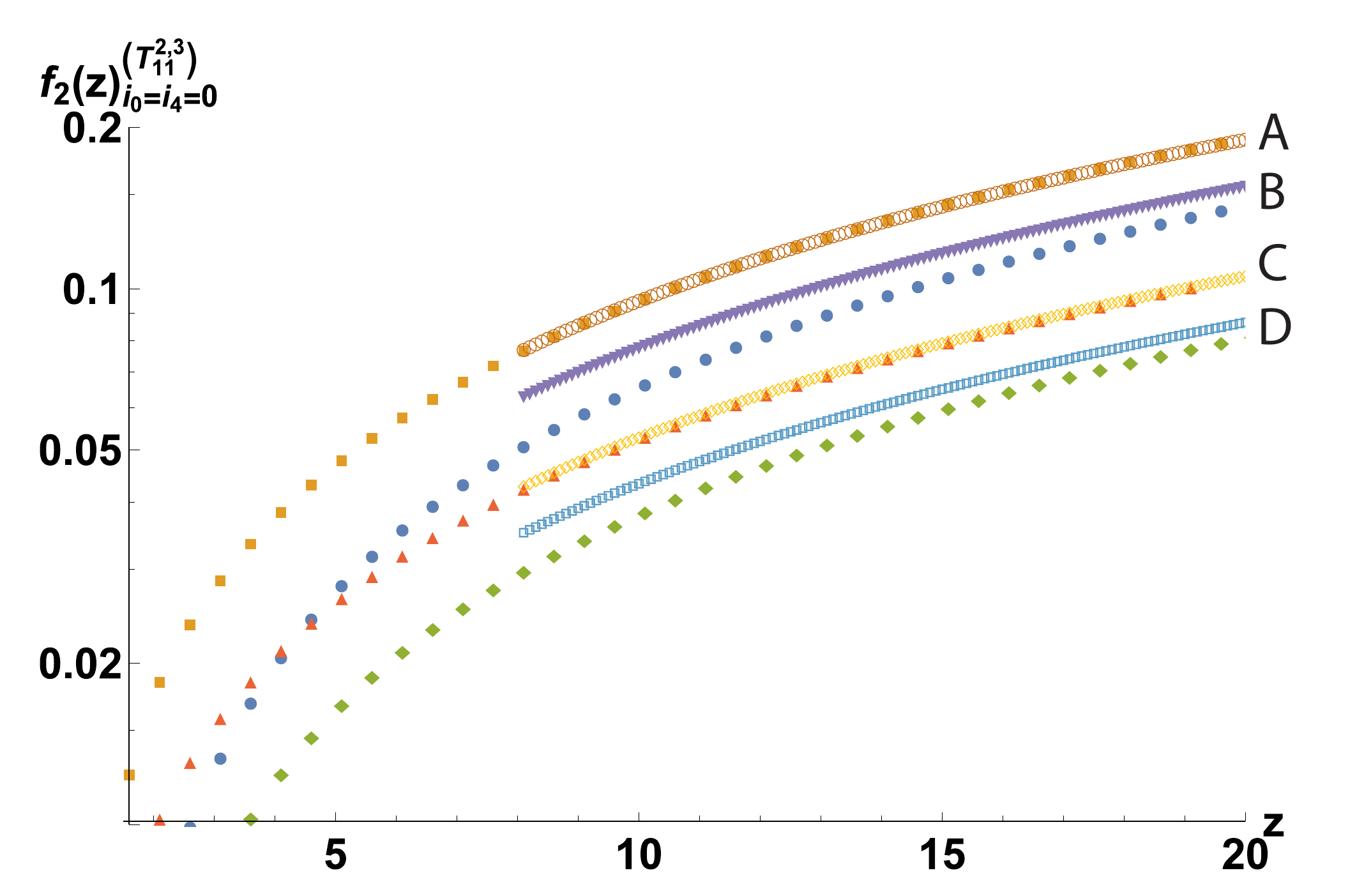}
    \caption{Comparison of full numerical calculations with the asymptotic, 
	large-virtual-time $z \rightarrow \infty$ approximation to the 
	second-order dual-bath cumulant amplitude in Eq.~(\ref{SecondOrderInitial}).
	This plot compares the contributions obtained by numerically integrating the terms involving 
	$T_{11}^s$ with $s \in \{2,3\}$ in Eq.~(\ref{SecondOrderInitial}) to the 
	asymptotic formulae quoted in Table~\ref{SecondOrderTable}, involving the amplitude
	functions $C_7(\beta,i_1,i_2,i_3)$ and $C_8(\beta,i_1,i_2,i_3)$. 
	Here we define $f_2(z)_{i_0=i_4=0}^{(T^s_{11})}$ to be the contribution to $f_2$ [Eqs.~(\ref{Scalingfs}) and (\ref{SecondOrderInitial})]
	from $T^s_{11}$ [Eq.~(\ref{GeneralTKernel})], at energy $i_0=i_4=0$, with $i_1,i_2,i_3$ left unspecified.
	This plot demonstrates that the asymptotic form for the $T^s_{11}$ contribution 
	with Airy prime summand labels pinned at $i_0 = i_4 = 0$ and $s = 2,3$, as specified in Table~\ref{SecondOrderTable}, is correct. 
	Above, $i_j$ gives the energy level of the $j^{th}$ energy and $z = \eta D/a^2$. 
	In this plot, we set $i_1 = i_3 = 0$, vary the Airy summand label $i_2$ from $0$ to $1$, plot both sectors $s = 2$ and $s=3$, and we set $\beta = 1$ [Eq.~(\ref{BetaDef})]. 
	A: $i_2 = 0, s = 3$, 
	B: $i_2 = 0, s = 2$, 
	C: $i_2 = 1, s = 3$, 
	D: $i_2 = 1, s = 2$.
	The solid curves are the asymptotic approximation in Table~\ref{SecondOrderTable}, $2zC_{7,8}(1,0,i_2,0)$. 
	The symbols obtain from the full numerical integration of Eq.~(\ref{SecondOrderInitial}), using the 
	exact expression for $T^{2,3}_{11}(z,1,k_1^2,k_2^2,\alpha'_0,\alpha_0,\alpha'_{i_2},\alpha_0,\alpha'_0)$ [Eq.~(\ref{GeneralTKernel})].
	}
    \label{T11_2,3_Comparison}
\end{figure}

The coefficient functions $\{C_i,D_i\}$ listed in Table~\ref{SecondOrderTable} are 
defined explicitly in Appendix~\ref{MoreFirstAndSecondOrderDetails}. 
The $D_1$ and $D_2$ amplitudes in Table~\ref{SecondOrderTable} are nontrivial functions of $z$ that 
grow via power laws with exponents approximately equal to $-1/2$ and  $1/2$, respectively.
However, these terms do not contribute to the final rate due to cancellations. 
The total surviving asymptotic contributions to the second order result are
then [via Eq.~(\ref{Scalingfs})]
\bsub
\begin{align}
	\langle S^2_I \rangle_0^{s=1} 
	&\rightarrow  
	\left(\frac{\Gamma_t}{\Gamma_M}\right)^2 
	\left\{
	\begin{aligned}
		z^2 \mathcal{C}_1^2 + 2 z \mathcal{C}_1\big[ 2 \mathcal{C}_3 - 2\mathcal{C}_2\big]
	\\
		+
		2z\mathcal{C}_4
		- 
		2z\mathcal{C}_1\big[\mathcal{C}_5 - \mathcal{D}_1(z)\big]
    \end{aligned}
    \right\},
\\
	\langle S^2_I \rangle_0^{s=2} 
	&\rightarrow 
	\left(\frac{\Gamma_t}{\Gamma_M}\right)^2 
	2z\mathcal{C}_7,
\\
	\langle S^2_I \rangle_0^{s=3} 
	&\rightarrow 
	\left(\frac{\Gamma_t}{\Gamma_M}\right)^2
	2z\mathcal{C}_8,
\end{align}
\esub
where the superscript $s$ denotes the topological sector [Fig.~\ref{SecondOrderDiagrams}].
When we subtract the square of the first order contribution, we find several nontrivial cancellations. 
Using Eq.~(\ref{FullFirstOrder}), the final second order contribution to the cumulant expansion 
[as in Eq.~(\ref{CumulantDefinition})] 
is
\begin{align}\label{SI2}
	\langle S_I^2 \rangle_0 - \langle S_I \rangle_0^2 
	&= 
	2z\left(\frac{\Gamma_t}{\Gamma_M}\right)^2
	\left[
	\begin{aligned}
		\mathcal{C}_4 - \mathcal{C}_1\mathcal{C}_2 
	\\ 
		+\mathcal{C}_7 + \mathcal{C}_8 
	\end{aligned}
	\right]
	+ \mathcal{O}(1).
\end{align}
This is the basis of the results in Eqs.~(\ref{FinalCoexistCooperon}) and (\ref{CT2Def}), quoted in Sec.~\ref{CoexistingDephasingResults}. 
In Eq.~(\ref{SI2}), the quadratic $z^2$ terms have canceled exactly. 
This cancellation is the key difference between this dual diffusive- and Markovian-bath result,
and the pure-diffusive bath expansion studied in Sec.~\ref{DiffusiveDephasing} [Eq.~(\ref{FinalAnswerDiffusive})]. 
By killing off the higher-order $\eta$ dependence, this cancellation stabilizes the cumulant expansion at long virtual 
times and determines a well-defined dephasing rate via Eq.~(\ref{DephasingTimeDefinition}). 
In Eq.~(\ref{SI2}), the amplitudes $\mathcal{C}_3$, $\mathcal{C}_5$, and $\mathcal{D}_1$ cancel out as well. 
This is notable, because the energy-level-resolved amplitudes in Table~\ref{SecondOrderTable},
$C_3(\beta,i_0,i_1)$, 
$C_5(\beta,i_0,i_1)$, 
and 
$D_1(z;\beta,i_3)$ contribute with level $\{i_j\}$-dependent
signs, so that the overall sign of the total contribution 
would require numerically precise summation over many Airy energy levels. 

The asymptotic results for $T^{2,3}_{11}$ summarized in Table~\ref{SecondOrderTable}
are compared to the direct numerical integration of Eq.~(\ref{SecondOrderInitial}) 
in Fig.~\ref{T11_2,3_Comparison}.

The major dephasing contribution at second order comes from the 
$-\mathcal{C}_1 \mathcal{C}_2$ term in Eq.~(\ref{SI2}). 
We can evaluate this with series acceleration. 
Summing $\mathcal{C}_4$, $\mathcal{C}_7$, and $ \mathcal{C}_8$ to the first four excited states 
gives the plot in Fig.~\ref{SecondOrderCoexisitingPlot}, which definitively shows that the net contribution is \textit{positive}, 
and thus that the final second order result is \textit{rephasing}, as discussed 
in Sec.~\ref{CoexistingDephasingResults}.

\section{Discussion and Conclusion}\label{Conclusions}

\subsection{Dephasing and rephasing}

We have investigated the dephasing of quasi-1D systems with diffusive noise baths as a possible analytical window into the physics of the many-body localization (MBL) transition. 
The diffusive noise bath is self-generated by short-ranged interactions at intermediate temperatures in an isolated fermion system with quenched disorder. 

In Sec.~\ref{DiffusiveDephasing}, we studied a system with a purely diffusive noise bath. This describes a quasi-1D system of ultracold fermions with contact 
interactions, which could be a potential realization for MBL. 
We calculated the Cooperon through second-order perturbation theory around the noninteracting result. 
This procedure gives a well-defined, divergence-free short-time expansion, but the latter breaks down at long times and fails to yield a meaningful result for the dephasing time.

To better understand our results, we also considered in Secs.~\ref{CoexistingDephasingResults}--\ref{CalculationDetails} a physical regularization of the previous problem, 
in which the diffusive bath coexists with the Markovian noise bath that arises due to screened Coulomb interactions. This corresponds to an SU(2) spin-symmetric many-channel 
quantum wire with Coulomb and short-range spin-triplet-exchange interactions. Treating the Coulomb bath exactly 
(via an extension of the AAK technique \cite{AAK}) and the diffusive bath perturbatively, we 
found that the presence of the Coulomb interaction stabilizes the perturbation theory, giving well-defined corrections to the dephasing rate due to the Coulomb interaction. 
Reminiscent of our results in Sec.~\ref{DiffusiveDephasing}, we find that the second-order term in this expansion is \emph{rephasing}.
Rephasing corrections are consistent with RG results showing that vertex corrections can suppress the Cooperon-noise coupling strength \cite{Liao18}. 

We demonstrated that commonly used (in higher dimensional dephasing calculations) self-consistent 
approaches that bootstrap lowest-order perturbation theory fail to capture the correct physics 
of dephasing due to the diffusive bath, in both the absence and presence of an additional Markovian bath. 

A key goal for future work is to obtain a nonperturbative understanding for dephasing due to the diffusive bath in isolation.
As articulated in Ref.~\cite{Liao18}, this can be cast as a type of self-interacting polymer problem, with a gyration radius
that sets the dephasing length in the long-virtual-time limit.

\subsection{Enhancement of dephasing in spin SU(2)-symmetric quantum wires via itinerate spin-exchange interactions \label{Sec:TripletEnh}}

\begin{figure}[b!]
    \centering
    \includegraphics[width=0.40\textwidth]{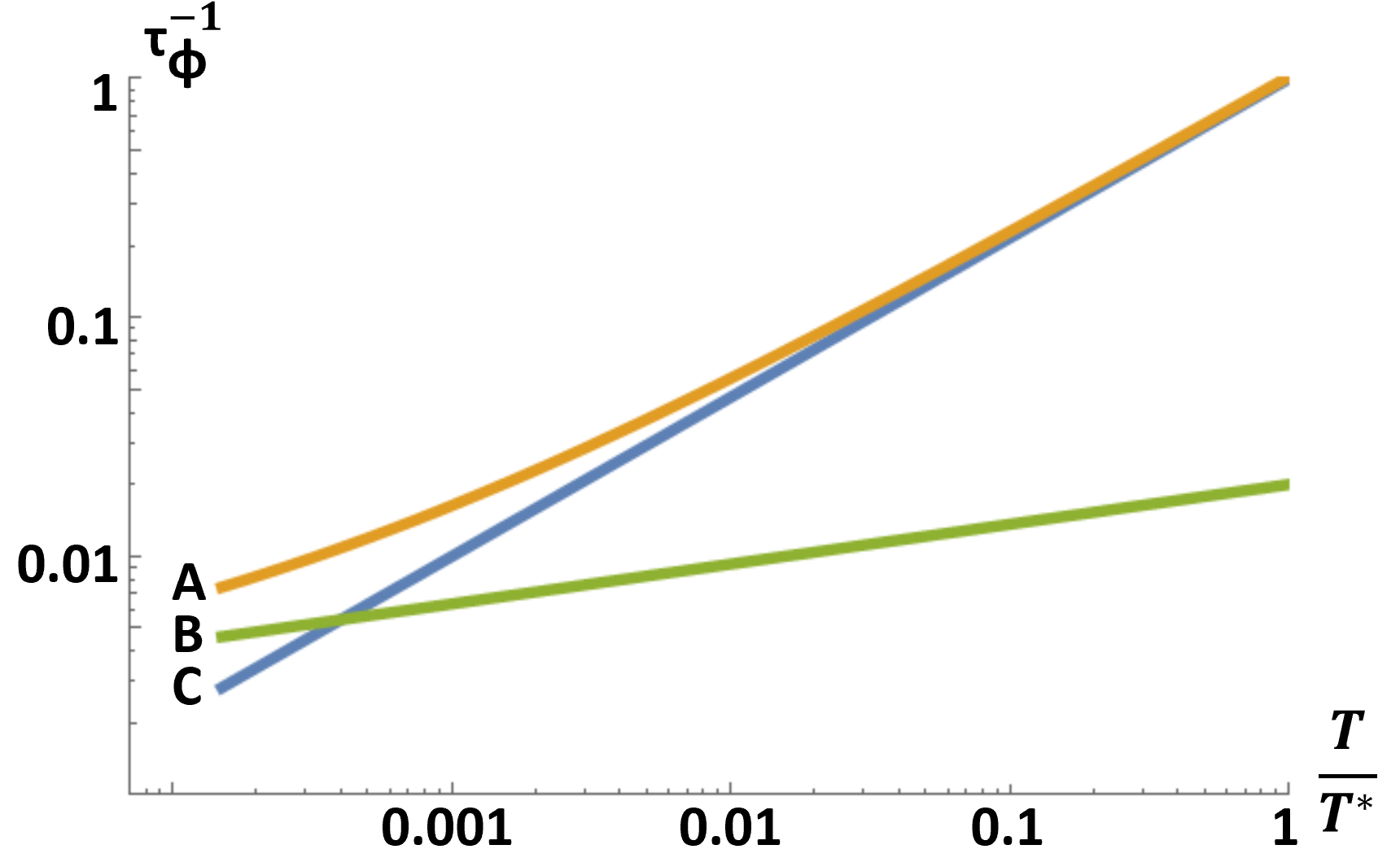}
	\caption{Schematic log-log plot of the temperature dependence of the Cooperon dephasing rate 
	for a quasi-1D quantum wire with coexisting short-ranged spin-triplet-exchange and long-ranged Coulomb interactions. 
	The temperature scale is set by $T^*$, as defined in Eq.~(\ref{TemperatureCorrectionEquation}). 
	A: Dephasing rate due to coexisting interactions including the first quantum corrections to $\gamma_t$, as given by Eq.~(\ref{TemperatureCorrectionEquation}). 
	Here we have set $\alpha^* = 0.02.$
	B: Subleading correction to the dephasing rate due to temperature-dependent quantum corrections to $\gamma_t$.
	C: Classical AAK scaling result for the dephasing rate, neglecting temperature-dependent quantum corrections to $\gamma_t$.
	By plotting the experimentally measured dephasing rate (A) on a log-log scale, it should be possible to extract the prefactor of the AAK scaling ($T^*$). 
	The leading-order contribution (C) could then be subtracted off and the subleading temperature dependence (B) could be plotted directly.
}
    \label{TemperatureCorrection}
    \hfill
\end{figure}

In the theory of the interacting, disordered (``Anderson-Mott'' \cite{BK}) zero-temperature metal-insulator transition (MIT), 
it has long been appreciated that short-ranged, spin-triplet-exchange interactions are enhanced \cite{BK,50YearsFinkel,PunnooseFinkelstein}
due to the presence of quenched disorder. Such enhancements are generically expected due to the confluence
of wave function criticality near a MIT (multifractality \cite{Feigelman07,Baturina08,Yazdani10}) and the smooth scaling of matrix elements with energy (Chalker scaling \cite{Foster14}).
The enhancement of the spin interaction strength in spin SU(2)-symmetric (orthogonal class AI) systems \cite{BK,50YearsFinkel,PunnooseFinkelstein}
means that dephasing due to this channel should also be enhanced, which could play a role in the physics of weakly disordered, 
many-channel 1D quantum wires. 
This enhancement might provide an additional mechanism for the apparent ``saturation'' of the dephasing rate 
(deviation from the $\tau_\phi \sim T^{-2/3}$ AAK prediction for Markovian-Coulomb dephasing \cite{AAK}) 
observed in experiments on such systems \cite{Webb1, Webb2, Webb3, GZ, AAG, VDelftPartI, VDelftPartII, KondoTheory1, KondoTheory2, Natelson, KondoMeasure1, KondoMeasure2, ICS}.

The enhancement of the ferromagnetic exchange interaction $\gamma_t < 0$ due to the first quantum correction in class AI takes
the form 
\begin{align}\label{TripletCorrect}
	\delta|\gamma_t| 
	= 
	g(\gamma_t)
	\left(\frac{T_0}{T}\right)^{1/2},
\end{align}
where $g(\gamma_t) \geq 0$ is a well-behaved function of $\gamma_t$ that vanishes in the $\gamma_t \rightarrow 0,1$ 
limits \cite{BK,Liao17}.
The temperature scale $T_0 \propto 1 / (\sigma_0 \nu_0)$, where $\sigma_0$ is the classical 
conductivity and $\nu_0$ is the density of states. 
In the context of dephasing, this leads to a $T^{-1/2}$ enhancement of the noise coupling strength $\Gamma_t$
[Eq.~(\ref{Gammat})] at intermediate temperatures. The first-order correction to the dephasing due to the diffusive
bath in Eq.~(\ref{FinalCoexistCooperon}) thus contains an additional boost due to this enhancement, leading
to a slowing of the dephasing rate relative to the AAK result $\tau_\phi \sim T^{-2/3}$ \cite{AAK}.  
Expanding $\Gamma_t$ through the first quantum correction [Eq.~(\ref{TripletCorrect})] in our expression 
for the Cooperon [Eq.~(\ref{FinalCoexistCooperon})], we find the result
\begin{align}
	\nonumber
	\frac{1}{\tau_\phi} 
	\simeq&\,
	D\left(\frac{2 k_B}{\kappa_0 D^2}\right)^{2/3}
	\left[
			F_0(\gamma_t)
			\,
			T^{2/3} 
			+ 
			F_1(\gamma_t)
			\,
			T_0^{1/2}
			\,
			T^{1/6}
	\right]
	\\
\label{TemperatureCorrectionEquation}
	\propto&\, 
	\left(\frac{T}{T^*}\right)^{2/3} 
	+ 
	\alpha^* 
	\left(\frac{T}{T^*}\right)^{1/6},
\end{align}
where we have defined $T^*$ and $\alpha^*$ as necessary. 
In Eq.~(\ref{TemperatureCorrectionEquation}), $\{F_{0,1}(\gamma_t)\}$ denote 
dimensionless functions of the dimensionless triplet coupling constant $\gamma_t$;
pure Coulomb scattering corresponds to the limit 
$F_0(\gamma_t \rightarrow 0) > 0$ [$F_1(\gamma_t \rightarrow 0) = 0$]. 
On the second line of this equation, the magnitude of the dimensionless parameter 
$\alpha^* \propto T_0^{1/2}$ is suppressed by the large bare conductance, as discussed
below Eq.~(\ref{TripletCorrect}). 
We see that the diffusive bath contributes a subleading correction to the dephasing rate that decays as $T^{1/6}$. 
Fig.~\ref{TemperatureCorrection} compares this prediction to the classic AAK result.
At even lower temperatures, Eqs.~(\ref{FinalCoexistCooperon}) and (\ref{TripletCorrect}) predict a suppression of dephasing relative to AAK,
due to the second-order rephasing correction.
The full interplay of enhanced or suppressed dephasing (real processes) with all virtual quantum corrections 
could in principle be tackled using the dynamical version of the Finkel'stein nonlinear sigma model \cite{Liao17}.

\acknowledgments

We thank Doug Natelson, Jan von Delft, and Igor Burmistrov for helpful discussions. 
This research was supported by NSF CAREER Grant No.~DMR-1552327, 
and by the Welch Foundation Grant No.~C-1809.


\appendix

\begin{widetext}


\section{Vertex operator correlators}\label{Correlators}

In this appendix we record results for the various charge-neutral vertex operator correlators 
that arise throughout the paper. These are Gaussian and obtain from Wick's theorem. 
For the first-order pure diffusive bath calculation [Eqs.~(\ref{S_I}) and (\ref{FirstOrderExpectationResult})], we have
\begin{align}
	\langle e^{i  k  r(\tau_b)} \, e^{- i  k  r(\tau_a)}\rangle_0^r
	= 
	e^{-D h_r(\eta,k,\tau_a,\tau_b)},
	\quad
	h_r(\eta,k,\tau_a,\tau_b) 
	= 
	\frac{k^2}{2}
	\left|\tau_b - \tau_a\right|	
	\left[
		1
		-
		\frac{1}{2\eta}
		\left|\tau_b - \tau_a\right|
	\right].
\end{align}
The second-order generalization is more complicated and depends on the time ordering. 
We can use symmetry to reduce to the case where 
$\tau_{1a} < \tau_{1b}$, 
$\tau_{2a} < \tau_{2b}$, 
and
$\tau_{1a} < \tau_{2a}$ 
[Eq.~(\ref{SecondOrderTimeSectors}) and Fig.~\ref{SecondOrderDiagrams}].
We have
\begin{align}
	\langle 
	e^{i  k_1  r(\tau_{1b})}
	\,
	e^{-i k_1  r(\tau_{1a})}
	\,
	e^{i  k_2  r(\tau_{2b})}
	\,
	e^{-i k_2  r(\tau_{2a})}
	\rangle_0^r 
	=&\, 
	e^{-D h_r(\eta,k_1,\tau_{1a},\tau_{1b})}
	\,
	e^{-D h_r(\eta,k_2,\tau_{2a},\tau_{2b})}
	\,
	e^{D \phi_r(\eta,k_1,k_2,\tau_{1a},\tau_{1b},\tau_{2a},\tau_{2b})},
\end{align}
where 
\begin{align}
	\phi_r(\eta,k_1,k_2,\tau_{1a},\tau_{1b},\tau_{2a},\tau_{2b}) 
	= 
	\frac{k_1 k_2}{2\eta}(\tau_{2b}-\tau_{2a})(\tau_{1b}-\tau_{1a}) 
	- 
	k_1 k_2
	\begin{cases}
	0, 				&\quad \tau_{1a} < \tau_{1b} < \tau_{2a} < \tau_{2b},\\
	\tau_{1b} - \tau_{2a}, 		&\quad \tau_{1a} < \tau_{2a} < \tau_{1b} < \tau_{2b},\\
	\tau_{2b} - \tau_{2a}, 		&\quad \tau_{1a} < \tau_{2a} < \tau_{2b} < \tau_{1b}.
    \end{cases}
\end{align}

The expectation values over the center-of-time path integral $R(\tau)$ 
used in the coexisting bath calculation presented in Secs.~\ref{CoexistingDephasingResults} and \ref{CalculationDetails} 
are simpler due to the averaging over the endpoint $R_0$ [Eq.~(\ref{GeneralPathIntegralFolded})]. 
We find that 
\begin{align}
	\langle e^{i k R(\tau_b)} \, e^{-i k R(\tau_a)}\rangle^R_0 
	= 
	e^{-D h_R(k_,\tau_a,\tau_b)},
	\quad 
	h_R(k,\tau_a,\tau_b) 
	= 
	\frac{k^2}{4}
	\left|\tau_b - \tau_a\right|. 
\end{align}
At second order, we have
\begin{align}
    \langle 
	e^{i k_1 R(\tau_{1b})}
	\,
	e^{-i k_1 R(\tau_{1a})}
	\,
	e^{i k_2 R(\tau_{2b})}
	\,
	e^{- i k_2 R(\tau_{2a})}
	\rangle_0^R 
	=&\, 
	e^{-D h_R(k_1,\tau_{1a},\tau_{1b})}
	e^{-D h_R(k_2,\tau_{2a},\tau_{2b})}
	e^{D \phi_R(k_1,k_2,\tau_{1a},\tau_{1b},\tau_{2a},\tau_{2b})},
\end{align}
with
\begin{align}
	\phi_R(k_1,k_2,\tau_{1a},\tau_{1b},\tau_{2a},\tau_{2b}) 
	= 
	\frac{k_1 k_2}{2}
	\begin{cases}
	0, 				&\quad \tau_{1a} < \tau_{1b} < \tau_{2a} < \tau_{2b},\\
	(\tau_{1b} - \tau_{2a}),	&\quad \tau_{1a} < \tau_{2a} < \tau_{1b} < \tau_{2b},\\
	(\tau_{2b} - \tau_{2a}),	&\quad \tau_{1a} < \tau_{2a} < \tau_{2b} < \tau_{1b}.	
	\end{cases}		
\end{align}


\section{Further details on the purely diffusive bath calculation}\label{MoreDiffusionDetails}

We explicitly define the functions $G^s_2(\beta)$ used in Eq.~(\ref{SecondOrderGs}). 
The time sectors $\{\Omega_{1,2,3}\}$ are defined via Eq.~(\ref{SecondOrderTimeSectors}),
illustrated by the diagrams in Fig.~\ref{SecondOrderDiagrams}.
We find that 
\begin{align}
	G_2^s(\beta) 
	=&\, 
	\frac{8}{\pi}\int\limits_{\Omega_s}d^4\vex{\tau}
	\left[
	\begin{aligned}
		\frac{1}{\sqrt{\Theta^{s}_{11}(\beta,\vex{\tau})}}
		-
		\frac{1}{\sqrt{\Theta^{s}_{21}(\beta,\vex{\tau})}}
		-
		\frac{1}{\sqrt{\Theta^{s}_{12}(\beta,\vex{\tau})}}
		+
		\frac{1}{\sqrt{\Theta^{s}_{22}(\beta,\vex{\tau})}}
	\end{aligned}
	\right],
\end{align}
where 
\begin{align}
	\Theta^{1}_{ij}(\beta,\tau_{1a},\tau_{1b},\tau_{2a},\tau_{2b}) 
	=&\, 
	\bigg\{
		4g_i(\beta,\tau_{1a},\tau_{1b}) \, g_j(\beta,\tau_{2a},\tau_{2b}) - \big[(\tau_{2b}-\tau_{2a})(\tau_{1b}-\tau_{1a})\big]^2
	\bigg\},
\\
	\Theta^{2}_{ij}(\beta,\tau_{1a},\tau_{1b},\tau_{2a},\tau_{2b}) 
	=&\, 
	\bigg\{
		4g_i(\beta,\tau_{1a},\tau_{1b}) \, g_j(\beta,\tau_{2a},\tau_{2b}) - \big[(\tau_{2b}-\tau_{2a})(\tau_{1b}-\tau_{1a}) - 2(\tau_{1b}-\tau_{2a})\big]^2
	\bigg\},
\\
	\Theta^{3}_{ij}(\beta,\tau_{1a},\tau_{1b},\tau_{2a},\tau_{2b}) 
	=&\, 
	\bigg\{
		4g_i(\beta,\tau_{1a},\tau_{1b}) \, g_j(\beta,\tau_{2a},\tau_{2b}) - \big[(\tau_{2b}-\tau_{2a})(\tau_{1b}-\tau_{1a}) - 2(\tau_{2b}-\tau_{2a})\big]^2
	\bigg\},
\end{align}
and the functions $\{g_i\}$ are defined by Eq.~(\ref{gFunctionsDefinition}).


\section{Further details on the coexisting bath calculation}\label{MoreCoexistingDetails}

\subsection{General calculation}\label{MoreGeneralDetails}

We compute the functional integral over $\rho(\tau)$ [Eq.~(\ref{rhoVertexCorr})] via the Coulomb-Markovian 
eigenfunctions defined in Eqs.~(\ref{EigenfunctionsEven}) and (\ref{EigenfunctionsOdd}). 
Let $\tilde{\tau}_j$ (where $j$ runs over $\{1,...,2n\}$) be the time-ordering of all the time variables 
$\{\tau_{ia},\tau_{ib}\}$ appearing in Eq.~(\ref{rhoVertexCorr}). 
(E.g., restricting so that 
$\tau_{ia} < \tau_{ib}$ 
and 
$\tau_{ia} < \tau_{ja}$ for $i<j$, 
one has that 
$\tilde{\tau}_{1} = \tau_{1a}$ [see Fig.~\ref{SecondOrderDiagrams}]. 
The rest of the $\tilde{\tau}_j$ will depend on the time-sector topology, as discussed in Secs.~\ref{DiffusiveDephasing} 
and Appendix~\ref{Correlators}). 
In general, we then have 
(after the scaling applied in Eq.~(\ref{VariableScaling}))
\begin{align}
    F^n_{\rho}(k's,\tau's) 
	\equiv&\,
	\left\langle
	\sin\left[
	\frac{k_1\rho(\tau_{1a})}{2}
	\right]
	\sin\left[
	\frac{k_1\rho(\tau_{1b})}{2}
	\right]
	\times
	\ldots
	\times
	\sin\left[
	\frac{k_n\rho(\tau_{na})}{2}
	\right]
	\sin\left[
	\frac{k_n\rho(\tau_{nb})}{2}
	\right]
	\right\rangle_0^\rho
	\\
	=&\,
	\frac{1}{f_0(z)}
	\sum_{i_0,\ldots,i_{2n} = 0}^\infty
	\frac{
	\tilde{S}_{i_{2n}i_{2n-1}}[k_{r(2n)}]
	\times
	\ldots
	\times
	\tilde{S}_{i_{1}i_{0}}[k_{r(1)}]	
	}{
	\sqrt{\alpha'_{i_0} \alpha'_{i_{2n}}}}
	\exp\left\{
		-
		\eta
		\left[ 
			\e_{2n} 
			+ 
			\sum_{j=1}^{2n} \tilde{\tau}_j(\e_{j-1}-\e_{j})
		\right]
	\right\},
\end{align}
where the $\e_j$ are given by Eq.~(\ref{Energies}) and $f_0(z)$ is given by Eq.~(\ref{MarkovianCooperon}).
Above, the subscript $k_{r(j)}$ indicates that one has to be careful about allocating the momenta to the 
expectation values. The ordering of the momenta is dependent on the topology of the time sector being considered. 
We formalize this by defining a function $r:\{1,...,2n\} \rightarrow \{1,...,n\}$, as follows: 
if 
$\tilde{\tau}_j = \tau_{ia}$ or $\tau_{ib}$, 
then 
$r(j) = i$. 
The order of the momenta depends on the time-ordering of the $\tau$ variables.


\subsection{First and second order}\label{MoreFirstAndSecondOrderDetails}

We give the coefficients defined in the first-order asymptotic analysis in Table~\ref{FirstOrderTable}.
\begin{align}
    \label{C1}
    C_1(\beta,i_1) &=
    2\int\frac{dk}{2\pi}
    \frac{\left[\tilde{S}_{i_10}(k)\right]^2}{(\alpha'_0-\alpha_{i_1}+\theta k^2)},
    \\
    \label{C2}
    C_2(\beta,i_1) &= 
    2\int\frac{dk}{2\pi}
    \frac{\left[\tilde{S}_{i_10}(k)\right]^2}{(\alpha'_0-\alpha_{i_1}+\theta k^2)^2},
    \\
    \label{C3}
    C_3(\beta,i_2,i_1) &= 
    2\left|\frac{\alpha'_{0}}{\alpha'_{i_2}}\right|^{1/2}\frac{1}{\alpha'_0-\alpha'_{i_2}}
    \int\frac{dk}{2\pi}
    \frac{\tilde{S}_{i_10}(k) \, \tilde{S}_{i_2i_1}(k)}{(\alpha'_0-\alpha_{i_1}+\theta k^2)},
\end{align}
\begin{align}
    \label{C5}
    C_5(\beta,i_0,i_1) &= 
    2\left|\frac{\alpha'_{0}}{\alpha'_{i_0}}\right|^{1/2}
    \int\frac{dk}{2\pi}
    \frac{\tilde{S}_{i_10}(k) \, \tilde{S}_{i_0i_1}(k)}{(\alpha'_0 - \alpha'_{i_0}+\beta k^2)(\alpha'_0-\alpha_{i_1}+\theta k^2)}.
\end{align}
Similarly, the coefficients introduced in the second-order asymptotic analysis Table~\ref{SecondOrderTable} 
are
\begin{align}
    \label{C4}
    C_4(\beta,i_1,i_2,i_3) &= 
    \frac{4}{\alpha'_0-\alpha'_{i_2}}
    \left[
    \int\frac{dk}{2\pi}
    \frac{\tilde{S}_{i_10}(k) \, \tilde{S}_{i_1i_2}(k)}{(\alpha'_0-\alpha_{i_1}+\theta k^2)}
    \right]
    \left[
    \int\frac{dk}{2\pi}
\frac{\tilde{S}_{i_30}(k) \, \tilde{S}_{i_3i_2}(k)}{(\alpha'_0-\alpha_{i_3}+\theta k^2)}
    \right],
    \\
    \label{C7}
    C_7(\beta,i_1,i_2,i_3) &= 
    8\int\frac{dk_1}{2\pi}\int\frac{dk_2}{2\pi}
    \frac{
	\tilde{S}_{i_10}(k_1) \, \tilde{S}_{i_2i_1}(k_2) \, \tilde{S}_{i_2i_3}(k_1) \, \tilde{S}_{i_30}(k_2)
	}{
	(\alpha'_0-\alpha_{i_1}+\theta k_1^2)(\alpha'_0-\alpha_{i_3}+\theta k_2^2)(2\alpha'_0-2\alpha'_{i_2}+2\theta k_1^2+2\theta k_2^2 + k_1k_2)
	},
    \\
    \label{C8}
    C_8(\beta,i_1,i_2,i_3) &= 
    8\int\frac{dk_1}{2\pi}\int\frac{dk_2}{2\pi}
    \frac{
	\tilde{S}_{i_10}(k_1) \, \tilde{S}_{i_2i_1}(k_2) \, \tilde{S}_{i_2i_3}(k_2) \, \tilde{S}_{i_30}(k_1)
	}{
	(\alpha'_0-\alpha_{i_1}+\theta k_1^2)(\alpha'_0-\alpha_{i_3}+\theta k_1^2)(2\alpha'_0-2\alpha'_{i_2}+2\theta k_1^2+2\theta k_2^2 + k_1k_2)
	}.
\end{align}
Tables~\ref{FirstOrderTable} and \ref{SecondOrderTable} also contain ``anomalous'' contributions $D_1$ and $D_2$, 
which arise in the $T_2$, $T^1_{12}$, and $T^1_{21}$ terms when all ``even'' energies ($i_0,i_2,i_4$) are pinned at the ground state.
These are defined by
\begin{align}
	D_1(z;\beta,i_1) 
	=&\, 
	\frac{2z^{1/2}}{\beta}\frac{1}{\alpha'_0-\alpha_{i_1}}\int\frac{dk}{2\pi}\frac{1}{k^2}e^{-\beta k^2}\left[\tilde{S}_{i_10}\left(\frac{k}{\sqrt{z}}\right)\right]^2,
	\\
	D_2(z;\beta,i_1) 
	=&\, 
	\frac{4z^{3/2}}{\beta^2}\frac{1}{\alpha'_0-\alpha_{i_1}}\int\frac{dk}{2\pi}\frac{1}{k^4}(1-e^{-\beta k^2})\left[\tilde{S}_{i_10}\left(\frac{k}{\sqrt{z}}\right)\right]^2.
\end{align}
In each case, we can expand $\tilde{S}_{ij}(k)$ in powers of $k/\sqrt{z}$ for large $z$, 
since $k/\sqrt{z}$ will be small for the dominant portion of the integrand. 
Because the leading order contribution to $\tilde{S}_{ij}(k/\sqrt{z})$ is linear in $k/\sqrt{z}$, 
the leading order asymptotic contributions to $D_1$ and $D_2$ are $z^{-1/2}$ and $z^{1/2}$, as seen numerically. 
We note again that both $D_1$ and $D_2$ cancel from the final dephasing expressions.
\end{widetext}


\subsection{Numerical comparison with asymptotics}\label{NumericalVsAsymptotic}

\begin{figure}[t!]
        \includegraphics[width=0.35\textwidth]{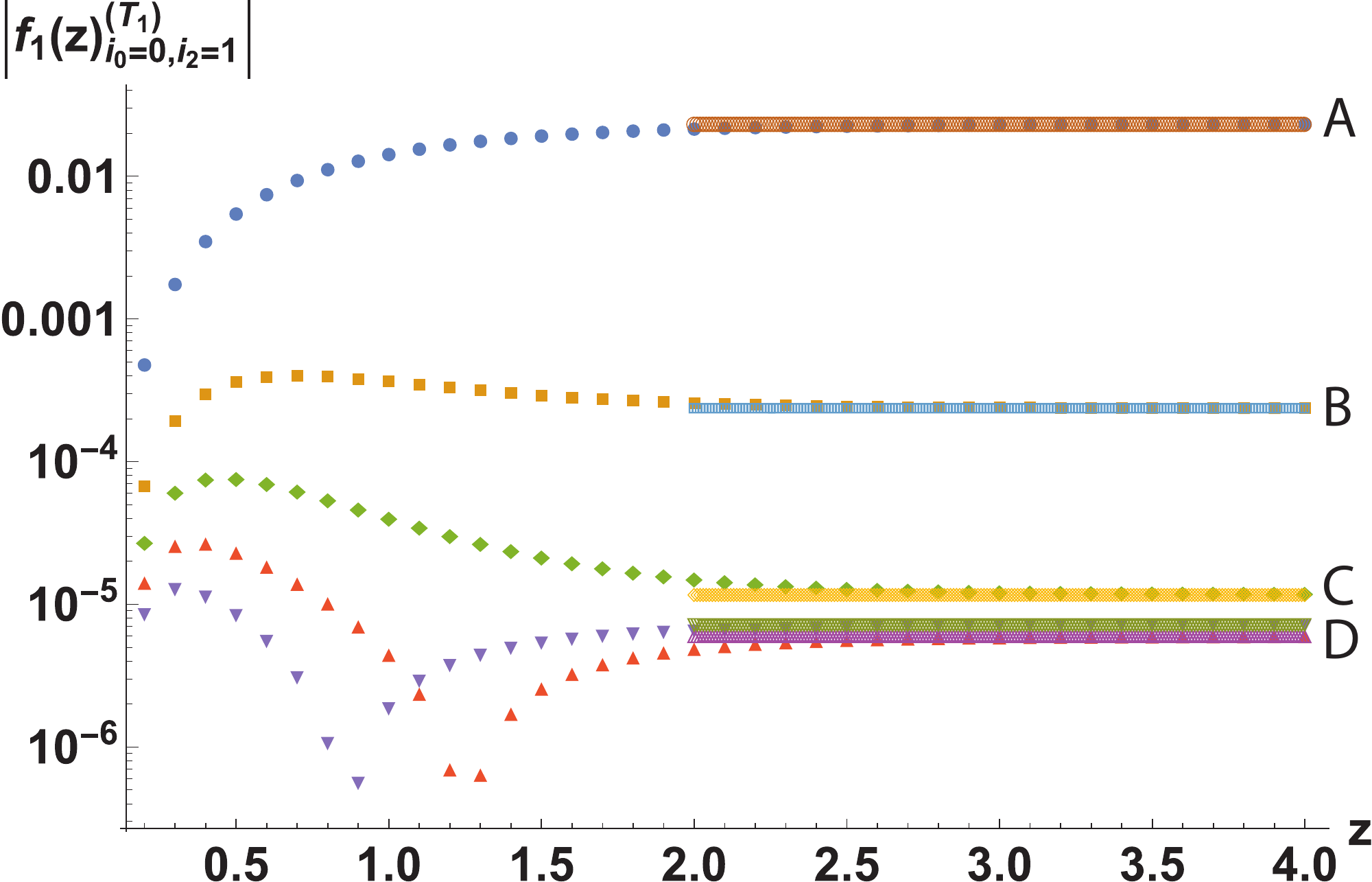}
        \caption{
	Comparison of full numerical calculations with the asymptotic, 
	large-virtual-time $z \rightarrow \infty$ approximation to the 
	first-order dual-bath cumulant amplitude in Eq.~(\ref{FirstOrderInitialExpression}).
	Here we define $f_1(z)_{i_0=0,i_2=1}^{(T_1)}$ to be the contribution to $f_1$ 
	[Eqs.~(\ref{Scalingfs}) and (\ref{FirstOrderInitialExpression})] from $T_1$ [Eq.~(\ref{T1Def})], 
	at energy $i_0=0, i_2=1$, with $i_1$ left unspecified.
	This plot demonstrates that the asymptotic form for this $T_1$ contribution 
	with Airy prime summand labels pinned at $i_0 = 0, i_2 = 1$, as specified in Table~\ref{FirstOrderTable}, is correct. 
	Above, $i_j$ gives the energy level of the $j^{th}$ energy and $z = \eta D/a^2$. 
	In this plot, we vary the Airy summand label $i_1$ from $0$ to $4$, and we set $\beta = 1$ [Eq.~(\ref{BetaDef})]. 
	A: $i_1 = 0$, 
	B: $i_1 = 1$, 
	C: $i_1 = 2$, 
	D: $i_1 = 3,4$.
	The solid curves are the asymptotic approximation in Table~\ref{FirstOrderTable} and Eq.~(\ref{FullFirstOrder}). 
	The asymptotic formula in this case is given by $C_3$, Eq.~(\ref{C3}).
	The symbols obtain from the full numerical integration of Eq.~(\ref{FirstOrderInitialExpression}), using the 
	exact expression for $T_1(z,1,k^2,\alpha'_1,\alpha_{i_1},\alpha'_0)$ from Eq.~(\ref{T1Def}).
	}
	\label{ComparisonPlotF1i00i21}
\end{figure}

\begin{figure}[t!]
        \includegraphics[width=0.35\textwidth]{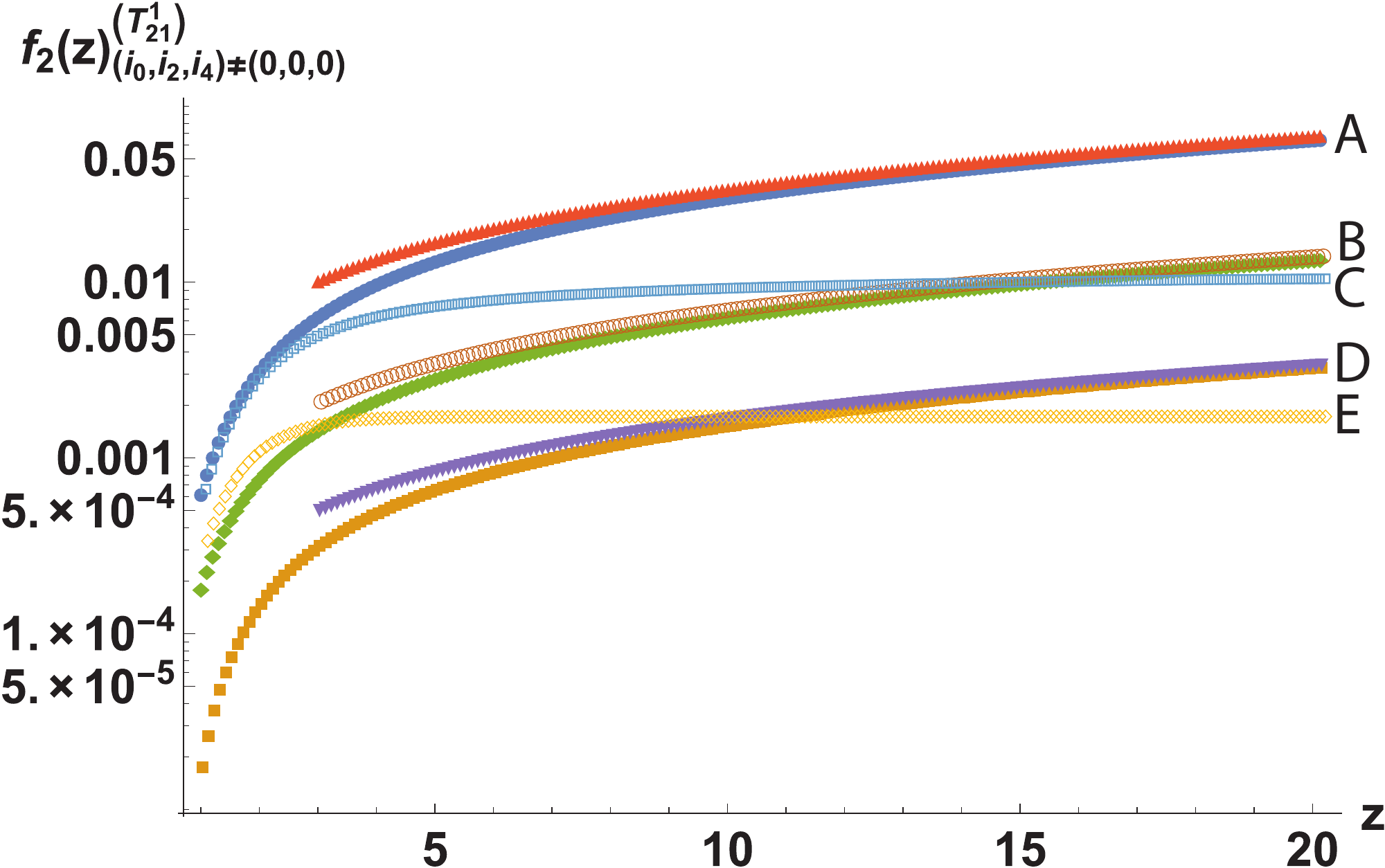}
        \caption{
        We define $f_2(z)_{(i_0,i_2,i_4)\neq(0,0,0)}^{(T^1_{21})}$ to be the contributions to $f_2$ 
	[Eqs.~(\ref{Scalingfs}) and (\ref{SecondOrderInitial})] from $T^1_{21}$ at energies away from $i_0 = i_2 = i_4 = 0$. 
	(No $i_j$ explicitly specified.)
	This plot demonstrates, for several sets of energies, that the asymptotic form for $T^1_{21}$ given in Table~\ref{SecondOrderTable} is correct. 
	In this plot, we vary the Airy summand labels $i_j$ $0$ to $1$. 
	A: $(i_0,i_1,i_2,i_3,i_4) = (1,0,0,0,0)$,
	B: $(i_0,i_1,i_2,i_3,i_4) = (1,1,0,0,0)$,
	C: $(i_0,i_1,i_2,i_3,i_4) = (0,0,1,0,0)$,
	D: $(i_0,i_1,i_2,i_3,i_4) = (1,0,0,1,0)$,
	E: $(i_0,i_1,i_2,i_3,i_4) = (0,0,0,0,1)$.
	The asymptotic approximations from Table~\ref{SecondOrderTable} start at $z=2$, 
	while the full numerical results start at $z=0$.
	The full numerical results obtain from the integration of Eq.~(\ref{SecondOrderInitial}), 
	using the exact expression for $T^1_{21}(z,1,k_1^2,k_2^2,\alpha'_{i_4},\alpha_{i_3},\alpha'_{i_2},\alpha_{i_1},\alpha'_{i_0})$.
        }
        \label{ComparisonPlotF2T21Excited}
\end{figure}

In this appendix subsection we provide a provide a collection of plots 
[Figs.~\ref{ComparisonPlotF1i00i21}--\ref{ComparisonPlotF2T21Excited}] demonstrating the accuracy of the 
asymptotic expressions given in Tables~\ref{FirstOrderTable} and \ref{SecondOrderTable}, 
similar to Figs.~\ref{T1GroundComparison}--\ref{T11_2,3_Comparison}. 
In Figs.~\ref{ComparisonPlotF1i00i21}--\ref{ComparisonPlotF2T21Excited}, we numerically calculate the 
full contributions from the terms listed in the rows of Tables~\ref{FirstOrderTable} and \ref{SecondOrderTable}, 
(i.e.\ $T_j,T^s_{jk}$), for some specific choices of the energy levels, as functions of $z = \eta D/a^2$. 
This is done via the method explained in Sec.~\ref{CalculationDetails}. These numerical results are then compared 
to the asymptotic forms listed in Tables~\ref{FirstOrderTable} and \ref{SecondOrderTable}. As in 
Figs.~\ref{T1GroundComparison}--\ref{T11_2,3_Comparison}, we define 
$f_1(z)_{i's}^{(T_j)}$ $\left(f_2(z)_{i's}^{(T^s_{jk})}\right)$ to be the contribution to $f_1$ $(f_2)$ 
from term $T_j$ $\left(T^s_{jk}\right)$ at the energy levels specified by the subscript ``$i's$''. 
The plots all show quick convergence to the expected behavior.

\begin{figure}[t!]
        \includegraphics[width=0.35\textwidth]{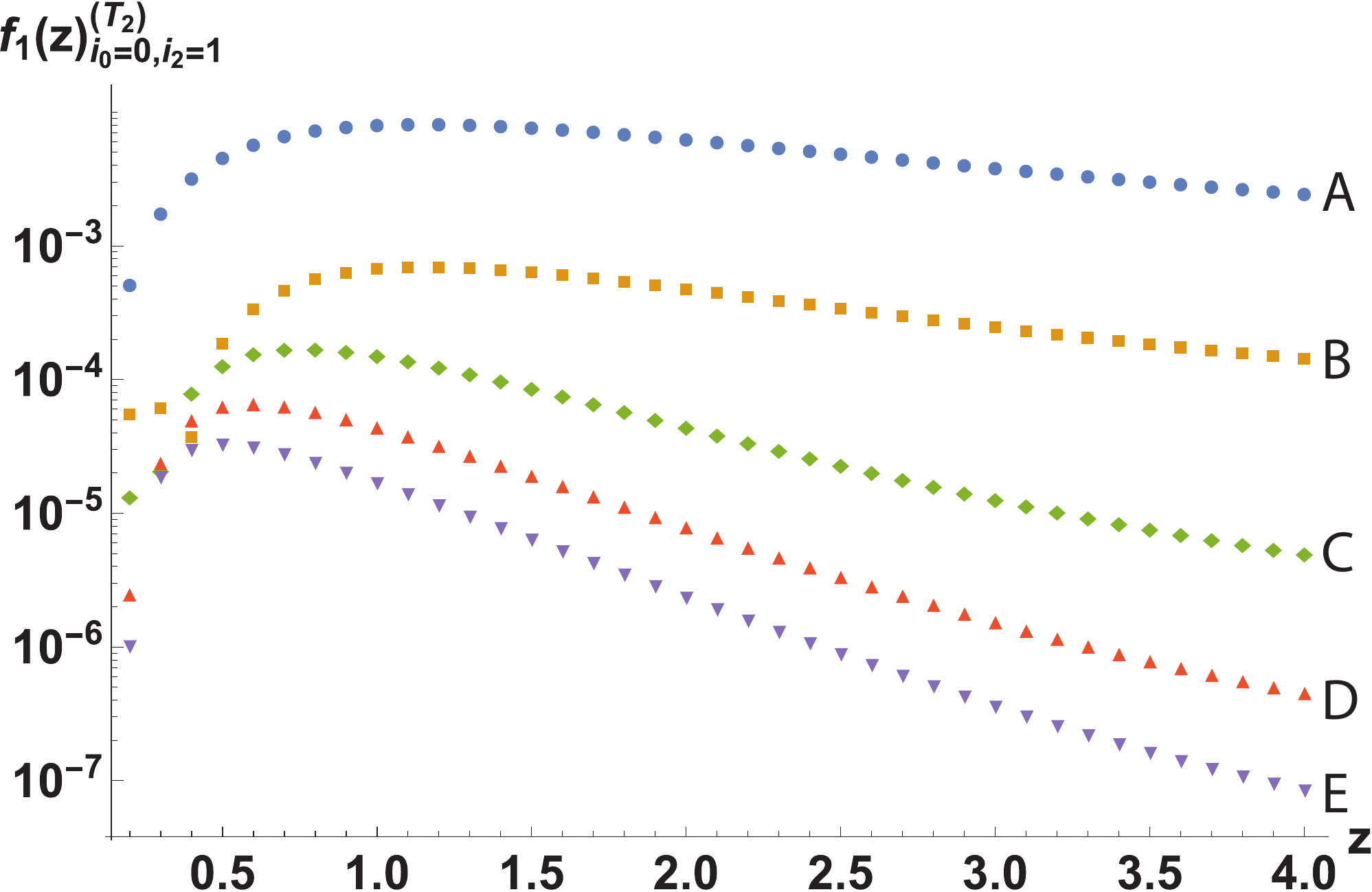}
        \caption{
	We define $f_1(z)_{i_0=0,i_2=1}^{(T_2)}$ to be the contribution to $f_1$ [Eqs.~(\ref{Scalingfs}) and 
	(\ref{FirstOrderInitialExpression})] from $T_2$ [Eq.~(\ref{T2Def})], at energy $i_0=0, i_2=1$, with $i_1$ left unspecified.
	This plot gives the asymptotic form for this $T_2$ contribution 
	with Airy prime summand labels pinned at $i_0 = 0, i_2 = 1$, as specified in Table~\ref{FirstOrderTable}. 
	Above, $i_j$ gives the energy level of the $j^{th}$ energy and $z = \eta D/a^2$. 
	In this plot, we vary the Airy summand label $i_1$ from $0$ to $4$, and we set $\beta = 1$ [Eq.~(\ref{BetaDef})]. 
	A: $i_1 = 0$, 
	B: $i_1 = 1$, 
	C: $i_1 = 2$, 
	D: $i_1 = 3$,
	E: $i_1 = 4$.
	The decay of these functions shows that these contributions asymptotically vanish, as indicated in Table~\ref{FirstOrderTable}.
	The symbols obtain from the full numerical integration of Eq.~(\ref{FirstOrderInitialExpression}), using the 
	exact expression for $T_2(z,1,k^2,\alpha'_1,\alpha_{i_1},\alpha'_0)$ from Eq.~(\ref{T2Def}).
	}
\end{figure}


\section{Perturbation theory for Coulomb dephasing}\label{MarkovianPT}

The exact solution for the Cooperon in the screened Coulomb (Markovian) case is given in Eq.~(\ref{MarkovianCooperon}). 
Here we instead treat this case perturbatively, via the cumulant expansion method 
employed throughout this paper.

The cumulant expansion requires the evaluation of the perturbing action defined by Eqs.~(\ref{AAKAction}) and (\ref{EffCoulKer}). 
We have
\begin{align}\label{SM-Moment}
	\langle S_M^n \rangle_0 
	&= 
	\frac{\Gamma_M^n}{D^n}
	\int\limits_0^\eta d\tau_n ... \int\limits_0^\eta d\tau_1
	\langle|\rho(\tau_n)|...|\rho(\tau_1)|\rangle_0\\
	&= 
	I_n \left(\frac{2 \Gamma_M}{\sqrt{\pi D}}\eta^{3/2}\right)^n,
\end{align}
where $I_n$ is a numerical prefactor. 
It is given by
\begin{align}
	\label{InDef}
	I_n 
	=&\, 
	(n!)
	\int\limits_0^1 d\tau_n 
	\int\limits_0^{\tau_n} d\tau_{n-1} \ldots \int\limits_0^{\tau_2} d\tau_1
	\int\limits_{-\infty}^{\infty}d\rho_n \ldots
	\int\limits_{-\infty}^{\infty}d\rho_1 
\nonumber\\
	&\,
	\times
	\frac{|\rho_n|\ldots|\rho_1|}
	{\sqrt{(1-\tau_n)(\tau_n-\tau_{n-1})\ldots
	(\tau_2-\tau_1)\tau_1}}
\nonumber\\
	&\,
	\times
	\exp\left[\frac{-\rho_n^2}{(1-\tau_n)}\right]
	\exp\left[\frac{-(\rho_n-\rho_{n-1})^2}{(\tau_n-\tau_{n-1})}\right]
\nonumber\\
	&\,
	\times
	\ldots
\nonumber\\
	&\,
	\times
	\exp\left[\frac{-(\rho_2-\rho_1)^2}{(\tau_2-\tau_1)}\right]
	\exp\left[\frac{-\rho_1^2}{\tau_1}\right].
\end{align}
We compute $I_n$ and the resulting moments numerically; the results are collected in Table~\ref{InNumerical}. 
In the cumulant expansion [Eq.~(\ref{CumulantDefinition})], the coefficient of the $n^{th}$ order term is given by the $n^{th}$ cumulant $\equiv \kappa_n$, 
formed from the first $n$ moments of the perturbing action. Using the moments in Table~\ref{InNumerical}, we 
explicitly calculate the first four cumulants and list them in Table~\ref{knNumerical}. 
We actually tabulate $\bar{\kappa}_n \equiv \kappa_n/n!$, which is the full numerical coefficient for the $n^{th}$ 
order term in the expansion, so that 
\begin{align}\label{ccMCumulantExp}
	\cc_M(\eta)
	=&\, 
	\cc_0(\eta)
	\,
	\exp\left[
	\begin{aligned}
	&\,
		-
		\bar{\kappa}_1
		\left(\frac{2 \Gamma_M}{\sqrt{\pi D}}\eta^{3/2}\right)		
	\\&\,
		+ 
		\bar{\kappa}_2
		\left(\frac{2 \Gamma_M}{\sqrt{\pi D}}\eta^{3/2}\right)^2		
		+ 
		\ldots
	\end{aligned}
	\right].
\end{align}
Table~\ref{knNumerical} shows that the perturbative cumulant expansion for the Markovian gives alternating dephasing and rephasing terms.

\begin{figure}[t!]
        \includegraphics[width=0.35\textwidth]{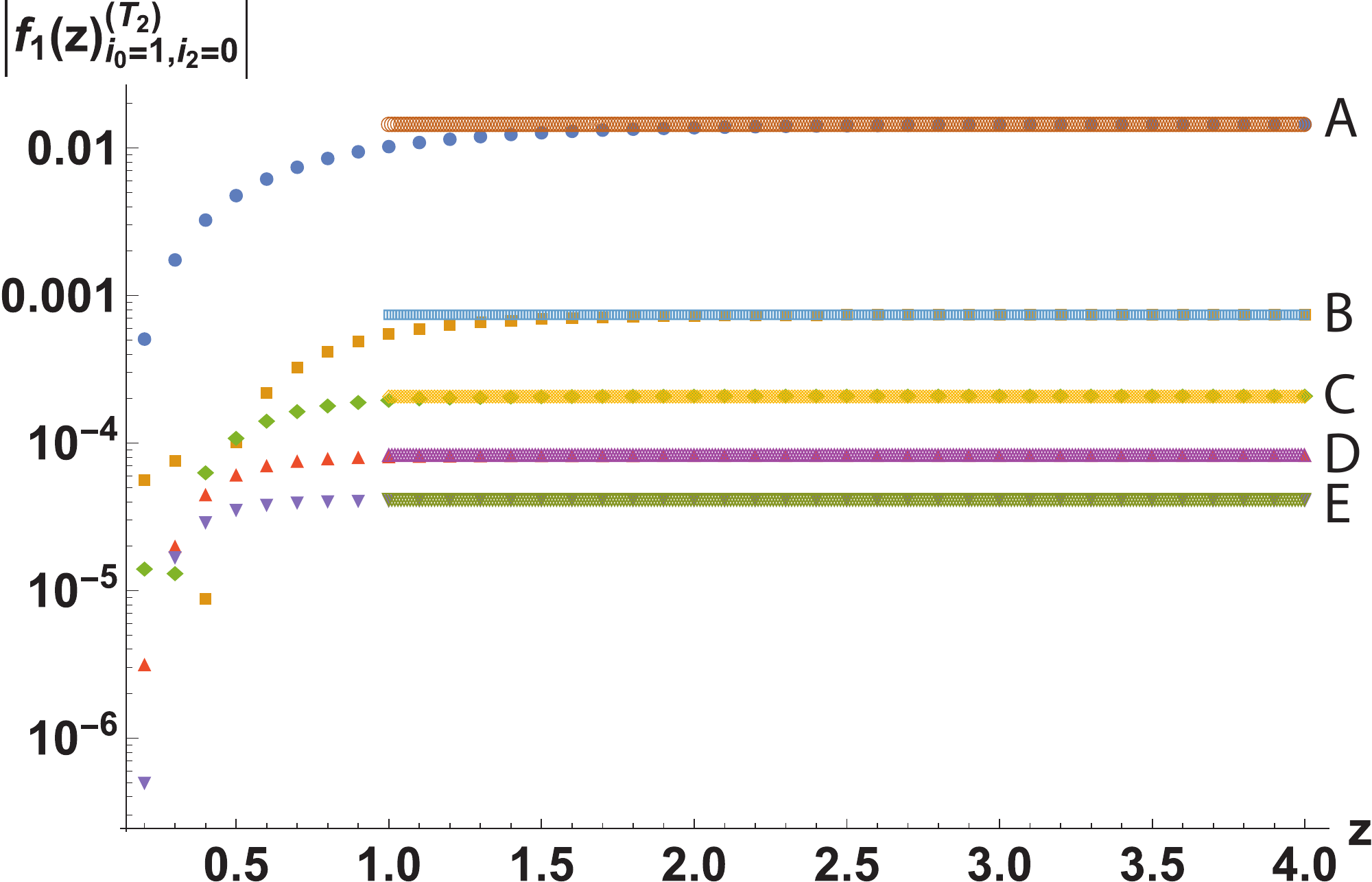}
        \caption{
	We define $f_1(z)_{i_0=1,i_2=0}^{(T_2)}$ to be the contribution to $f_1$ 
	[Eqs.~(\ref{Scalingfs}) and (\ref{FirstOrderInitialExpression})] from $T_2$ [Eq.~(\ref{T2Def})], at energy $i_0=1, i_2=0$, with $i_1$ left unspecified.
	This plot demonstrates that the asymptotic form for this $T_2$ contribution 
	with Airy prime summand labels pinned at $i_0 = 1, i_2 = 0$, as specified in Table~\ref{FirstOrderTable}, is correct. 
	Above, $i_j$ gives the energy level of the $j^{th}$ energy and $z = \eta D/a^2$. 
	In this plot, we vary the Airy summand label $i_1$ from $0$ to $4$, and we set $\beta = 1$ [Eq.~(\ref{BetaDef})]. 
	A: $i_1 = 0$, 
	B: $i_1 = 1$, 
	C: $i_1 = 2$, 
	D: $i_1 = 3$,
	E: $i_1 = 4$.
	The solid curves are the asymptotic approximation in Table~\ref{FirstOrderTable} and Eq.~(\ref{FullFirstOrder}). 
	The asymptotic formula in this case is given by $C_5$, Eq.~(\ref{C5}).
	The symbols obtain from the full numerical integration of Eq.~(\ref{FirstOrderInitialExpression}), using the 
	exact expression for $T_2(z,1,k^2,\alpha'_0,\alpha_{i_1},\alpha'_1)$ from Eq.~(\ref{T2Def}).}
\end{figure}

\begin{figure}[t!]
	\includegraphics[width=0.35\textwidth]{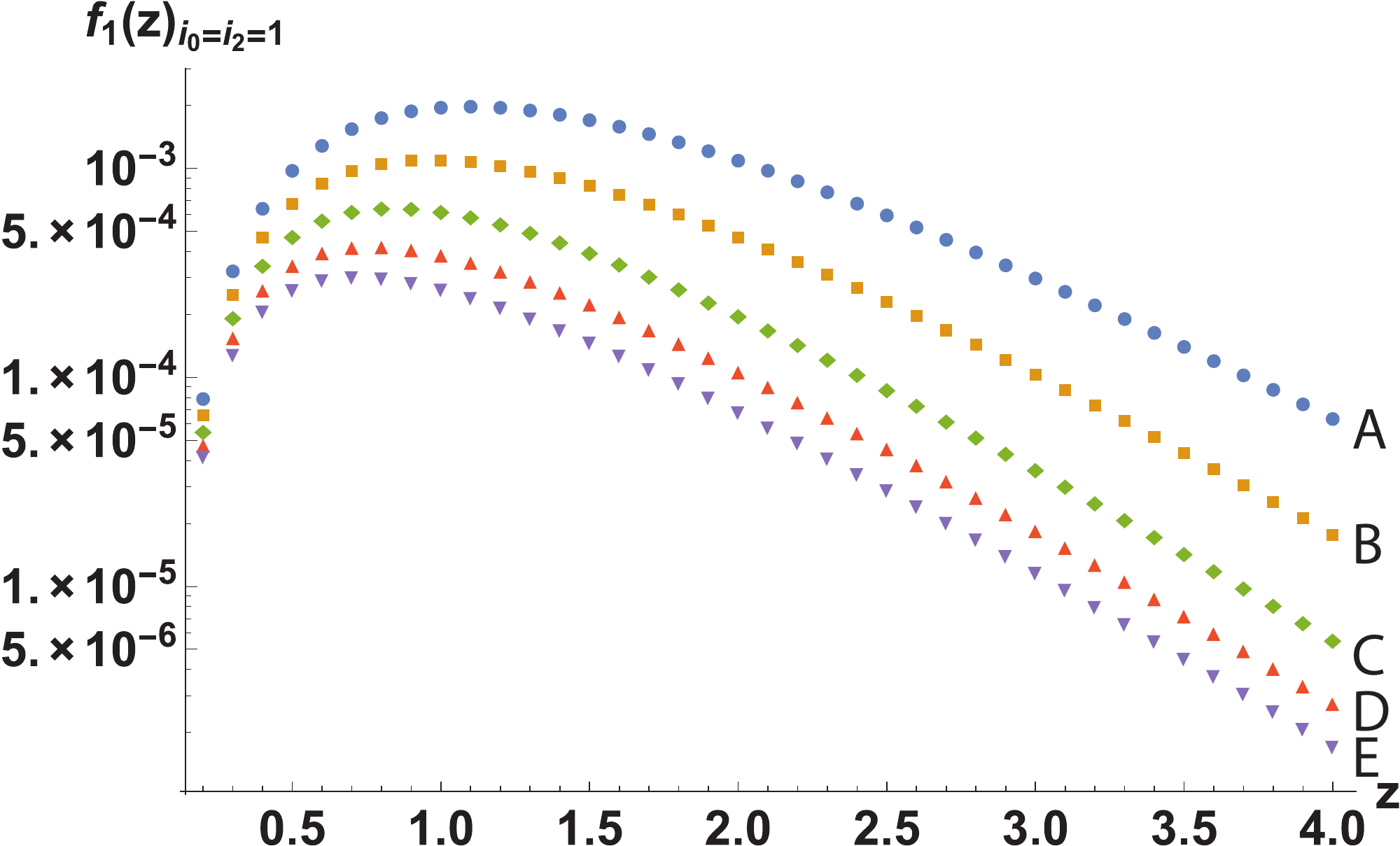}
	\caption{
	We define $f_1(z)_{i_0=i_2=1}$ to be the contribution to $f_1$ [Eqs.~(\ref{Scalingfs}) and 
	(\ref{FirstOrderInitialExpression})] [from both $T_1$ and $T_2$, Eqs.~(\ref{T1Def}) and (\ref{T2Def})], 
	at energy $i_0=i_2=1$, with $i_1$ left unspecified.
	This plot gives the asymptotic form for these contributions 
	with Airy prime summand labels pinned at $i_0 = i_2 = 1$, as specified in Table~\ref{FirstOrderTable}.
	Above, $i_j$ gives the energy level of the $j^{th}$ energy and $z = \eta D/a^2$. 
	In this plot, we vary the Airy summand label $i_1$ from $0$ to $4$, and we set $\beta = 1$ [Eq.~(\ref{BetaDef})]. 
	A: $i_1 = 0$, 
	B: $i_1 = 1$, 
	C: $i_1 = 2$, 
	D: $i_1 = 3$,
	E: $i_1 = 4$.
	Here we see that these contributions asymptotically vanish, as indicated in Table~\ref{FirstOrderTable} (``Else'').
	The symbols obtain from the full numerical integration of Eq.~(\ref{FirstOrderInitialExpression}), using the 
	exact expressions for $T_1(z,1,k^2,\alpha'_1,\alpha_{i_1},\alpha'_1)$ and $T_2(z,1,k^2,\alpha'_1,\alpha_{i_1},\alpha'_1)$ 
	from Eqs.~(\ref{T1Def}) and (\ref{T2Def}).}
\end{figure}

\begin{figure}[t!]
        \includegraphics[width=0.35\textwidth]{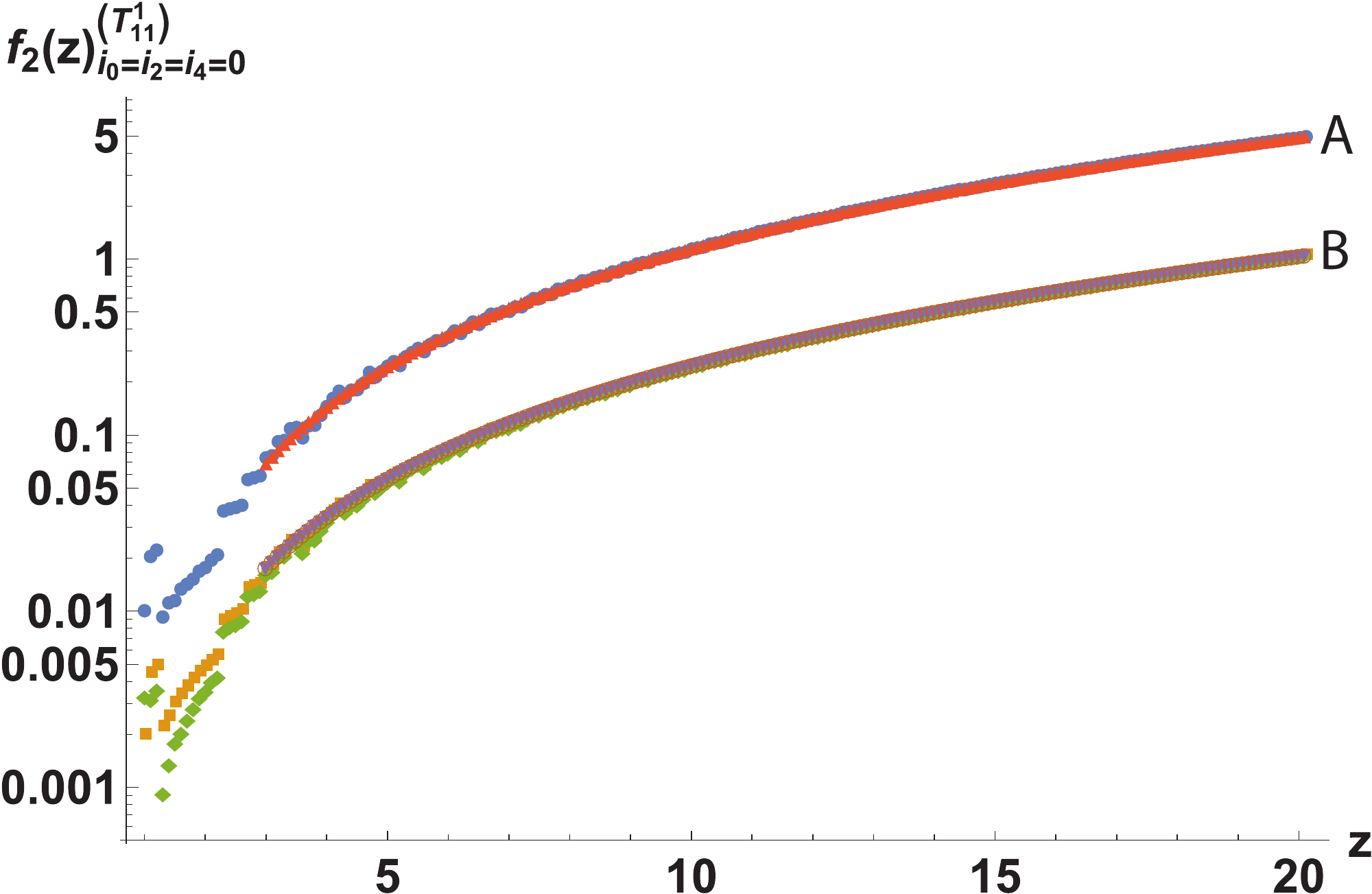}
        \caption{
	We define $f_2(z)_{i_0=i_2=i_4=0}^{(T^1_{11})}$ to be the contribution to $f_2$ 
	[Eqs.~(\ref{Scalingfs}) and (\ref{SecondOrderInitial})] from $T^1_{11}$ at energy 
	$i_0 = i_2 = i_4 = 0$, with $i_1$ and $i_3$ left unspecified.
	This plot demonstrates that the asymptotic form for this $T^1_{11}$ contribution with 
	Airy prime summand labels pinned at $i_0 = i_2 = i_4 = 0$, as specified in Table~\ref{SecondOrderTable}, is correct. 
	In this plot, we vary the Airy summand labels $i_1,i_3$ from $0$ to $1$. 
	A: $(i_1,i_3) = (0,0)$,
	B: $(i_1,i_3) = (1,0),(0,1)$. 
	The asymptotic approximations from Table~\ref{SecondOrderTable} start at $z=2$, while the full numerical results start at $z=0$.
	The full numerical results obtain from the integration of Eq.~(\ref{SecondOrderInitial}), 
	using the exact expression for $T^1_{11}(z,1,k_1^2,k_2^2,\alpha'_0,\alpha_{i_3},\alpha'_0,\alpha_{i_1},\alpha'_0)$.}
\end{figure}

\begin{figure}[t!]
        \includegraphics[width=0.35\textwidth]{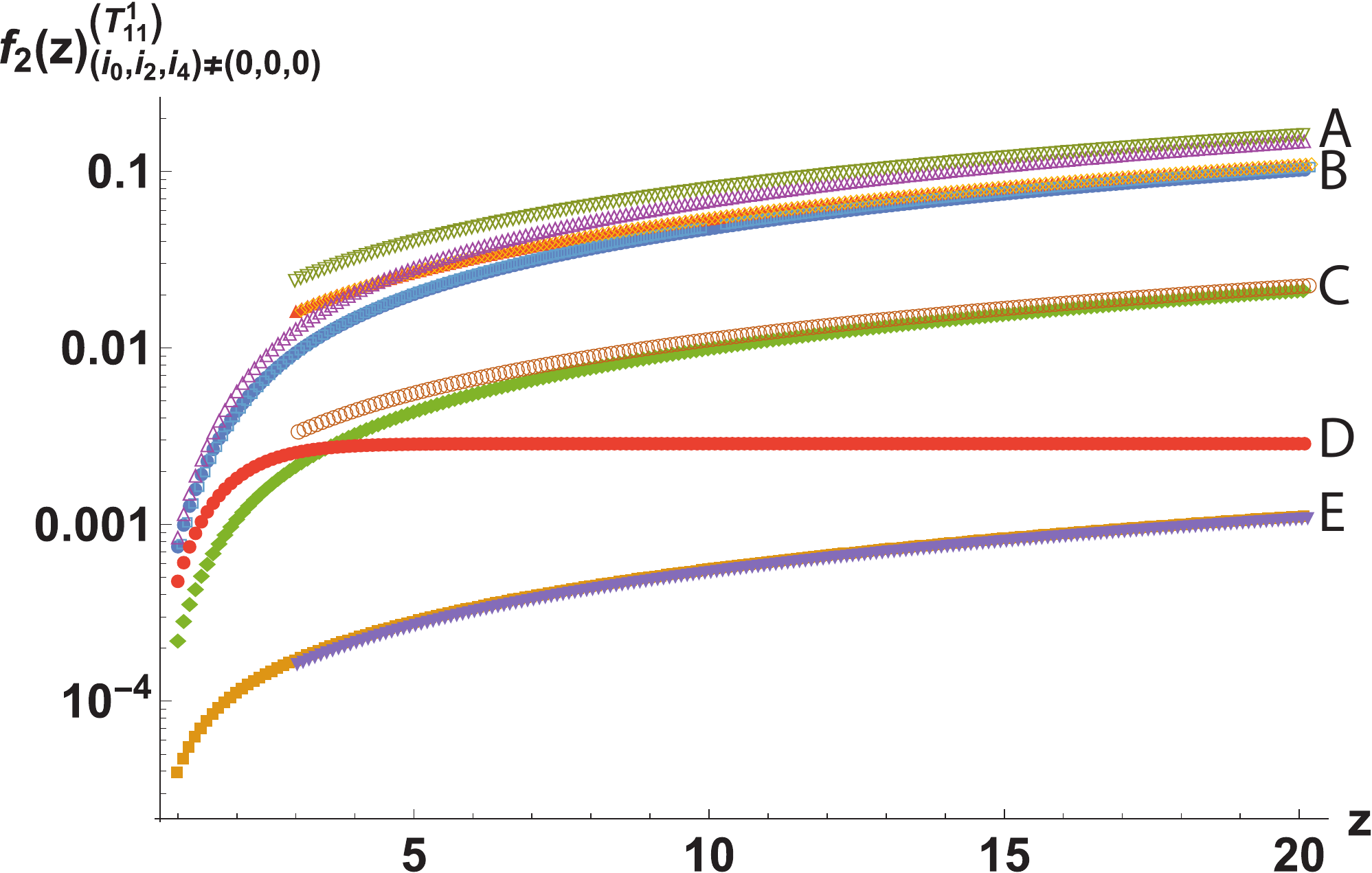}
        \caption{
        We define $f_2(z)_{(i_0,i_2,i_4)\neq(0,0,0)}^{(T^1_{11})}$ to be the contributions to 
	$f_2$ [Eqs.~(\ref{Scalingfs}) and (\ref{SecondOrderInitial})] from $T^1_{11}$ at energies 
	away from $i_0 = i_2 = i_4 = 0$. (No $i_j$ explicitly specified.)
	This plot demonstrates, for several sets of energies, that the asymptotic form for $T^1_{11}$ given in Table~\ref{SecondOrderTable} is correct. 
	In this plot, we vary the Airy summand labels $i_j$ $0$ to $1$. 
	A: $(i_0,i_1,i_2,i_3,i_4) = (1,0,0,0,0)$,
	B: $(i_0,i_1,i_2,i_3,i_4) = (1,1,0,0,0)$,
	C: $(i_0,i_1,i_2,i_3,i_4) = (1,0,0,1,0)$,
	D: $(i_0,i_1,i_2,i_3,i_4) = (0,0,1,0,0)$,
	E: $(i_0,i_1,i_2,i_3,i_4) = (0,0,0,0,1)$.
	The asymptotic approximations from Table~\ref{SecondOrderTable} start at $z=2$, 
	while the full numerical results start at $z=0$.
	The full numerical results obtain from the integration of Eq.~(\ref{SecondOrderInitial}), using the exact 
	expression for $T^1_{11}(z,1,k_1^2,k_2^2,\alpha'_{i_4},\alpha_{i_3},\alpha'_{i_2},\alpha_{i_1},\alpha'_{i_0})$.}
\end{figure}

\begin{figure}[t!]
        \includegraphics[width=0.35\textwidth]{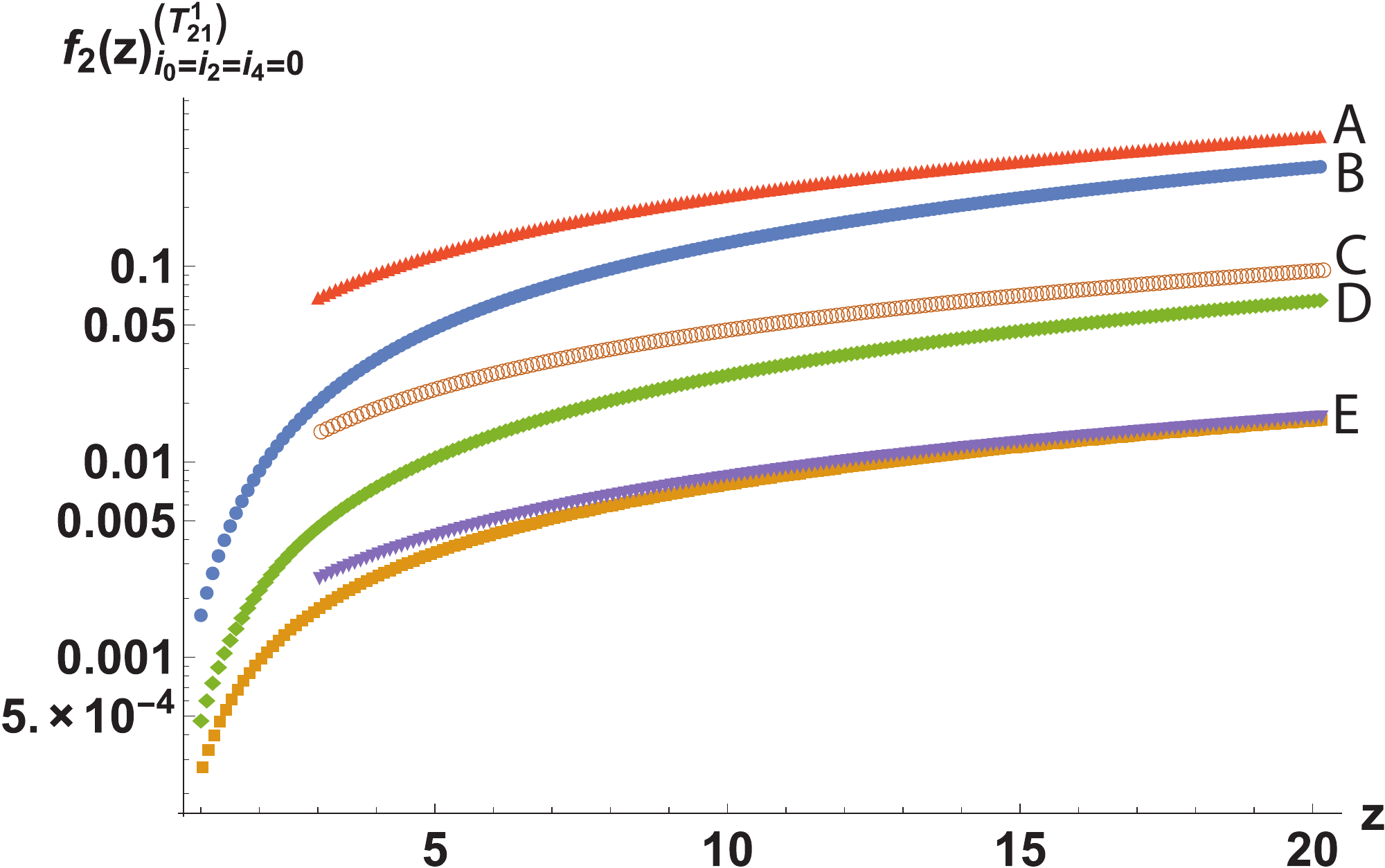}
        \caption{
	We define $f_2(z)_{i_0=i_2=i_4=0}^{(T^1_{21})}$ to be the contribution to $f_2$ 
	[Eqs.~(\ref{Scalingfs}) and (\ref{SecondOrderInitial})] from $T^1_{21}$ at energy 
	$i_0 = i_2 = i_4 = 0$, with $i_1$ and $i_3$ left unspecified.
	This plot demonstrates that the asymptotic form for this $T^1_{21}$ contribution with Airy prime 
	summand labels pinned at $i_0 = i_2 = i_4 = 0$, as specified in Table~\ref{SecondOrderTable}, is correct. 
	In this plot, we vary the Airy summand labels $i_1,i_3$ from $0$ to $1$. 
	A: $(i_1,i_3) = (0,0)$ (asymptotic),
	B: $(i_1,i_3) = (0,0)$ (exact),
	C: $(i_1,i_3) = (1,0)$ (asymptotic),
	D: $(i_1,i_3) = (1,0)$ (exact),
	E: $(i_1,i_3) = (0,1)$ (both asymptotic and exact).
	The asymptotic approximations from Table~\ref{SecondOrderTable} start at $z=2$, 
	while the full numerical results start at $z=0$.
	The full numerical results obtain from the integration of Eq.~(\ref{SecondOrderInitial}), 
	using the exact expression for $T^1_{21}(z,1,k_1^2,k_2^2,\alpha'_0,\alpha_{i_3},\alpha'_0,\alpha_{i_1},\alpha'_0)$. 
	The exact and asymptotic results here differ slightly due to the anomalous $D_1$ term, 
	which is dropped from the asymptotic expression used here.}
\end{figure}

\begin{table}[b!]
    \centering
    \begin{tabular}{|c|c|c|c|}
    \hline
    \multicolumn{4}{|c|}{Numerical results for $I_n$ coefficients} \\
    \hline
    $I_1$ & $I_2$ & $I_3$ & $I_4$ \\
        \hline
    $0.3927 = \pi/8$ & $0.183$ & $0.1017$ & $0.0724$ \\    
        \hline
    \end{tabular}
	\caption{Table collecting numerical results for the $I_n$ moment coefficients,
	defined via Eqs.~(\ref{SM-Moment}) and (\ref{InDef}).}
	\label{InNumerical}
\end{table}

\begin{table}[b!]
    \centering
    \begin{tabular}{|c|c|c|c|}
    \hline
    \multicolumn{4}{|c|}{Numerical results for $\bar{\kappa}$} \\
    \hline
    $\bar{\kappa}_1$ & $\bar{\kappa}_2$ & $\bar{\kappa}_3$ & $\bar{\kappa}_4$ \\
        \hline
    $0.3927 = \pi/8$ & $0.0145$ & $0.00116$ & $0.000346$\\    
        \hline
    \end{tabular}
	\caption{Numerical results for the cumulant coefficients $\bar{\kappa}_n \equiv \kappa_n/n!$,
	which determine the dephasing and rephasing terms in the cumulant expansion for the Markovian
	bath, Eq.~(\ref{ccMCumulantExp}).
	}
	\label{knNumerical}
\end{table}


\section{Diagram folding}\label{DiagramFolding}

To treat the Markovian noise kernel exactly, we need to fold the time integrations from the region $(-\eta,\eta)$ to $(0,\eta)$ 
[Eqs.~(\ref{Change}) and (\ref{GeneralPathIntegralFolded-0})]. 
In general, this procedure folds the different diagram topologies 
associated with time-ordering
into one another. 
This should be considered if one wants to study the effects of a specific class of diagrams, or to make contact with 
the field theoretic formulation of the dephasing problem \cite{Liao18}, [Appendix~\ref{FieldTheory}]. 
At first order there is no issue, since there is only a single diagram topology. 
At second order, however, we find a nontrivial mixing of the distinct topologies 
(``double,'' ``crossed,'' and ``nested''), 
defined by Eq.~(\ref{SecondOrderTimeSectors}) and shown in Fig.~\ref{SecondOrderDiagrams}. 

To ``unfold'' an $n^{th}$-order diagram, any subset of the $2n$ time variables can be flipped to the negative 
side of the time interval. For each topological sector, we have $2^n$ distinct preimages under the 
$\eta$-folding map to consider. Table~\ref{DiagramFoldingTable} maps out this inverse-folding for the second-order calculation.

From Table~\ref{DiagramFoldingTable}, we see that the (folded) double diagram (sector 1) is really a 50-50 combination of the 
(unfolded) double and nested diagrams. The (folded) nested and crossed diagrams (sectors 2 and 3) are both 25-25-50 combinations of the 
(unfolded) double, nested, and crossed diagrams, respectively. However, the terms contributing non-trivial asymptotics are marked 
in Table~\ref{DiagramFoldingTable} in bold, and we see that most (but not all) of the contributions come from diagrams that are originally 
double. Interestingly, the ``nontrivial'' contributions in the 
coexisting bath calculation [Secs.~\ref{CoexistingDephasingResults},\ref{CalculationDetails}] that give rise to $\mathcal{C}_7$ and $\mathcal{C}_8$ 
[Table~\ref{DiagramFoldingTable}, Eqs.~(\ref{FinalCoexistCooperon}) and (\ref{CT2Def})]
are actually split equally between 
trivial (double) and nontrivial (crossed, nested) diagrams in the unfolded framework.

\begin{table}
    \centering
    \begin{tabular}{|c||c||c|c|c|}
        \hline
        \multicolumn{5}{|c|}{Diagram unfolding} \\
        \hline
        \multicolumn{2}{|c|}{Folded diagrams} & Double & Nested & Crossed\\
        \hline
        Flipped $\tau$'s & $T$-type & \multicolumn{3}{|c|}{Unfolding results}\\
        \hline
        \hline
        \{\} & $T_{11}$ & \textbf{Double} & \textbf{Nested} & \textbf{Crossed} \\
        \{1\} & $T_{21}$ & \textbf{Double} & Nested & Crossed \\
        \{2\} & $T_{21}$ & \textbf{Double} & Crossed & Double \\
        \{3\} & $T_{12}$ & Nested & Crossed & Nested \\
        \{4\} & $T_{12}$ & Nested & Double & Crossed \\
        \{1,2\} & $T_{11}$ &\textbf{ Double} & \textbf{Double} & \textbf{Double} \\
        \{1,3\} & $T_{22}$ & Nested & Crossed & Nested \\
        \{1,4\} & $T_{22}$ & Nested & Crossed & Crossed \\
        \{2,3\} & $T_{22}$ & Nested & Crossed & Crossed \\
        \{2,4\} & $T_{22}$ & Nested & Crossed & Nested \\
        \{3,4\} & $T_{11}$ & \textbf{Double} & \textbf{Double} & \textbf{Double} \\
        \{1,2,3\} & $T_{12}$ & Nested & Crossed & Crossed \\
        \{1,2,4\} & $T_{12}$ & Nested & Crossed & Nested \\
        \{1,3,4\} & $T_{21}$ & \textbf{Double} & Double & Double \\
        \{2,3,4\} & $T_{21}$ &  \textbf{Double} & Nested & Crossed \\
        \{1,2,3,4\} & $T_{11}$ & \textbf{Double} & \textbf{Nested} & \textbf{Crossed} \\
        \hline
    \end{tabular}
	\caption{
	Table tracing the inverse folding of diagrams. The bold entries are the 
	terms contributing meaningfully to dephasing or rephasing in the asymptotic limit, 
	see Table~\ref{SecondOrderTable}. We are interested in taking a ``folded'' diagram of a given 
	shape and $T$-kernel type, and diagnosing its diagram shape in the ``unfolded'' path integral. 
	The right half of the top row lists the three possible topologies for the folded diagrams. 
	The far left column lists which time variables are to be flipped to the negative side of the time interval 
	$(-\eta,\eta)$ and the second column lists the type of contribution (``$T$-kernel'') of the folded diagram.}
    \label{DiagramFoldingTable}
\end{table}


\section{Series acceleration}\label{SeriesAcceleration}

\begin{figure}[t!]
        \includegraphics[width=0.4\textwidth]{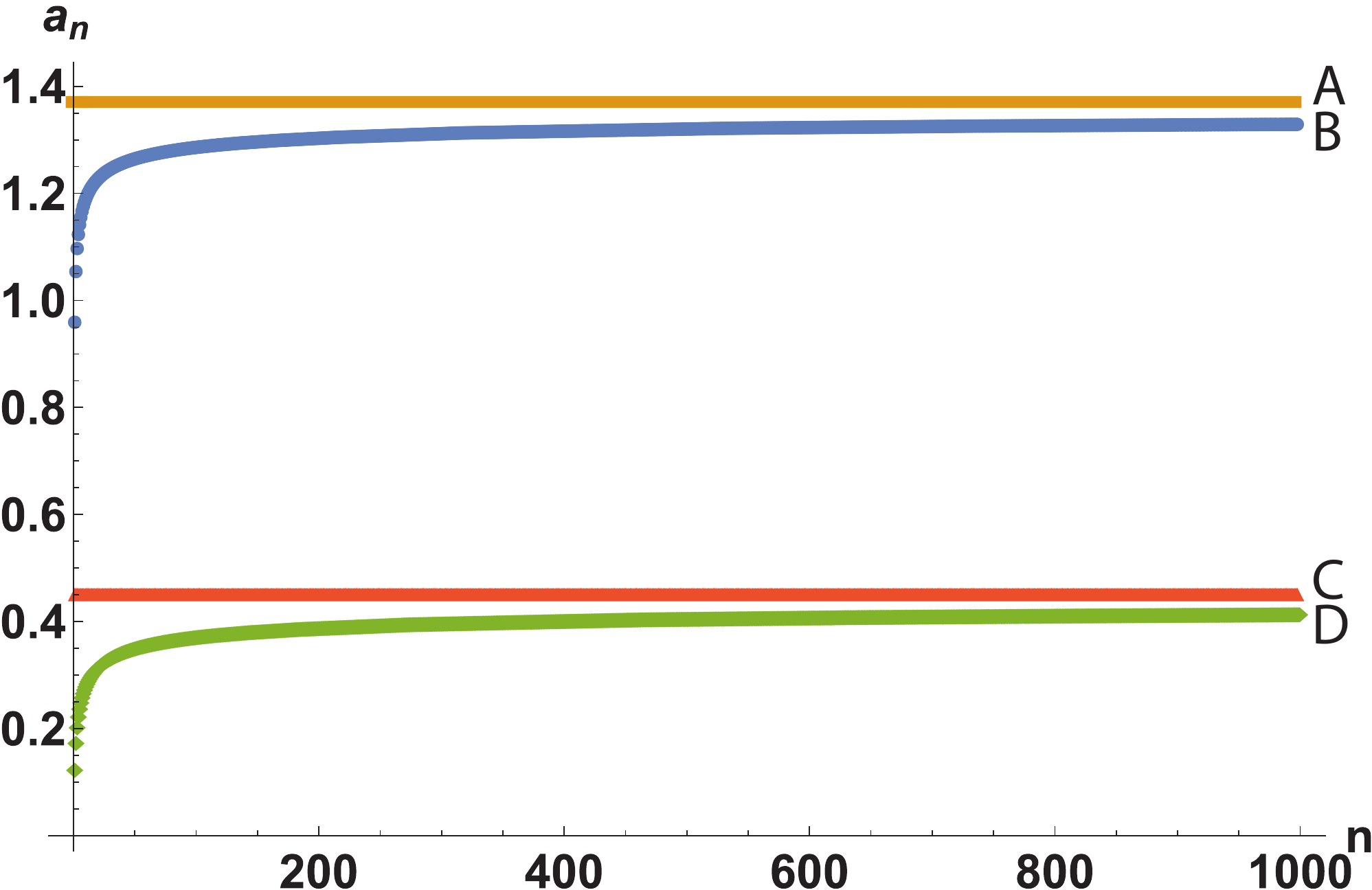}
        \caption{Depiction of the series acceleration technique demonstrated on the 
	Airy summation Eq.~(\ref{TestSum}). 
	A: The true answer given by The RHS of Eq.~(\ref{TestSum}). 
	B: Approximations by series truncation of the LHS of Eq.~(\ref{TestSum}) after $n$ terms. 
	We note that the series is slowly converging, and nontrivial error exists after 1000 terms have been summed. 
	C: Known analytical limit of the $p$-series summation, $C \zeta(p)$. 
	D: Partial sums of the best $p$-series approximation in the decomposition Eq.~(\ref{AiryPSeries}). 
	This plot shows that the slowly-converging nature of the original sum can be approximated well by a $p$-series with similar convergence properties.}
        \label{AirySumIllustration}
\end{figure}

\begin{figure}[b!]
        \includegraphics[width=0.4\textwidth]{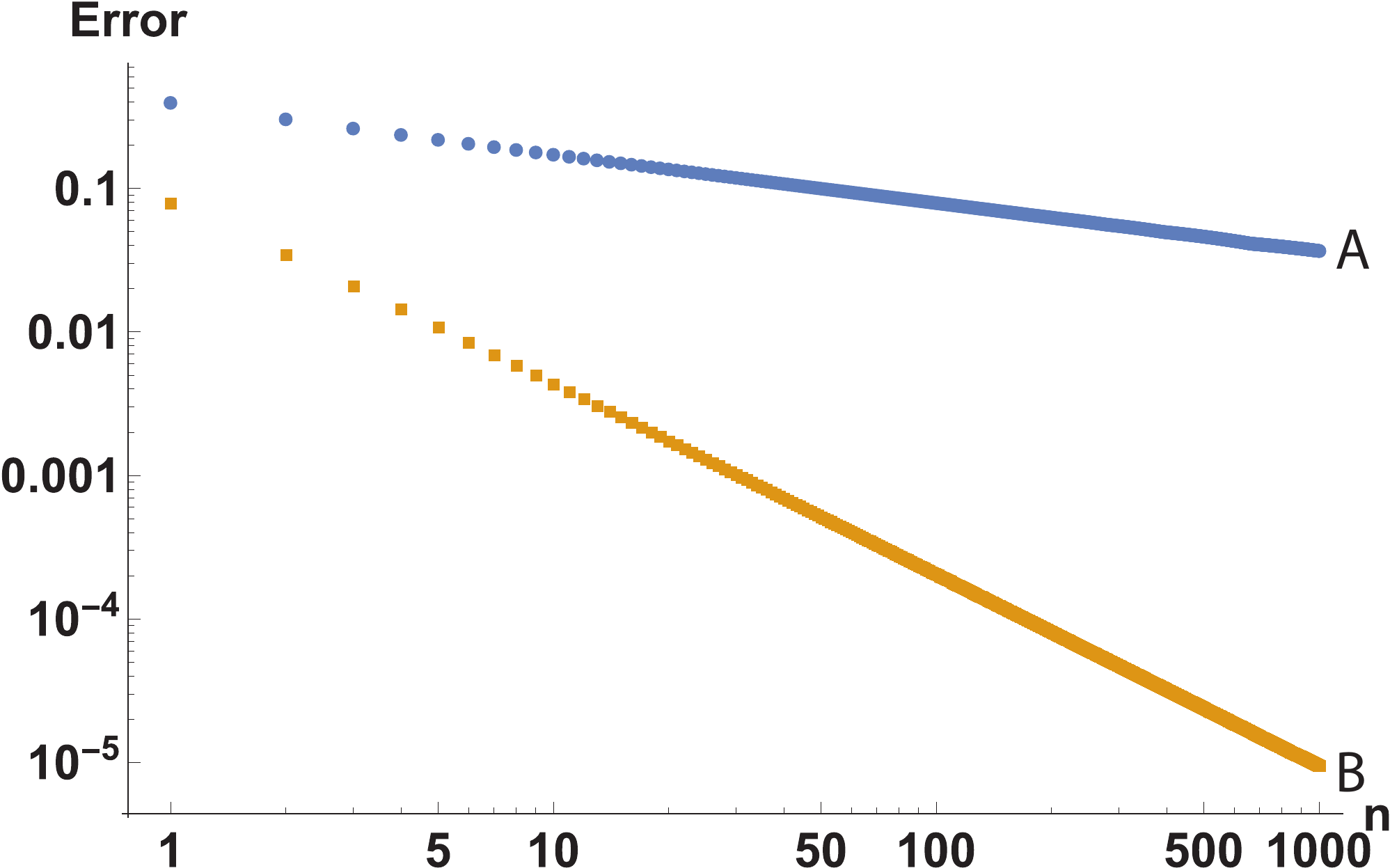}
        \caption{This plot compares convergence speeds for the direct summation of 
	Eq.~(\ref{TestSum}) and the accelerated sum. 
	A: direct summation---LHS of Eq.~(\ref{SumTrick}) applied to Eq.~(\ref{TestSum}). 
	B: Accelerated sum---RHS of Eq.~(\ref{SumTrick}) applied to Eq.~(\ref{TestSum}). 
	We see that the accelerated sum converges several orders of magnitude faster than the original sum.}
        \label{AirySumConvergence}
\end{figure}

\begin{figure}[t!]
    \centering
    \includegraphics[width=0.4\textwidth]{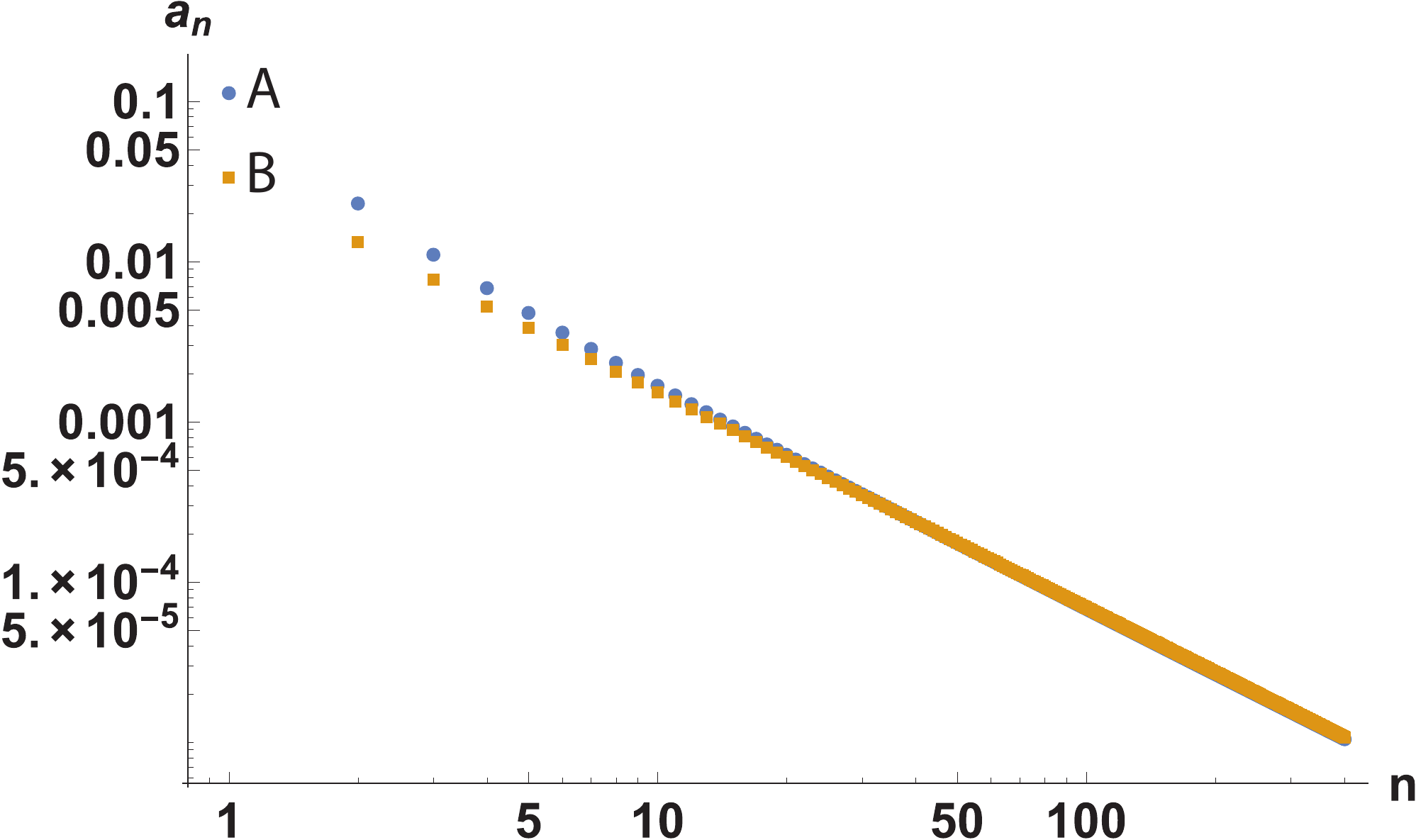}
    \caption{
	The acceleration technique discussed in this section requires the summands of the series in 
	question to be well-approximated by a $p$-series. This can be checked empirically by fitting a line 
	to the summands in a log-log scale. This plot uses this method to demonstrate that the $C_1(\beta)$ series, 
	defined in Eq.~(\ref{C1}) is well-approximated by a $p$-series. 
	A: $C_1(\beta = 1,i_1 = n)$, 
	B: $C n^{-p}$, where $C$ and $p$ are extracted from a linear fit.}
    \label{C1PSeries}
    \end{figure}

    \begin{figure}[b!]
    \includegraphics[width=0.4\textwidth]{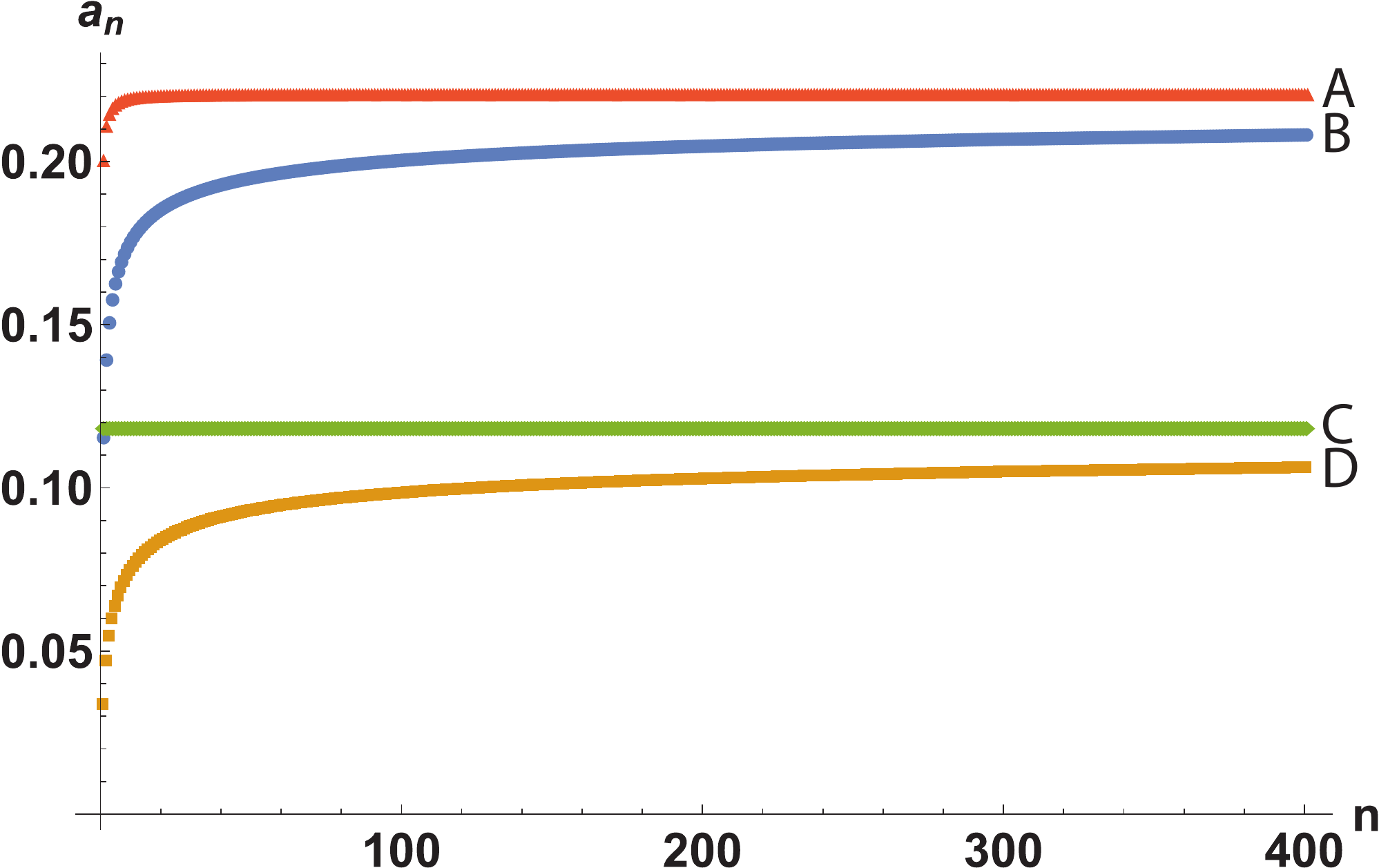}
    \caption{
	Depiction of the series acceleration of $\mathcal{C}_1(\beta)$, defined via Eq.~(\ref{C1}), at $\beta = 1$. 
	A: Partial sums of the accelerated series. We see that these converge rapidly. 
	B: Partial sums of the original series, which is slowly converging. 
	C: Known analytical limit of the $p$-series summation, $C \zeta(p)$. 
	D: Partial sums of the best $p$-series approximation in the decomposition Eq.~(\ref{SumTrick}). 
	We see that $\mathcal{C}_1(1) \approx 0.22$.}
    \label{C1Summation}
\end{figure}

We summarize a series acceleration technique used to estimate the dephasing coefficients, 
which are defined by slowly converging sums. For an absolutely convergent sum $\sum_{n=0}^\infty a_n$ 
and constants $p>1$, $C>0$, we have in general that
\begin{align}\label{SumTrick}
    \sum_{n=0}^\infty a_n = C\zeta(p) + \sum_{n=0}^\infty (a_n-Cn^{-p}),
\end{align}
where $\zeta$ is the Riemann Zeta function.
We call the sum over $C n^{-p}$ a ``$p$-series''.
In the case that $a_n \rightarrow Cn^{-p}$ rapidly, Eq.~(\ref{SumTrick}) can be used to efficiently estimate the sum: 
\begin{align}\label{SumTrick2}
    \sum_{n=0}^\infty a_n \simeq C\zeta(p) + \sum_{n=0}^N(a_n-Cn^{-p}),
\end{align}
for some sufficiently large $N$. This procedure works by packaging the slow convergence of $a_n$ into the zeta-function. 

To illustrate the method, we estimate the sum from Eq.~(\ref{AAKConductivity}),
\begin{align}
    \label{TestSum}
    \sum_n \frac{1}{(\alpha'_n)^2} = \frac{2\pi}{3^{5/6}\Gamma(2/3)^2},
\end{align}
which was used to derive the exact conductivity correction for the screened Coulomb noise bath. 
Since the zeros of $\Ai'(x)$ are asymptotically given by $\alpha_n \simeq -(3\pi/2)^{2/3}n^{2/3}$, 
the series in Eq.~(\ref{TestSum}) tends to a $p$-series with 
\begin{align}\label{AiryPSeries}
    \frac{1}{(\alpha'_n)^2} \simeq \left(\frac{2}{3\pi}\right)^{4/3}n^{-4/3}.
\end{align}
In Fig.~\ref{AirySumIllustration}, we plot the convergence of the left-hand-side of 
Eq.~(\ref{TestSum}) to the right-hand-side of Eq.~(\ref{TestSum}) alongside the convergence of 
$
	\sum_{n=0}^{\infty}Cn^{-p}
$ 
to 
$
	C\zeta(p),
$ 
with $C = \left(3\pi/2\right)^{-4/3}$ and $n=4/3$. 
We see that both series are slowly converging, but that the convergence rate is extremely similar. 
In Fig.~\ref{AirySumConvergence}, we compare the convergence speeds of the original and boosted summations, 
given by truncating the left- and right-hand sides of Eq.~(\ref{SumTrick}), respectively. 
We see that the accelerated sum converges several orders of magnitude faster than the original sum.

We can use the series acceleration technique to approximate the sums $\mathcal{C}_1$ and $\mathcal{C}_2$ 
[Eqs.~(\ref{FullFirstOrder}), (\ref{CSum}), (\ref{C1}) and (\ref{C2})]
to all energies, giving us the full first order correction to the dephasing rate. This gives 
Figs.~\ref{FirstOrderCoexisitingPlot} and \ref{SecondOrderCoexisitingPlot}. 
Fig.~\ref{C1PSeries} shows that the $\mathcal{C}_1$ summation is well-approximated by a $p$-series, 
and 
Fig.~\ref{C1Summation} compares the convergences of the original and accelerated series.


\section{Fermionic field theory}\label{FieldTheory}

\subsection{Field theory for Cooperon}\label{Cooperon}

We reviewed in Sec.~\ref{MarkovianDephasing} how the single-particle path integral and relative-time coordinates provide a powerful tool 
for the non-perturbative treatment of Markovian noise kernels \cite{AAK}. The fluctuation-averaged Cooperon studied in this 
paper can also be calculated in terms of a replicated fermionic field theory framework \cite{Liao18, Cardy}. 
The generating function of the theory is
\begin{align}
\label{PartitionFuction}
    \mathcal{Z} &= \int\mathcal{D}\bar{\Psi}\mathcal{D}\Psi\mathcal{D}\phi_{\text{cl}} \, e^{-S_\Psi[\bar{\Psi},\Psi] - S_\phi[\phi_{\text{cl}}]-S_c[\bar{\Psi},\Psi,\phi_{\text{cl}}]},
\end{align}
with the action components
\begin{align}
	\label{ActionPsi}	  
	S_\Psi[\bar{\Psi},\Psi]
	=&\, 
	\int\limits_{\vex{k},\omega}\bar{\Psi}^a(\omega,\vex{k})\left[\frac{D}{2}k^2-i\omega\right]\Psi^a(\omega,\vex{k}),\!\!
\\
	\label{ActionPhi}
	S_\phi[\phi_{\text{cl}}] 
	=&\, 
	\frac{1}{2}\Gamma\int\limits_{\vex{k},\omega}\frac{\phi_{\text{cl}}(\omega,\vex{k})  \, \phi_{\text{cl}}(-\omega,-\vex{k})}{\Delta(\omega,\vex{k})},
\\
	\!\! 
	S_c[\bar{\Psi},\Psi,\phi_{\text{cl}}] 	
	=&\, 
	\frac{i}{2}\sqrt{\Gamma}\int\limits_{\vex{k},\omega}\int\limits_{\vex{q},\Omega}\phi_{\text{cl}}(\Omega,\vex{q})
\nonumber
\\
	&\,
	\times
	\left[
	\begin{aligned}
	\bar{\Psi}^a(\omega+\frac{\Omega}{2},\vex{k}+\vex{q})\\
	-\bar{\Psi}^a(\omega-\frac{\Omega}{2},\vex{k}+\vex{q})
	\end{aligned}
	\right]
	\Psi^a(\omega,\vex{k}).\!\!
	\label{ActionInt}
\end{align}
Above, $\Delta(\omega,\vex{k})$ is the noise kernel for the theory, 
$D$ is the classical diffusion constant due to elastic scattering [Eq.~(\ref{CooperStoch})], 
and 
$\Gamma$ is the coupling to the bath [as in Eqs.~(\ref{GammaM}) and (\ref{Gammat}) in the main text]. 
We choose to embed the Cooperon using the replicated fermion field $\Psi^a$, where $a \in \{1,2,\ldots,n\}$ 
and we take $n \rightarrow 0$ at the end \cite{Liao18};
the doubly-repeated replica index is Einstein summed in Eqs.~(\ref{ActionPsi})--(\ref{ActionInt}).
(Equivalently, we could employ the 
Keldysh formalism to normalize the generating function $\mathcal{Z} = 1$; this is 
natural in the full dynamical sigma model \cite{AlexAlex,KamBook,Liao17}. 
We use replicas here only to lighten the notation.)

\begin{figure}
    \includegraphics[width=0.4\textwidth]{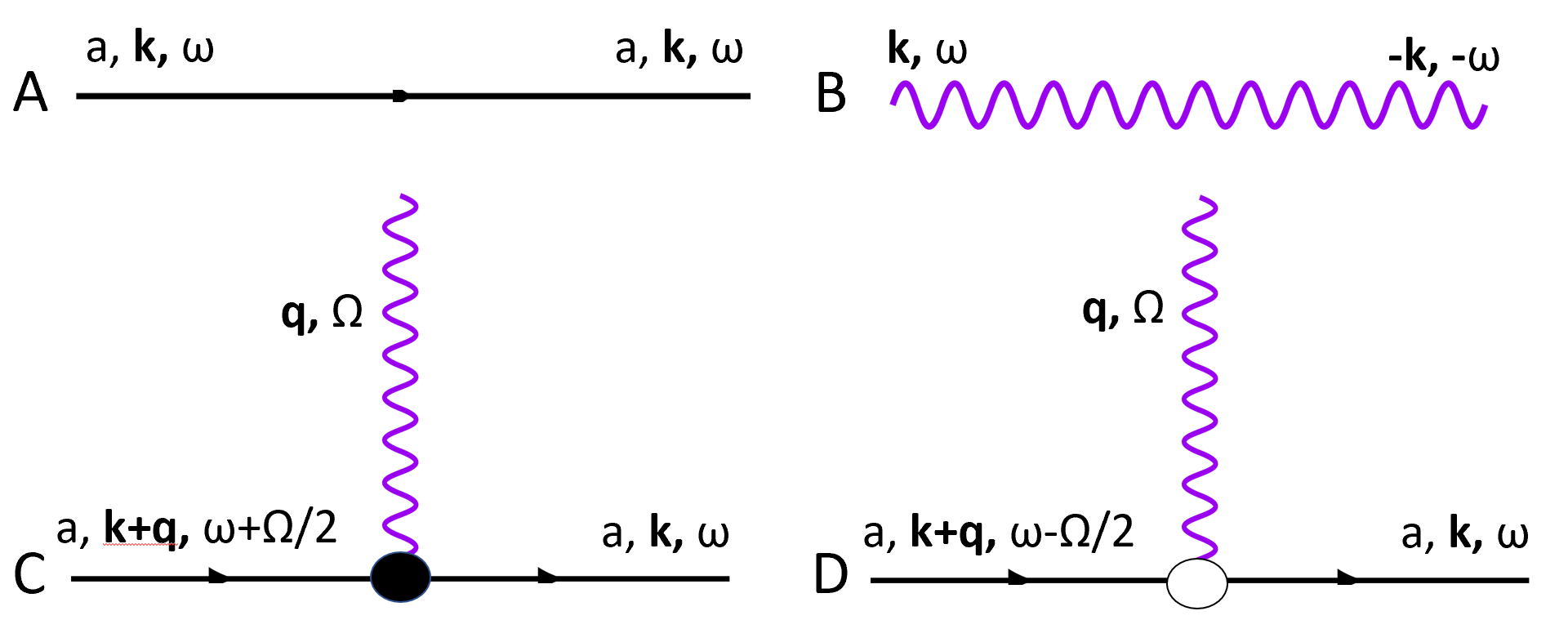}
	\caption{
	The Feynman rules for the field theory of the fluctuation-averaged Cooperon, 
	defined in Eq.~(\ref{PartitionFuction}). 
	Diagrams (A) and (B) represent the bare propagators for the $\Psi^a$ and $\phi_{\text{cl}}$ fields, respectively. 
	Diagrams (C) and (D) depict the two types of interaction vertices coupling the fields. 
	The vertices in (C) and (D) are the ``causal'' and ``anticausal'' vertices, respectively.}
    \label{FieldTheoryFeynmanRules}
\end{figure}

The full fluctuation-averaged Cooperon is obtained in the replica limit as the correlation function 
\begin{align}
	\langle \cc^t_{\omega',\omega}(\vex{k})\rangle_{\phi_{\text{cl}}} 
	&= 
	\frac{D}{2}
	\tilde{\cc}(\omega',\omega,\vex{k})
\\
	&\equiv
	\frac{D}{2}
	\langle\Psi^a(\omega',\vex{k}) \, \bar{\Psi}^a(\omega,\vex{k})\rangle_{\mathcal{Z}},
\end{align}
where $\langle\cdots\rangle_{\mathcal{Z}}$ denotes a functional average over the full partition function $\mathcal{Z}$,
and the ``reduced Cooperon'', $\tilde{\cc}_R\left(\omega,\vex{k}\right)$, follows from averaging over relative frequency $\Omega$,
\begin{align}
	\tilde{\cc}_R\left(\omega,\vex{k}\right) 
	&= 
	\frac{1}{2}
	\int\limits_{\Omega}
	\tilde{\cc}\left(\frac{\omega - \Omega}{2},\frac{\omega + \Omega}{2},\vex{k}\right),
\\
	\cc(\eta) 
	&= 
	\frac{D}{2}
	\int\limits_{\omega,\vex{k}}
	e^{-i\eta\omega}
	\,
	\tilde{\cc}_R\left(\omega,\vex{k}\right).
\end{align}
The Feynman rules for the theory are given in Fig.~\ref{FieldTheoryFeynmanRules}. 
The corresponding bare propagators for the theory are given by 
\bsub\label{FieldTheoryPropagators}
\begin{align}
	\langle\Psi^a(\omega,\vex{k})\bar{\Psi}^b(\omega,\vex{k})\rangle_{0} 
	&= 
	\delta^{ab}
	\left[\frac{D}{2}k^2 - i\omega\right]^{-1}
\nonumber\\
	&\equiv \delta^{ab}
	\tilde{\cc}_0(\omega,\vex{k}),
\\
	\langle\phi_{\text{cl}}(\omega,\vex{k})\phi_{\text{cl}}(-\omega,-\vex{k})\rangle_{0} 
	&= 
	\Delta(\omega,\vex{k})/\Gamma.
\end{align}
\esub

\begin{figure}[b]
    \includegraphics[width=0.4\textwidth]{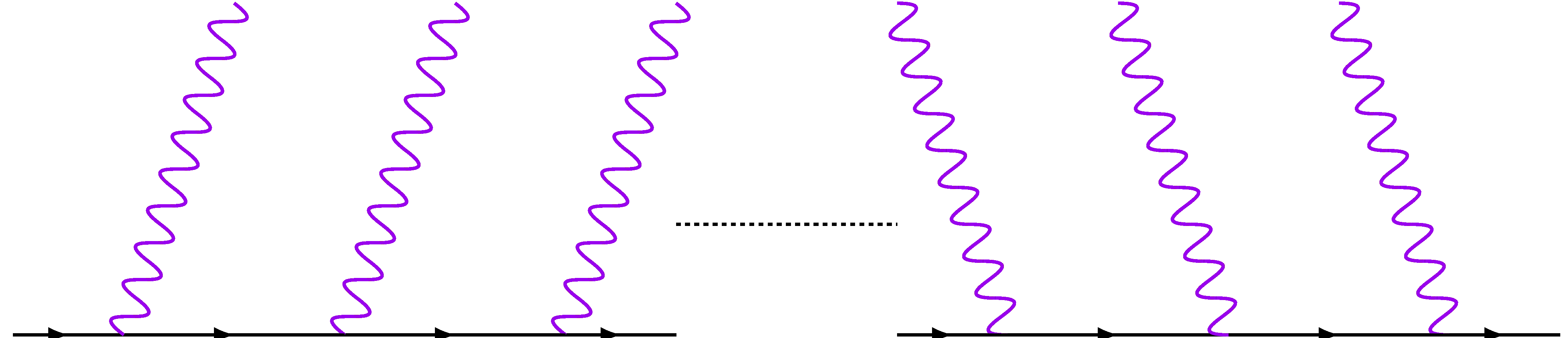}
	\caption{
	The topology of a generic Feynman diagram contributing to the Cooperon, before averaging over the bath.
	The replica limit removes all diagrams with closed Fermion loops, so all relevant diagrams 
	contain a single Fermion line dressed with some number of noise phonons.}
    \label{GenericFieldTheoryDiagram}
\end{figure}

All diagrams with closed fermion loops vanish in the replica limit, and so the only contributing diagrams 
to the full Cooperon contain a 
single fermion line dressed with noise propagators. This restriction on the diagram topology leaves 
$(2n)!/(n!2^n)$ topologically distinct diagrams at $n^{th}$ order, with a generic diagram depicted in 
Fig.~\ref{GenericFieldTheoryDiagram}.

Fig.~\ref{FieldTheoryFeynmanRules} demonstrates the interaction vertices coupling between the Cooperon $\Psi^a$ and noise 
$\phi_{\text{cl}}$ fields. The two distinct vertices arise from the first [diagram C] and second [diagram D] terms in 
the noise bath action in 
Eq.~(\ref{ActionInt}).
We will refer to these as causal and anticausal vertices, 
respectively, and they contribute factors of $\pm i\sqrt{\Gamma}/2$ to the diagram's overall prefactor.

\begin{figure}
    \includegraphics[width=0.4\textwidth]{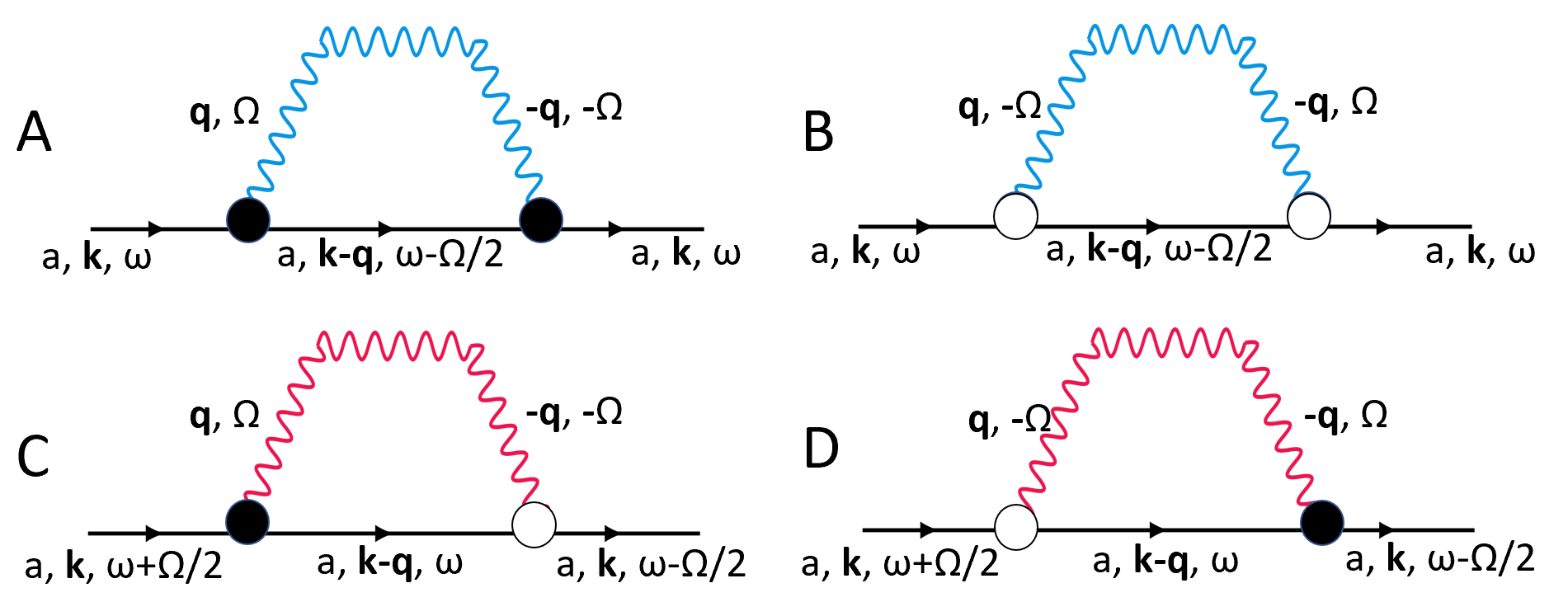}
	\caption{Labeling convention that allows us to forget about the causal and anti-causal vertices shown in Fig.~\ref{FieldTheoryFeynmanRules}(C,D) 
	and instead consider only ``type I'' (colored blue) and ``type II'' (colored red) noise phonons (as defined in the main text). 
	We choose the frequency of the noise phonon to be $\omega_j$ ($-\omega_j$) if the leftmost vertex is causal (anticausal). 
	Type I phonons (A,B) give frequency-diagonal contributions, while type II phonons (C,D) give frequency-off-diagonal contributions.}
	\label{DiagramLabeling}
\end{figure}

We introduce some useful terminology and convention. A noise phonon connecting two vertices of the same type (causal-causal or anticausal-anticausal) 
will be called 
``type I,'' 
while a noise propagator connecting two vertices of the opposite type will be called 
``type II.'' 
By choosing the momenta and frequencies for internal noise phonons as shown in Fig.~\ref{DiagramLabeling}, we can forget about the 
causal and anti-causal vertices and work directly with type I and II noise phonons. We note that while type I phonons contribute 
frequency-diagonal terms to the Cooperon, the type II phonons introduce frequency non-diagonal terms. The only frequency-diagonal 
diagrams are purely type I. While type I diagrams control the RG flow in higher dimensions \cite{Liao18}, 
both type I and II diagrams play important roles in the perturbative dephasing calculation. We note that type I and II phonons contribute factors of $-\Gamma/2$ and $\Gamma/2$ respectively to 
the overall diagram prefactor. Each $n^{th}$ order diagram will have $2^n$ colorings of its noise phonons as type I or II, so that each diagram topology generally has competition between exponentially many 
opposite-sign contributions.

\subsection{Diffusive bath and connection with cumulant expansion}\label{DiffusiveBath}

Re-expanding the cumulant expansion, Eq.~(\ref{CumulantDefinition}), directly in $\Gamma$ gives us 
\begin{align}
    \cc(\eta) &= \cc_0({\eta})\left[1 - \langle S_1 \rangle + \frac{1}{2}\langle S^2_1 \rangle + \ldots\right].
\end{align}
We can also directly expand the Cooperon in powers of $\Gamma$ in the replicated fermionic field theory:
\begin{align}
    \cc(\eta) &= \cc_0({\eta}) + \cc_1({\eta}) + \cc_2({\eta}) + \ldots
\end{align}
Formally equating terms of the two power series, we find that 
\begin{align}
    \langle S_1^n \rangle_0 = (-1)^n n! \frac{\cc_n({\eta})}{\cc_0({\eta})}.
\end{align}
We can thus compute terms in the cumulant expansion ---originally framed as expectations in a path integral--- 
directly in the field theory.

In particular, calculating the first order type I 
diagrams
depicted in Fig.~\ref{DiagramLabeling} 
for the diffusive noise kernel ($ \equiv D^{\text{diff.}}_{1,\text{type I}}$), we find 
\begin{align}
	\label{Ddiff1IEval}
	D^{\text{diff.}}_{1,\text{type I}} 
	=&\, 
	-
	\left(\frac{D}{2}\frac{1}{\sqrt{4\pi D\eta}}\right)
	\left(\Gamma_t\frac{\eta^{3/2}}{\sqrt{D}}\right)\tilde{G}^1_1(\beta)
\\
	\tilde{G}^1_1(\beta) 
	\equiv&\, 
	\sqrt{\frac{2}{\pi}}
	\int\limits_{-1}^1 d\tau_a 
	\int\limits_{\tau_a}^1 d\tau_b\ 
	g_1(\beta,\tau_a,\tau_b)^{-1/2},
\end{align}
in line with the results of Eqs.~(\ref{FirstOrderExpectationResult}) and (\ref{FirstOrderGFunc}). 
Evaluating the first order type II diagrams in Fig.~\ref{DiagramLabeling} 
then gives the other contribution to Eq.~(\ref{FirstOrderGFunc}). 
We note that in the field theory framework, the parametric integration defining $G_1(\beta)$ 
arises from a Feynman parameter.


\subsection{Markovian Coulomb bath and connection to AAK}\label{MarkovianBath}

\subsubsection{General remarks and divergence regularization}

In the case of a Markovian noise kernel, we have seen that a coordinate change in the path integral 
formalism [Eq.~(\ref{Change})] allows for a massive reduction in complexity, recasting the fluctuation-averaged 
Cooperon as the Green's function of a single-particle quantum mechanics problem, Eq.~(\ref{AAKReductionIII}). 
In the Markovian limit of the field theory description, we find that all but a special class of diagrams vanish exactly, 
and that summing the remaining diagrams to all orders recovers the AAK reduction formula for the propagator. Thus, 
Eqs.~(\ref{AAKReductionII}) and (\ref{AAKReductionIII}) can be alternatively derived as an infinite-order field-theoretic 
summation. 

In the Markovian limit, we find IR divergences in the momentum integration due to the divergence of the noise kernel 
at zero momentum. This is regularized by a perfect cancellation of all IR divergences between all the diagrams at a given order. 
This is easily seen at first order and can be shown at arbitrary order. This cancellation is fundamentally related to the 
cancellation of IR divergences in the Fourier transform that gave us Eq.~(\ref{EffCoulKer}):
\begin{align}
\label{IRCancellation}
    \tilde{\Delta}_M(0)-\tilde{\Delta}_M(\vex{\rho}) = \frac{2\Gamma_M}{D}\int\limits_{\vex{k}} \frac{1}{k^2}(1-e^{i\vex{k}\cdot\vex{\rho}}).
\end{align}
Note that individually, $\tilde{\Delta}_M(0)$ and $\tilde{\Delta}_M(\vex{\rho})$ are IR divergent in one or two spatial dimensions, 
but their difference is finite in 1D (and UV divergent in higher dimensions). 
We will see that the UV contributions of the type I phonons are responsible for the 
``$\tilde{\Delta}_M(0)$'' term, while the type II phonons are responsible for the ``$\tilde{\Delta}_M(\vex{\rho})$'' 
term in Eq.~(\ref{AAKReductionIII}).

In Eqs.~(\ref{FirstOrderCisEvaluation})--(\ref{SelfConsistentII}) (below), we will sum all the diagrams in the perturbative 
expansion via a two-step framework. We first sum the type I phonons into a dressed propagator and then translate the type II 
phonon contributions into a self-consistent integral equation. This calculation will require us to treat the type I and type II 
phonon lines on separate footing, obscuring the fact that the IR divergences between the various diagrams cancel out order-by-order 
in perturbation theory. We must therefore cancel out the IR divergent portions of the diagrams in the beginning, \textit{before} 
the type I resummation. In the type I resummation and later in the self-consistent equation, we then drop the IR-divergent portion 
of the momenta integrations. We will use the integral superscript ``$(-IR)$'' to indicate that it is necessary to remove by hand the 
infrared divergences in the momenta integrations. We also define
\begin{align}
	\tilde{\Delta}_M^{(-IR)}(\vex{\rho})
	\equiv
	\int\limits_{\vex{k}}^{(-IR)} \Delta_M(\vex{k}) \, e^{i\vex{k}\cdot\vex{\rho}}.
\end{align}
We point out that in 1D, $\tilde{\Delta}_M^{(-IR)}(0) = 0,$ while in higher dimensions, $\tilde{\Delta}_M^{(-IR)}(0)$ 
is dependent on a UV cutoff.

\begin{figure}[t!]
    \includegraphics[width=0.4\textwidth]{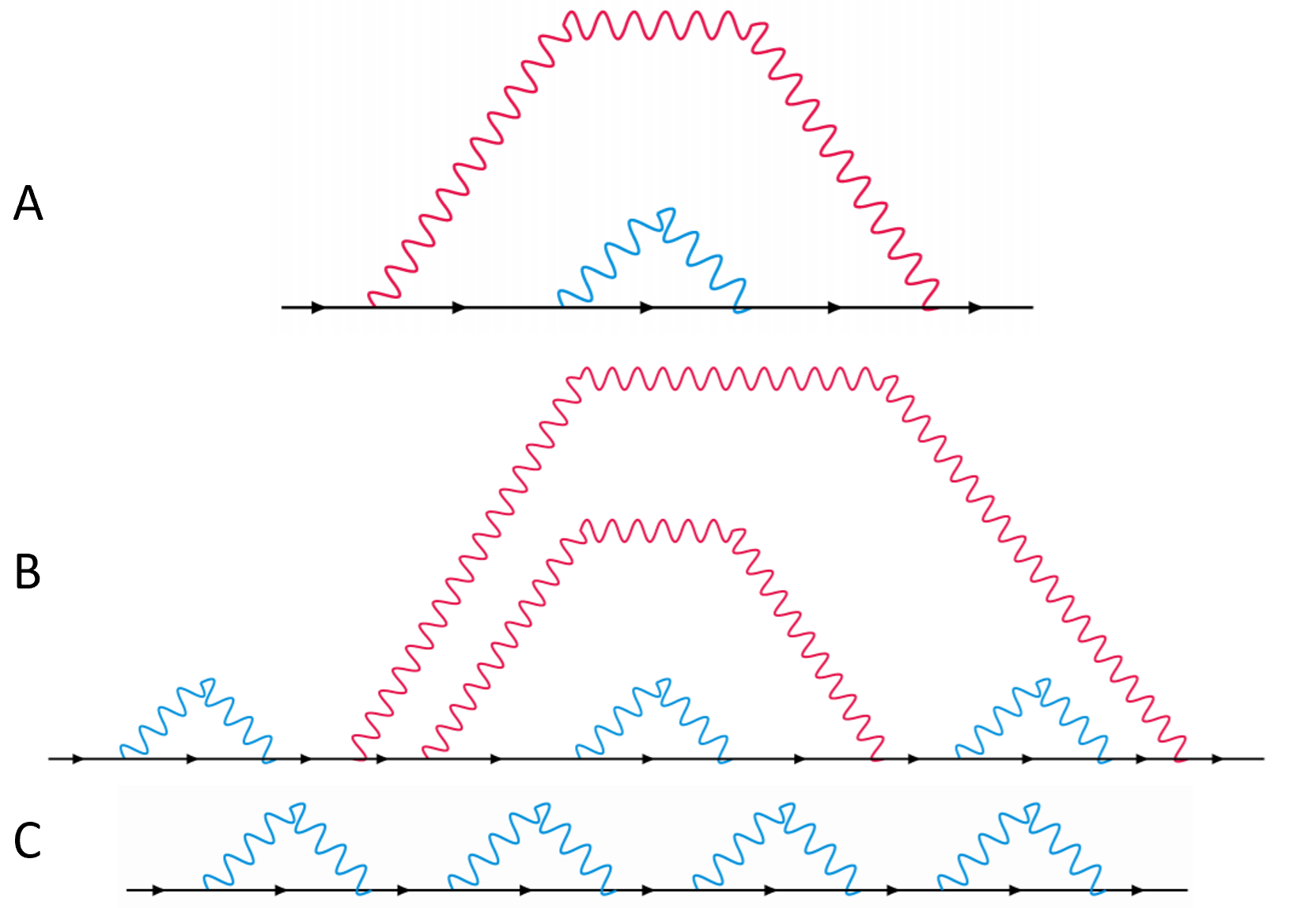}
	\caption{Examples of non-vanishing diagrams that follow the rules outlined in Sec.~\ref{Rules}. 
	In all the above diagrams, type I noise phonons are colored blue and type II noise phonons are colored red. 
	As required to be non-vanishing, the type I noise phonons pass over no vertices, and the type II noise phonons are in a nested rainbow configuration.}
    \label{NonVanishingDiagramExamples}
\end{figure}

\begin{figure}[b!]
    \includegraphics[width=0.4\textwidth]{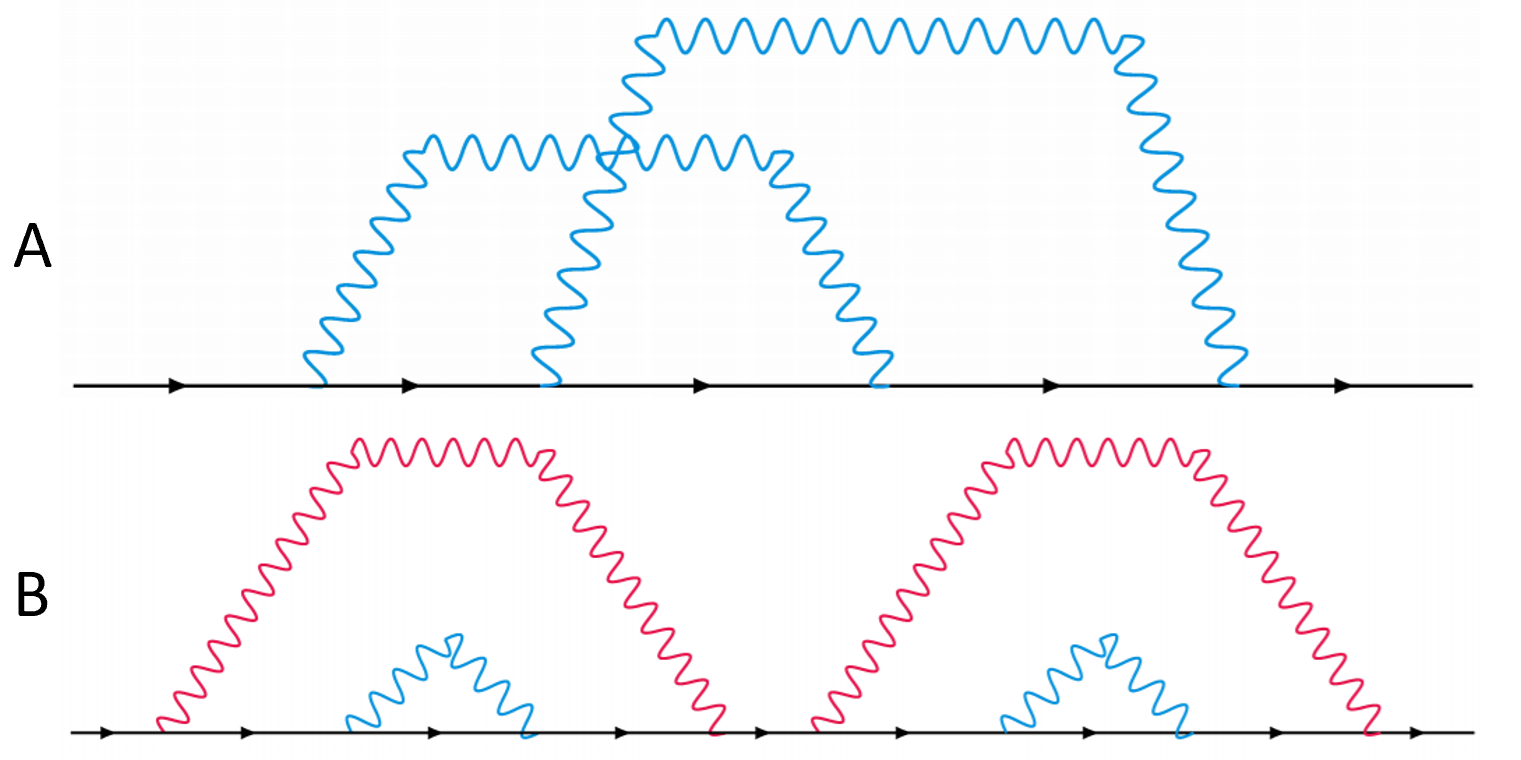}
	\caption{Two examples of vanishing diagrams, which fail to follow the rules outlined in Sec.~\ref{Rules}. 
	In all the above diagrams, type I noise phonons are colored blue and type II noise phonons are colored red. 
	Diagram (A) fails to follow the first rule, since each type I phonon passes over a vertex on the fermion line. 
	Diagram (B) fails to follow the second rule, since its two type II phonons are not in the required nested rainbow configuration. 
	We note that the fact that diagram (A) in Fig.~\ref{NonVanishingDiagramExamples} cannot be resummed as a self-energy for the 
	Cooperon can be understood by noting that Diagram (B) in this figure vanishes.}
    \label{VanishingDiagramExamples}
\end{figure}

\begin{figure}[t!]
\includegraphics[angle=0,width=.4\textwidth]{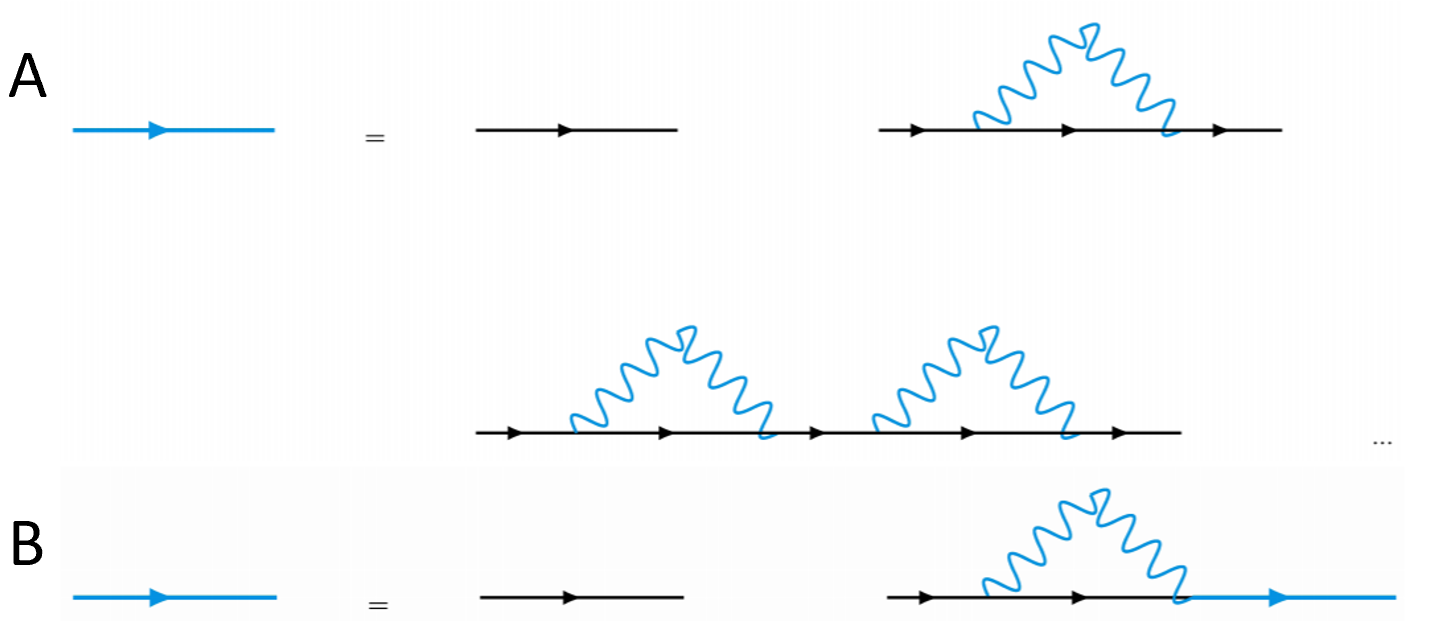}
\caption{The diagrammatic infinite-order summation of the type I noise phonons via a Dyson's equation with a 
``pseudo self-energy''. In (A), we define a ``type I dressed propagator'' (blue fermion line) to be a the sum of all Cooperon diagrams dressed only by type I phonons. 
In (B) we re-write the summation self-consistently as a Dyson equation. In this case, the role of the self-energy is played by the diagram with 
a single type I phonon, which we evaluate directly in Eq.~(\ref{FirstOrderCisEvaluation}).}
\label{CisSummation}
\end{figure}


\subsubsection{Diagram non-vanishing requirements}\label{Rules}

Performing the frequency integrations analytically, we find that most of the Feynman diagrams 
one can draw vanish in the Markovian case. We can codify this into two rules that 
a diagram contributing to the Green's function must satisfy. The rules are as follows:
\begin{enumerate}
\item There can be no vertices under a type I phonon. If a type I phonon leaves the fermion line, it must return at the very next vertex. 
\item If a diagram contains \textit{any} type II phonons, they must all be nested in a non-crossing rainbow configuration. 
\end{enumerate}

To prove the vanishing rules, one can perform the frequency contour integrations 
in the complex plane and show that if either rule is violated, the diagram can be labeled so 
that there is at least one frequency integration for which all the poles lie in one half of the plane. 
Closing the frequency integration contour in the opposite half of the plane shows that the integral vanishes. 
Diagrams illustrating the vanishing rules are given in Figs.~\ref{NonVanishingDiagramExamples} and \ref{VanishingDiagramExamples}.

The non-vanishing rules for this theory require alteration of the usual paradigms of field theory. For 
example, the self-energy cannot be thought of in usual terms. A diagram with nested type II rainbows, 
as in Fig.~\ref{NonVanishingDiagramExamples}(A) above, does not vanish and is 1-particle irreducible. 
However, it cannot be resummed into a self-energy, because the diagram consisting of two sequential copies of it, 
as in Fig.~\ref{VanishingDiagramExamples}(B), vanishes due to violation of rule 2. Thus, no diagrams containing 
type II phonons can be resummed into a self-energy, for they only appear exactly once in the expansion of the Green's function. 
On the other hand, we \textit{can} sum the  type I diagrams to all orders into a ``pseudo self-energy'' 
---we take this up in the next subsection.


\subsubsection{Type I phonon resummation}

The type I diagrams can be resummed to all orders into a ``pseudo self-energy''. This is simple and 
can be done exactly since the type I phonons appear in isolation and cannot cross; the only diagram that enters into the 
pseudo self-energy is the first order type I diagram ($\equiv~D^{\text{Coul.}}_{1,\text{type I}}$) [Fig.~\ref{DiagramLabeling} (A, B)]. We can then define a type I dressed propagator as shown diagrammatically in 
Fig.~\ref{CisSummation}.

We can perform this summation via the self-consistent equation given by the diagrams in Fig.~\ref{CisSummation}, 
letting $\tilde{\cc}_\text{I}$ denote the ``type I dressed propagator''. We find
\begin{equation}\label{CisDressedPropagator}
	\tilde{\cc}_{\text{I}}\left(\frac{\omega-\Omega}{2},\frac{\omega+\Omega}{2},\vex{k}\right) 
	= 
	\frac{4\pi\delta(\Omega)}{Dk^2 - i\omega + \tilde{\Delta}_M(0)},
\end{equation}
because 
\begin{align}\label{FirstOrderCisEvaluation}
	D^{\text{Coul.}}_{1,\text{type I}} 
	&= \frac{-1}{2}\int\limits_{\nu,\vex{q}}^{(-IR)}\frac{2 \Delta_M(\vex{q})}{D(\vex{k}-\vex{q})^2-i\omega-i\nu}
\nonumber\\
	&= \frac{-1}{2}\int\limits_{\vex{q}}^{(-IR)}\Delta_M(\vex{q}) 
\nonumber\\
	&= \frac{-1}{2}\tilde{\Delta}_M^{(-IR)}(0).
\end{align}


\begin{figure}[t!]
\includegraphics[angle=0,width=.4\textwidth]{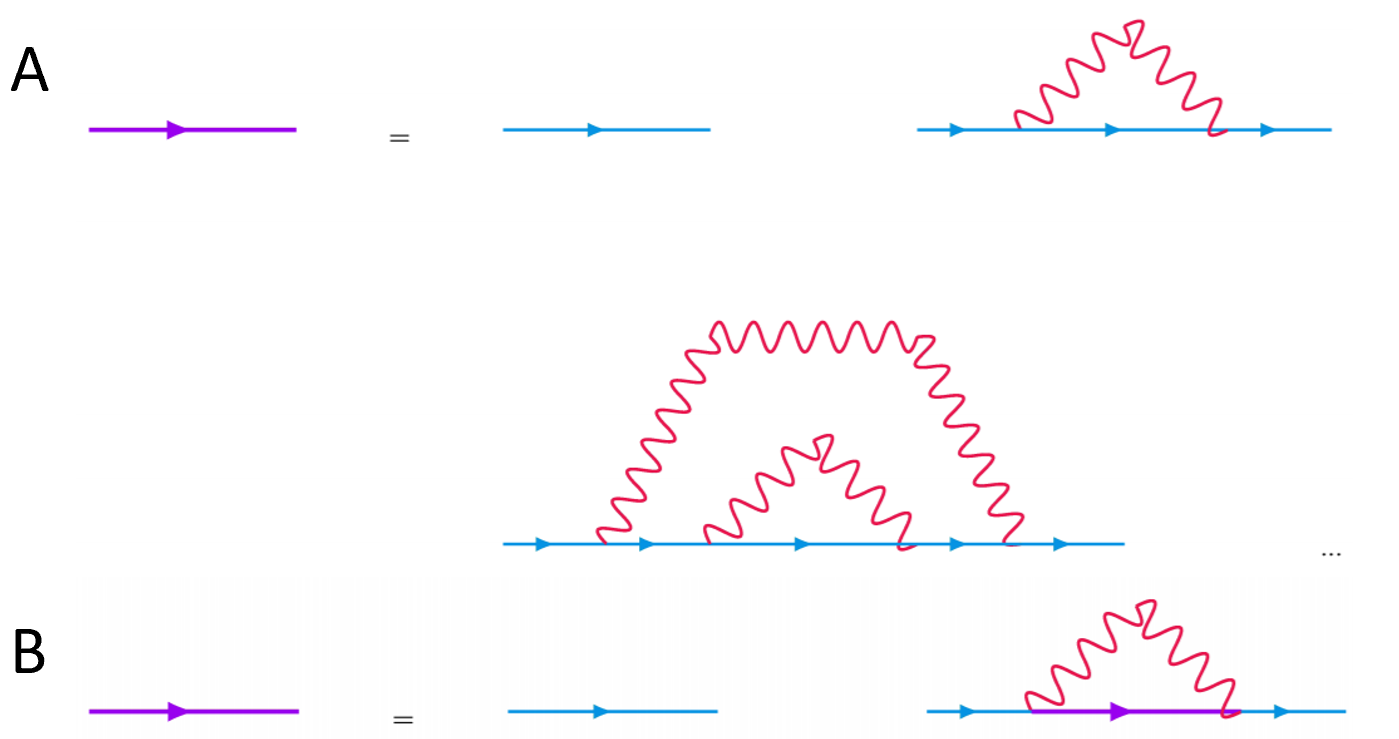}
\caption{The diagrammatic infinite-order summation of the type II noise phonons via a self-consistent equation. 
In (A), we express the fully dressed propagator (purple fermion line) as the ``type I dressed propagator'' (blue fermion line) dressed by the summation of all maximally-nested 
rainbow configurations of type II noise phonons. In (B) we re-write the expansion in (A) as a self-consistent equation. The structure 
of the self-consistent equation is reminiscent of that of the self-consistent born approximation [Eq.~(\ref{SCBA1})], but here the LHS 
of the equation is the fully dressed propagator, rather than the self-energy. This self-consistent equation sums a proper subset of the 
usual SCBA diagrams. 
The diagrammatics are translated into a Fredholm integral equation [Eqs.~(\ref{SelfConsistentI}), (\ref{SelfConsistentII}), and (\ref{BackToAAK})] 
that turns out to be equivalent to the AAK result from the main text, Eq.~(\ref{AAKReductionIII}).}
\label{TransSummation}
\end{figure}

\subsubsection{Full propagator and connection to AAK solution}\label{FullPropAAKReduction}

With the type I propagators summed to infinite order via the pseudo self-energy, we can put the perturbative series for the full Green's function into a simpler form. The remaining diagrams to consider are the non-crossing rainbow diagrams with type I dressed fermion propagators and type II noise propagators, as shown in Fig.~\ref{TransSummation}.

In this framework there is a single diagram left at each order (in the type II phonon) in the series defining the Cooperon. The average and internal frequency integrations can be performed analytically, giving the $n^{th}$ order contribution to the reduced Cooperon as
\begin{align}\label{NthOrderPerturbationCooperon}
	\tilde{\cc}^{(n)}_R(\omega,\vex{k}) 
	=&\, 
	\frac{1}{2}\int\limits^{(-IR)}_{\vex{l_1}}
	\Delta_M(\vex{l_1}) 
	\ldots  
	\int\limits^{(-IR)}_{\vex{l_n}}
	\Delta_M(\vex{l_n})
\nonumber\\
	&\,
	\times
	\frac{2}{Dk^2 + \tilde{\Delta}_M^{(-IR)}(0) - i\omega}
\nonumber\\
	&\,
	\times
	\cdots
\nonumber\\
	&\,
	\times
	\frac{2}{D(\vex{k} - \ldots - \vex{l}_n)^2 + \tilde{\Delta}_M^{(-IR)}(0) - i\omega}.
\end{align}

We can treat the type II phonon diagrams to all orders by deriving a self-consistent integral equation for the full propagator, as shown diagrammatically in Fig.~\ref{TransSummation}. The diagrammatic result corresponds to the self-consistent equation for the full Cooperon.
\begin{align}
\label{SelfConsistentI}
	\tilde{\cc}&\left(\frac{\omega-\Omega}{2},\frac{\omega+\Omega}{2},\vex{k}\right) 
\nonumber\\
	=&\,\,
	\frac{4\pi\delta(\Omega)}{Dk^2+\tilde{\Delta}_M^{(-IR)}(0) - i\omega} 
\nonumber\\
	&\,\,
	+ 
\left\{
	\begin{aligned}
	&\,
	\frac{2}{\left[Dk^2+\tilde{\Delta}_M^{(-IR)}(0) - i\omega + i\Omega\right]}
\\
	&\,
	\phantom{+}
	\times
	\frac{2}{\left[Dk^2+\tilde{\Delta}_M^{(-IR)}(0) - i\omega - i\Omega\right]}
\\
	&\,
	\phantom{+}
	\times
	\int\limits^{(-IR)}_{\vex{l}}
	\Delta_M(\vex{l})
	\int\limits_{\nu}\tilde{\cc}
	\left(\frac{\omega-\nu}{2},\frac{\omega+\nu}{2},\vex{k}-\vex{l}\right)
	\end{aligned}
\right\}.
\end{align}
Integrating over $\Omega$ we find the reduced equation
\begin{align}
\label{SelfConsistentII}
	\tilde{\cc}_R(\omega,\vex{k}) 
	=&\, 
	\frac{1}{Dk^2+\tilde{\Delta}_M^{(-IR)}(0)-i\omega} 
\nonumber\\
	&\,
	+ 
	\frac{1}{Dk^2+\tilde{\Delta}_M^{(-IR)}(0)-i\omega}
\nonumber\\
	&\,
	\phantom{+}
	\times
	\int\limits^{(-IR)}_{\vex{l}}
	\,
	\Delta_M(\vex{l})
	\,
	\tilde{\cc}_R(\omega,\vex{k}-\vex{l}).
\end{align}
Defining
\begin{align}
    \cc_R(\eta,\vex{\rho}) \equiv \frac{D}{2}\int\limits_{\omega,\vex{k}}e^{-i\omega\eta}e^{i\vex{k}\cdot\vex{\rho}}\tilde{\cc}_R(\omega,\vex{k}),
\end{align}
the position space formulation of Eq.~(\ref{SelfConsistentII}) is
\begin{align}
\label{BackToAAK}
    \left[\partial_\eta - D\nabla_{\rho}^2 + \tilde{\Delta}_M(0)-\tilde{\Delta}_M(\vex{\rho})\right]\cc_R(\eta,\vex{\rho}) = \frac{D}{2} \delta(\eta)\delta(\vex{\rho}).
\end{align}
This states that $\cc_R$ is the imaginary-time propagator for the single-particle 
quantum mechanics Hamiltonian $\hat{h}$, recovering the AAK reduction in Eq.~(\ref{AAKReductionIII}). 
We see out that the ``$\tilde{\Delta}_M(0)$,'' term arises from the type-I ``pseudo self-energy'' while the ``$\tilde{\Delta}_M(\vex{\rho})$'' term arises from the self-consistent treatment of the type II phonon.


\subsection{Coexisting interaction baths}

Finally, we briefly note that the theory for the coexisting diffusive and screened Coulomb baths 
can also be treated in the field theory language. In this case, one defines two distinct species of noise phonon, 
one for each noise bath. (Each noise bath will have both type I and type II phonons.) 
As before, the replica limit enforces the topological constraints explained by Fig.~\ref{GenericFieldTheoryDiagram}; 
the Cooperon is given by all diagrams with a single fermion line dressed by any combination of the four distinct noise phonons. 
It turns out that the special vanishing rules discussed in Sec.~\ref{Rules} still apply in this more general scenario, 
though only to the phonons generated by the Markovian Coulomb bath. The perturbative calculations carried out in the main 
text thus correspond to an exact (though asymptotic) partially-infinite-order summation over the diagrams with arbitrarily 
many Coulomb phonons (restricted by the vanishing rules), but up to two diffusive phonons. 
We note that the AAK transformation of variables Eq.~(\ref{Change}) in the 
single-particle 
path integral formalism allows us to get this result 
directly in terms of the Airy eigenfunction summations exploited in Sec.~\ref{CalculationDetails}.



\begin{thebibliography}{99}
\bibitem{AAK}
	B. L. Altshuler, A. G. Aronov, and D. E. Khmelnitsky, 
	Effects of electron-electron collisions with small energy transfers on quantum localisation,
	J. Phys. C. {\bf 15}, 7367 (1982). 
\bibitem{LeeRama}
	P. A. Lee and T. V. Ramakrishnan,
	Disordered Electron Systems,
	Rev. Mod. Phys. \textbf{57}, 287 (1985).
\bibitem{AA85}
	B. L. Altshuler and A. G. Aronov,
	Electron-electron interaction in disordered conductors,
	in 
	\textit{Electron-Electron Interactions in Disordered Systems},
	edited by M. Pollak and A. L. Efros
	(North-Holland, Amsterdam, 1985).
\bibitem{AAG}
	I. L. Aleiner, B. L. Altshuler, and M. E. Gershenson, 
	Interaction effects and phase relaxation in disordered systems,
	Waves Random Media {\bf 9}, 201 (1999).
\bibitem{Webb1}
	P. Mohanty, E. M. Q. Jariwala and R. A. Webb,
	Intrinsic decoherence in mesoscopic systems,
	Phys. Rev. Lett. {\bf 78} 3366 (1997).
\bibitem{Webb2}
	P. Mohanty and R. A. Webb,
	Decoherence and quantum fluctuations,
	Phys. Rev. B {\bf 55} 13452 (1997).
\bibitem{Webb3}
	P. Mohanty and R. A. Webb,
	Low temperature anomaly in mesoscopic Kondo wires,
	Phys. Rev. Lett. {\bf 84} 4481 (2000).
\bibitem{GZ}
	D. S. Golubev and A. D. Zaikin, 
	Quantum decoherence in disordered mesoscopic systems,
	Phys. Rev. Lett. {\bf 81} 1074 (1998).
\bibitem{Natelson}
	D. Natelson, R. L. Willett, K. W. West, and L. N. Pfeiffer,
	Geometry-dependent dephasing in small metallic wires,
	Phys. Rev. Lett. {\bf 86} 1821 (2001).
\bibitem{VDelftPartI}
	F. Marquardt, J. von Delft, R. A. Smith, and V. Ambegaokar,
	Decoherence in weak localization. I. Pauli principle in influence functional,
	Phys. Rev. B. \textbf{76}, 195331, (2007).
\bibitem{VDelftPartII}
	F. Marquardt, J. von Delft, R. A. Smith, and V. Ambegaokar,
	Decoherence in weak localization. II. Bethe-Salpeter calculation of the cooperon,
	Phys. Rev. B. \textbf{76}, 195332, (2007).
\bibitem{ICS}
	G. Zarand, L. Borda, J. von Delft, and N. Andrei,
	Theory of inelastic scattering from magnetic impurities,
	Phys. Rev. Lett. {\bf 93} 107204 (2004).
\bibitem{KondoTheory1}
	T. Micklitz, A. Altland, T.A. Costi, and A. Rosch,
	Universal dephasing rate due to diluted Kondo impurities,
	Phys. Rev. Lett. {\bf 96} 226601 (2006).
\bibitem{KondoTheory2}
	T. Micklitz, T.A. Costi, and A. Rosch,
	Magnetic field dependence of dephasing rate due to diluted Kondo impurities,
	Phys. Rev. B {\bf 75} 054406 (2007).
\bibitem{KondoMeasure1}
	G. M. Alzoubi and N. O. Birge,
	Phase coherence of conduction electrons below the Kondo temperature,
	Phys. Rev. Lett. {\bf 97} 226803 (2006).    
\bibitem{KondoMeasure2}
	F. Mallet, J. Ericsson, D. Mailly, S. \"Unl\"ubayir, D. Reuter, A. Melnikov, 
	A. D. Wieck, T. Micklitz, A. Rosch, T. A. Costi, L. Saminadayar, and C. B\"auerle,
	Scaling of the low-temperature dephasing rate in Kondo systems,
	Phys. Rev. Lett. {\bf 97} 226804 (2006).
\bibitem{NZA}
	B. N. Narozhny, G. Zala, and I. L. Aleiner,
	Interaction corrections at intermediate temperatures: Dephasing time,
	Phys. Rev. B {\bf 65}, 180202(R) (2002).
\bibitem{BAA06}
	D. M. Basko, I. L. Aleiner, and B. L. Altshuler,
	Metal-insulator transition in a weakly interacting 
	many-electron system with localized single-particle states,
	Ann. Phys. (Amsterdam) {\bf 321}, 1126 (2006);	
	On the problem of many-body localization, 
	in 
	\textit{Problems of Condensed Matter Physics},
	edited by A. L. Ivanov and S. G. Tikhodeev
	(Oxford University Press, Oxford, England, 2007);
	arXiv:cond-mat/0602510.
\bibitem{Gornyi05}
	I. V. Gornyi, A. D. Mirlin, and D. G. Polyakov, 
	Interacting Electrons in Disordered Wires: Anderson Localization and Low-$T$ Transport,
	Phys. Rev. Lett. {\bf 95}, 206603 (2005).
\bibitem{NH}
	R. Nandkishore and D. A. Huse,
	Many-body localization and thermalization in quantum statistical mechanics,
	Annu. Rev. Condens. Matter Phys. {\bf 6}, 1538 (2015).
\bibitem{GopalParam}
	S. Gopalakrishnan and S. A. Parameswaran,
	Dynamics and Transport at the Threshold of Many-Body Localization,
	Phys. Rep. {\bf 862}, 1 (2020).
\bibitem{Liao18}
	Y. Liao and M. S. Foster,
	Dephasing Catastrophe in $4-\epsilon$ Dimensions: 
	A Possible Instability of the Ergodic (Many-Body-Delocalized) Phase,
	Phys. Rev. Lett. {\bf 120}, 236601 (2018).
\bibitem{Han18}
	Jae-Ho Han and Ki-Seok Kim,
	Boltzmann transport theory for many-body localization
	Phys. Rev. B {\bf 97}, 214206 (2018).
\bibitem{Prosen19}
	J. \v Suntajs, J. Bon\v ca, T. Prosen, and L. Vidmar,
	Quantum chaos challenges many-body localization,
	arXiv:1905.06345.
\bibitem{MBLDefense}
	D. A. Abanin, J. H. Bardarson, G. De Tomasi, S. Gopalakrishnan, V. Khemani, S. A. Parameswaran, F. Pollmann, A. C. Potter, M. Serbyn, and R. Vasseur,
	Distinguishing localization from chaos: challenges in finite-size systems,
	arXiv:1911.04501.
\bibitem{Panda19}
	R. K. Panda, A. Scardicchio, M. Schulz, S. R. Taylor, M. \v Znidari\v c
	Can we study the many-body localization transition?,
	Europhys. Lett. {\bf 128}, 67003 (2020).
\bibitem{PiotrSierant}
	P. Sierant, D. Delande, and J. Zakrzewski,
	Thouless time analysis of Anderson and many-body localization transitions,
	Phys. Rev. Lett. {\bf 124}, 186601 (2020).
\bibitem{Doggen20}
	E. V. H. Doggen, I. V. Gornyi, A. D. Mirlin, and D. G. Polyakov, 
	Slow many-body delocalization beyond one dimension,
	arXiv:2002.07635.	
\bibitem{Laumann14}
	N. Y. Yao, C. R. Laumann, S. Gopalakrishnan, M. Knap, M. M\"uller, E. A. Demler, and M. D. Lukin,
	Many-body localization in dipolar systems,
	Phys. Rev. Lett. {\bf 113}, 243002 (2014).
\bibitem{Liao17}
	Y. Liao, A. Levchenko, and M. S. Foster,
	Response theory of the ergodic many-body delocalized phase: 
	Keldysh Finkel'stein sigma models and the 10-fold way,
	Ann. Phys. (Amsterdam) {\bf 386}, 97 (2017).
\bibitem{BK}
	D. Belitz and T. R. Kirkpatrick,
	The Anderson-Mott transition,
	Rev. Mod. Phys. {\bf 66}, 261 (1994).
\bibitem{PunnooseFinkelstein}
	A. Punnoose and A. M. Finkel'stein,
	Metal-Insulator Transition in Disordered Two-Dimensional Electron Systems,
	Science \textbf{310}, 289 (2005).
\bibitem{50YearsFinkel}
	A. M. Finkel'stein,
	Disordered Electron Liquid with Interactions,
	in  
	\textit{50 Years of Anderson Localization}, 
	edited by E. Abrahams 
	(World Scientific, Singapore, 2010).
\bibitem{QPTKravchenko}
	S. V. Kravchenko,
	Metal-Insulator Transitions in Two-dimensional Electron Systems,
	in 
	\textit{Conductor-Insulator Quantum Phase Transitions}, 
	edited by V. Dobrosavljevi\'c, N. Trivedi, and J. M. Valles, Jr.
	(Oxford University Press, Oxford, 2012). 
\bibitem{CS}
	S. Chakravarty and A. Schmid, 
	Weak localization: The quasiclassical theory of electrons in a random potential,
	Phys. Rep. {\bf 140}, 193 (1986).
\bibitem{AleinerBlanter}
	I. L. Aleiner and Ya. M. Blanter, 
	Inelastic scattering time for conductance fluctuations,
	Phys. Rev. B {\bf 65}, 115317 (2002).	
\bibitem{AlexAlex}
	A. Kamenev and A. Levchenko, 
	Keldysh technique and non-linear $\sigma$-model: basic principles and applications,
	Adv. Phys. {\bf 58}, 197 (2009).
\bibitem{KamBook}
	A. Kamenev, 
	\textit{Field Theory of Non-Equilibrium Systems}
	(Cambridge University Press, Cambridge, England, 2011).
\bibitem{Footnote--NonChargeNeutral}
	The Dirichlet boundary conditions on $\vex{\rho}(\tau)$ in Eq.~(\ref{GeneralPathIntegralFolded-0}) 
	are important, since a perturbative expansion in $S_I$ [Eq.~(\ref{NoiseBathAction})] 
	produces non-charge-neutral vertex operator correlators in $\vex{\rho}$, 
	different from the usual situation \cite{BYB}.
\bibitem{BYB}
	See, e.g., Chapter 9 in 
	P. Di Francesco, P. Mathieu, and D. S\'en\'echal,
	\textit{Conformal Field Theory}
	(Springer-Verlag, New York, 1996).
\bibitem{GiamarchiBook}
	T. Giamarchi,
	\emph{Quantum Physics in One Dimension}
	(Clarendon Press, Oxford, England, 2003). 
\bibitem{AiryBook}
	O. Vall\'ee and M. Soares,
	\textit{Airy Functions and Applications to Physics}, 2nd ed.
	(Imperial College Press, London, England, 2010).
\bibitem{Feigelman07}
	M. V. Feigel'man, L. B. Ioffe, V. E. Kravtsov, and E. A. Yuzbashyan, 
	Eigenfunction Fractality and Pseudogap State near the Superconductor-Insulator Transition,
	Phys. Rev. Lett. {\bf 98}, 027001 (2007).
\bibitem{Baturina08}
	T. I. Baturina, A. Bilus\v i\'c, A. Yu. Mironov, V. M. Vinokur, M. R. Baklanov, and C. Strunk, 
	Quantum-critical region of the disorder-driven superconductor-insulator transition,
	Physica (Amsterdam) {\bf 468C}, 316 (2008).
\bibitem{Yazdani10}
	A. Richardella, P. Roushan, S. Mack, B. Zhou, D. A. Huse, D. D. Awschalom, and A. Yazdani, 
	Visualizing Critical Correlations Near the Metal-Insulator Transition in Ga$_{1-x}$Mn$_{x}$As,	
	Science {\bf 327}, 665 (2010).
\bibitem{Foster14}
	M. S. Foster, H.-Y. Xie, and Y.-Z. Chou, 
	Topological protection, disorder, and interactions: Survival at the surface of 3D topological superconductors,
	Phys. Rev. B {\bf 89}, 155140 (2014).
\bibitem{Cardy}
	J. Cardy, 
	\textit{Scaling and Renormalization in Statistical Physics} 
	(Cambridge University Press, Cambridge, England, 1996)
\end{thebibliography}
\end{document}